\documentclass[12pt]{amsart}
\usepackage{amssymb,amscd,verbatim}

\usepackage{amsmath,graphicx,mathrsfs}
\usepackage{xcolor}
\usepackage[colorlinks=true,allcolors=blue]{hyperref}\hypersetup{pdftitle={Root-of-Unity Limits of Trigonometric qKZ Connections},pdfauthor={Alexander Varchenko}}

\let\frak\mathfrak

\def\>{\relax\ifmmode\mskip.666667\thinmuskip\relax\else\kern.111111em\fi}
\def\<{\relax\ifmmode\mskip-.333333\thinmuskip\relax\else\kern-.0555556em\fi}
\def\vsk#1>{\vskip#1\baselineskip}
\def\vv#1>{\vadjust{\vsk#1>}\ignorespaces}
\def\vvn#1>{\vadjust{\nobreak\vsk#1>\nobreak}\ignorespaces}

\let\sssize\scriptscriptstyle \def\tfrac{\textstyle\frac}

\let\Maketitle\maketitle
\def\maketitle{\Maketitle\thispagestyle{empty}\let\maketitle\empty}

\newtheorem{thm}{Theorem}[section]
\newtheorem{cor}[thm]{Corollary}
\newtheorem{lem}[thm]{Lemma}
\newtheorem{prop}[thm]{Proposition}

\theoremstyle{definition}

\numberwithin{equation}{section}

\theoremstyle{definition}
\newtheorem*{rem}{Remark}
\newtheorem*{example}{Example}
\providecommand{\exno}{}

\let\mc\mathcal
\let\nc\newcommand

\let\al\alpha

\let\dl\delta
\let\Dl\Delta
\let\eps\varepsilon
\let\gm\gamma

\let\ka\kappa
\let\la\lambda
\let\La\Lambda

\let\phi\varphi
\let\si\sigma

\let\tht\theta

\let\om\omega
\let\Om\Omega

\let\ox\otimes

\let\ge\geqslant

\let\le\leqslant

\let\on\operatorname

\let\bs\boldsymbol

\def\C{{\mathbb C}}
\def\Z{{\mathbb Z}}
\def\R{{\mathbb R}}

\def\F{{\mc F}}

\def\End{\on{End}}

\def\id{\on{id}}
\def\Res{\on{Res}}

\def\tr{\on{tr}}

\def\0{{\sssize0}}

\def\beq{\begin{equation}}
\def\eeq{\end{equation}}
\def\be{\begin{equation*}}
\def\ee{\end{equation*}}

\nc{\bea}{\begin{eqnarray*}}
\nc{\eea}{\end{eqnarray*}}
\nc{\bean}{\begin{eqnarray}}
\nc{\eean}{\end{eqnarray}}

\nc{\Il}{{\mc I_{\bs\la}}}
\nc{\bla}{{\bs\la}}
\nc{\Fla}{\F_\bla}
\nc{\tfl}{{T^*\Fla}}
\nc{\GL}{{GL_n(\C)}}
\nc{\GLC}{{GL_n(\C)\times\C^*}}

\def\KZ/{{\slshape KZ\/}}
\def\qKZ/{{\slshape qKZ\/}}
\def\XXX/{{\slshape XXX\/}}

\nc{\A}{{\mc C}}

\def\slt{{\frak{sl}_2}}

\newcommand{\Uq}{U_q(\slt)}

\begin{document}

\title[Root-of-Unity Limits of Trigonometric \texorpdfstring{$\mathrm{\lowercase{q}}$KZ}{qKZ} Connections]
{Root-of-Unity Limits of Trigonometric \texorpdfstring{$\mathrm{\lowercase{q}}$KZ}{qKZ} Connections}
\author{Alexander Varchenko}
\address{Department of Mathematics, University of North Carolina at Chapel Hill,
Chapel Hill, NC 27599-3250, USA}
\email{anv@email.unc.edu}

\begin{abstract}
Over the field of complex numbers, we study root-of-unity
limits of trigonometric qKZ connections associated with tensor
products of $U_q(\mathfrak{sl}_2)$ Verma modules and of their
hypergeometric solutions. As the
multiplicative step approaches a primitive $m$-th root of unity
and the quantum-group and twist parameters approach $1$, the
$m$-step qKZ transport yields, on every fixed-degree subspace,
a differential connection. We show that this connection is the
pullback of the trigonometric KZ connection under the power map
$z\mapsto z^m$, with the KZ parameter rescaled by $m$, and
establish formal and $\ell$-adic versions.

In degree one, with all highest weights equal to $1$, we
determine the limits of normalized qKZ hypergeometric circle
integrals in arbitrary angular chambers. The limits are
expressed in terms of loaded hypergeometric integrals for the
trigonometric KZ equations and explicit chamber-dependent
upper-triangular matrices. We establish compatibility of the
resulting distinguished fundamental matrices under inward
analytic continuation, governed by the right weak Bruhat order,
and under power maps, including composition of root orders.
\end{abstract}

\maketitle 

\begingroup
\linespread{0.94}\selectfont
\tableofcontents
\endgroup

\section*{Introduction}

Over the field of complex numbers, we study root-of-unity limits
of the trigonometric qKZ connection associated with tensor products
of $\Uq$ Verma modules and of a distinguished family of its
hypergeometric solutions. The
multiplicative step $p$ approaches a root of unity, while the
quantum-group parameter $q$ and the twist parameter $\varkappa$
approach $1$. We identify the resulting differential connections,
calculate the limits of the hypergeometric solutions with their
exact normalizations, and describe how these limits are related
under power changes of variables and analytic continuation.

The hypergeometric integral solutions of the trigonometric qKZ
equations used here come from \cite[Section~5]{TV97}. They are
constructed from $q$-Pochhammer phase functions and trigonometric
and elliptic weight functions. Their classical counterparts are
hypergeometric solutions of the trigonometric KZ equations; see
\cite[Section~2.2]{TV00} for the connection and
\cite[Section~4]{MaV02} for the integral construction. We normalize the $R$-matrix to fix the tensor product of
highest-weight vectors and the classical $r$-operator to
annihilate the corresponding tensor product. The precise
relations with the conventions of these papers are recorded in
Sections~\ref{sec:connections}--\ref{sec:hyper-kz}.

The quasiclassical degeneration of rational qKZ
equations into KZ differential equations, together with
the corresponding limits of hypergeometric integral
solutions, was studied in
\cite[Sections~7--8]{TVInv97};
see also \cite[Appendix~A.1]{MV24}.
The present paper concerns root-of-unity limits of
the trigonometric qKZ connection and its
hypergeometric solutions.

Let $m$ be a positive integer and let $\zeta_m$ be a primitive
$m$th root of unity. Set
\[
 \pi_m(z)=z^m=(z_1^m,\ldots,z_n^m),\qquad
 \mc U_m=\{z\in(\C^\times)^n:z_i^m\ne z_j^m
                                  \text{ for }i\ne j\}.
\]
The first result concerns arbitrary complex highest weights
$\La_1,\ldots,\La_n$ and every fixed-degree subspace
$\mathbf V_r$ of the tensor product of classical Verma modules,
identified with its quantum counterpart. Let
$\ka\in\C^\times$ and $\mu\in\C$, and suppose that the qKZ
parameters depend holomorphically on $\delta$ near $0$, with
\[
 \begin{aligned}
 p&=\zeta_m\bigl(1+\delta+O(\delta^2)\bigr),\\
 q&=1-\frac{\delta}{2\ka}+O(\delta^2),\qquad
 \varkappa=1-\frac{2\mu\delta}{\ka}+O(\delta^2).
 \end{aligned}
\]
An $m$-fold iteration of the difference equation changes $z_i$
to $p^m z_i$, where $p^m=1+m\delta+O(\delta^2)$. Thus its
first-order term defines a differential equation. More precisely,
Theorem~\ref{thm:limit} gives the expansion
\[
 \widetilde K_i^{(m)}(z)
 =\id+\frac{m\delta}{\ka}H_i(z^m;\mu)+O(\delta^2),
 \qquad z\in\mc U_m,
\]
where $\widetilde K_i^{(m)}$ is the ordered $m$-step qKZ
transport and $H_i$ is the normalized trigonometric KZ Hamiltonian
defined in \eqref{eq:H}. Consequently the limiting connection is
\[
 \nabla_m(\ka,\mu)
 =\mathrm{d}-\frac1\ka\sum_{i=1}^n
 H_i(z^m;\mu)\frac{\mathrm{d}z_i}{z_i}
 =\pi_m^*\nabla_1(m\ka,\mu).
\]
The order of the root of unity therefore enters through the
coordinatewise power map and the rescaling $\ka\mapsto m\ka$.
The connection depends only on $m$, independently of the choice
of the primitive root $\zeta_m$. The proof combines the
semiclassical expansion of the normalized $R$-matrix with
cyclotomic averaging of the classical trigonometric $r$-operator.

This calculation also has a formal interpretation over any
field of characteristic zero containing $\zeta_m$ and the
parameters. Theorem~\ref{thm:formal-root-limit} establishes the
same first-order identity in the formal variable $\delta$.
When the field is a finite extension of $\mathbb Q_\ell$ and
the parameter functions are analytic, the identity yields an
$\ell$-adic limit of the finite transport matrices, as stated in
Corollary~\ref{cor:ell-adic-limit}.

For the analytic limits of hypergeometric solutions, we
specialize to the degree-one subspace with
$\La_1=\cdots=\La_n=1$, $n\ge2$, and the real parameter domain
\[
 \ka<-n,\qquad 0<\mu<\frac18.
\]
Let $\Psi=(\Psi_1,\ldots,\Psi_n)$ be the coordinate matrix of
the fixed circle-integral solutions of
Section~\ref{sec:hyper-qkz}. These integrals specify a particular
complex span of qKZ solutions. Their fixed
normalization is part of the limit problem: a general change of
fundamental matrix for a difference equation may involve
meromorphic quasiconstants, as explained in that section.
We take the radial limit
\[
 p=\zeta_m r,\qquad q=r^{-1/(2\ka)},\qquad
 \varkappa=r^{-2\mu/\ka},\qquad r\longrightarrow1^-.
\]
For every permutation $\si\in S_n$, the variables lie in the
angular chamber
\[
 \mc D_{m,\si}=\left\{z_a=e^{\mathrm{i}\vartheta_a}:\
 0<\vartheta_{\si(1)}<\cdots<\vartheta_{\si(n)}
 <\frac{2\pi}{m}\right\}.
\]
Theorem~\ref{thm:hyper-limit-all-chambers} proves that
\[
 \begin{aligned}
 B_{m,\si}(z;\ka,\mu)
 &:=\lim_{r\to1^-}\Psi(z;p,q,\varkappa)\\
 &=\mc I_\si(z^m;m\ka,\mu)\,
                         C_{m,\si}^{(n)}(\ka,\mu).
 \end{aligned}
\]
Here $\mc I_\si$ is the matrix of trigonometric KZ solutions
given by the loaded radial intervals from $0$ to $z_j^m$.
The loadings specify the branches of the multivalued factors
in these integrals. The comparison matrix
$C_{m,\si}^{(n)}$ is independent of $z$ and upper triangular
in the fixed numerical labels. Its entries are given explicitly
by trigonometric and exponential functions in
\eqref{eq:hyp-limit-C-sigma}; their dependence on the chamber
is expressed by the relative angular orders of the marked points.
They satisfy
\[
 C_{m,\si}^{(n)}(\ka,\mu)
 =C_{1,\si}^{(n)}(m\ka,\mu).
\]

The convergence is locally uniform in the angular chamber and
on compact subsets of sufficiently small complex neighborhoods
with the prescribed branches. Both $\mc I_\si$ and
$C_{m,\si}^{(n)}$ are invertible for every parameter pair in
the stated domain. The limiting columns therefore form a
fundamental matrix of local flat sections of
$\nabla_m(\ka,\mu)$. For the radial integral matrix,
invertibility follows from the determinant calculation in
Proposition~\ref{prop:kz-basis}; for the comparison matrix,
it follows from its explicit diagonal entries.

The proof of the integral limit retains the normalization of
the original circle integrals. We express these integrals as
sums of residues over inward pole strings and separate each
string into its cyclotomic components. Limits of
$q$-Pochhammer products and theta functions, together with
uniform Jackson beta estimates, identify the resulting sums
with the radial KZ integrals and give the comparison
coefficients. This also determines the branches needed to
compare different chambers and different root orders.

The distinguished bases in different angular chambers are
identified by prescribed inward analytic continuation. When two
consecutive entries in the angular order occur as $(a,b)$ with
$a<b$, we move the point $z_a$ slightly inside the unit circle,
past $z_b$, and back to the circle. These directed exchanges
give the covering relations of the right weak Bruhat order on
$S_n$. Theorem~\ref{thm:if-main} shows that, whenever
$\si\le_R\tau$, continuation along any directed chain of
the prescribed exchanges gives
\[
 \operatorname{Cont}_{\gm}B_{m,\si}=B_{m,\tau}.
\]
This is an equality of fundamental matrices of holomorphic
solution germs, and the identification is independent of the
chosen directed chain. The distinguished bases form a global
flat frame on the space $X$ obtained by joining the chambers
along the chosen inward paths. Section~\ref{sec:inward-formal}
also describes the topology of this space and its relation to
the permutohedron.

Finally, the limiting connections and the distinguished bases
satisfy parallel compatibility identities. For positive
integers $a,b$, on $\mc U_{ab}$ one has
\[
 \nabla_{ab}(\ka,\mu)
 =\pi_b^*\nabla_a(b\ka,\mu)
 =\pi_a^*\nabla_b(a\ka,\mu).
\]
For the degree-one solutions in the stated real parameter
domain, on $\mc D_{ab,\si}$ with the prescribed branches,
\[
 B_{ab,\si}(z;\ka,\mu)
 =B_{a,\si}(z^b;b\ka,\mu)
 =B_{b,\si}(z^a;a\ka,\mu).
\]
In particular,
\[
 B_{m,\si}(z;\ka,\mu)
 =B_{1,\si}(z^m;m\ka,\mu).
\]
These bases are independent of the choice of primitive root
of a fixed order. Any finite collection of root orders can be
compared on a common angular chamber, and the power-map
identities commute with the prescribed inward continuations.
These identities express functorial compatibility of the
root-of-unity limits: the limits form one compatible family of
differential connections with distinguished hypergeometric
bases.

A related phenomenon appears in the work of Garoufalidis and
Wheeler \cite{GW25}. In their study of root-of-unity expansions of
$q$-hypergeometric series in a $p$-adic setting, classical
hypergeometric functions arise through power changes of variables
and explicit differential operators; see
\cite[Section~1.4, (22)--(23)]{GW25}.

The paper is organized as follows.
Section~\ref{sec:connections} fixes the quantum-group,
$R$-matrix, and connection conventions.
Section~\ref{sec:connection-limits} proves the connection
limit and its formal and $\ell$-adic versions.
Sections~\ref{sec:hyper-qkz} and~\ref{sec:hyper-kz} describe
the qKZ and KZ hypergeometric solutions and their determinants.
Section~\ref{sec:hyper-limit} proves the limits in arbitrary
angular chambers.
Section~\ref{sec:inward-formal} establishes the inward
continuation theorem and studies the associated chamber space.
Section~\ref{sec:functorial-compatibility} collects the
power-map identities for the limiting connections and distinguished
bases, and their compatibility with inward continuation.

\section{Quantum groups and the qKZ and KZ connections}\label{sec:connections}

\subsection{The quantum group and its Verma modules}\label{sec:quantum}

Let $q$ belong to a sufficiently small punctured neighborhood of $1$ in $\C$.
The quantum group $\Uq$ is the algebra generated by
$e,f,q^h,q^{-h}$, subject to the relations
\begin{align}
 q^h q^{-h}&=q^{-h}q^h=1,\nonumber\\
 q^{\pm h}e&=q^{\pm2}e q^{\pm h},\qquad
 q^{\pm h}f=q^{\mp2}f q^{\pm h},\label{eq:quantum-rel}\\
 [e,f]&=\frac{q^h-q^{-h}}{q-q^{-1}}.\nonumber
\end{align}
We use the coproduct
\begin{align}
 \Dl(q^{\pm h})&=q^{\pm h}\ox q^{\pm h},\nonumber\\
 \Dl(e)&=e\ox q^{-h}+1\ox e,\label{eq:coproduct}\\
 \Dl(f)&=f\ox1+q^h\ox f.\nonumber
\end{align}
For $\La\in\C$, let $q^\La$ denote the holomorphic determination
near $q=1$ normalized by $1^\La=1$, and let $M_\La(q)$ be the
Verma module with highest-weight vector $v_\La$ satisfying
\[
 ev_\La=0,\qquad q^{\pm h}v_\La=q^{\pm\La}v_\La.
\]
A basis is $\{f^k v_\La:k\ge0\}$.  With
$[x]_q=(q^x-q^{-x})/(q-q^{-1})$, we have
\begin{align*}
 q^h f^k v_\La&=q^{\La-2k}f^k v_\La,\\
 e f^k v_\La&=[k]_q[\La-k+1]_q f^{k-1}v_\La,\qquad k\ge1.
\end{align*}

Fix a positive integer $n$ and weights $\bs\La=(\La_1,\ldots,\La_n)$.
Set $\mc M(q)=M_{\La_1}(q)\ox\cdots\ox M_{\La_n}(q)$,
with the tensor product action given by \eqref{eq:coproduct}.
For $r\in\Z_{\ge0}$, let $\mc M_r(q)$ be the subspace
spanned by the vectors
\beq\label{eq:basis}
 f^{\mathbf a}=f^{a_1}v_{\La_1}\ox\cdots\ox
 f^{a_n}v_{\La_n},\qquad
 a_i\in\Z_{\ge0},\quad a_1+\cdots+a_n=r.
\eeq
Thus $\mc M(q)=\bigoplus_{r\ge0}\mc M_r(q)$.
The tensor product operator $q^h\ox\cdots\ox q^h$
acts on $\mc M_r(q)$ by the scalar $q^{\sum_i\La_i-2r}$.
Each $\mc M_r(q)$ is finite dimensional.

For $1\le i\le n$, define the operator $h^{(i)}$
on $\mc M(q)$ by
\beq\label{eq:hi}
 h^{(i)}f^{\mathbf a}=(\La_i-2a_i)f^{\mathbf a}.
\eeq
\vsk1>
Let the Lie algebra $\slt$ have generators $\mathbf e,\mathbf f,\mathbf h$ with
$[\mathbf h,\mathbf e]=2\mathbf e$,
$[\mathbf h,\mathbf f]=-2\mathbf f$, and
$[\mathbf e,\mathbf f]=\mathbf h$.
Let $\mathbf M_\La$ be its Verma module with highest-weight
vector $\mathbf v_\La$.  In particular,
\[
 \mathbf h\,\mathbf f^a\mathbf v_\La
 =(\La-2a)\mathbf f^a\mathbf v_\La,
 \qquad
 \mathbf e\,\mathbf f^a\mathbf v_\La
 =a(\La-a+1)\mathbf f^{a-1}\mathbf v_\La.
\]
Set $\mathbf V=\bigotimes_{j=1}^n\mathbf M_{\La_j}$, and let
$\mathbf h^{(i)}$ denote the action of $\mathbf h$ on its
$i$th tensor factor.  Let
\[
 \mathbf V_r=\mathbf V[\La_1+\cdots+\La_n-2r]
\]
be the eigenspace of $\mathbf h^{(1)}+\cdots+\mathbf h^{(n)}$
with eigenvalue $\La_1+\cdots+\La_n-2r$.
We identify $\mc M_r(q)$ with $\mathbf V_r$ by matching
$f^{\mathbf a}$ with
$\mathbf f^{\mathbf a}
 =\mathbf f^{a_1}\mathbf v_{\La_1}\ox\cdots\ox
 \mathbf f^{a_n}\mathbf v_{\La_n}$.
Under this identification, $\mathbf h^{(i)}$ equals $h^{(i)}$.
As $q\to1$, the matrices of the quantum operators $e$ and $f$
(between the appropriate fixed-degree subspaces) tend to the
matrices of the classical operators $\mathbf e$ and $\mathbf f$,
respectively, under the identification above.  The operators
$e$ and $f$ change total degree by $-1$ and $+1$, respectively.
On each fixed degree of each individual tensor factor,
\[
 q^{\pm h}=\id\pm(q-1)\mathbf h+O((q-1)^2).
\]

\begin{samepage}
\subsection{Trigonometric \texorpdfstring{$R$}{R}-matrices and the qKZ connection}\label{sec:qkz}

\subsubsection{The normalized \texorpdfstring{$R$}{R}-matrix}

Fix $\La_a,\La_b\in\C$.  Let $R_{\La_a,\La_b}(x;q)$
denote the normalized trigonometric $R$-matrix on
$M_{\La_a}(q)\ox M_{\La_b}(q)$.  It preserves the degree
decomposition and, for generic $x$, is characterized by
\begin{align}
 R(x;q)\bigl(f\ox1+q^h\ox f\bigr)
   &=\bigl(f\ox q^h+1\ox f\bigr)R(x;q),\label{eq:R1}\\
 R(x;q)\bigl(f\ox q^h+x\,1\ox f\bigr)
   &=\bigl(f\ox1+x\,q^h\ox f\bigr)R(x;q),\label{eq:R2}\\
 R(x;q)(v_{\La_a}\ox v_{\La_b})
   &=v_{\La_a}\ox v_{\La_b}.\label{eq:R3}
\end{align}
\end{samepage}
On each fixed degree space, it is a rational function of $x$.
These are the conventions of \cite[Section~3, (3.3)--(3.4)]{TV97},
after the change of generators
$e=q^{-H}E$, $f=Fq^H$, $q^h=q^{2H}$, with
$\La_{\mathrm{here}}=2\La_{\mathrm{TV97}}$.
We write $R^{(i,j)}(x;q)$ for $R_{\La_i,\La_j}(x;q)$
acting in the $i$th and $j$th tensor factors in that order,
and as the identity on the remaining factors.

\subsubsection{The difference connection}

Let $z=(z_1,\ldots,z_n)\in(\C^\times)^n$ and
$p,\varkappa\in\C^\times$.  For $1\le i\le n$ define
\begin{align}
 K_i(z;p,q,\varkappa)
 &=R^{(i,i-1)}\!\left(\frac{p z_i}{z_{i-1}};q\right)
   \cdots R^{(i,1)}\!\left(\frac{p z_i}{z_1};q\right)
   \varkappa^{(\La_i-h^{(i)})/2}\nonumber\\[-2pt]
 &\qquad\times
   R^{(i,n)}\!\left(\frac{z_i}{z_n};q\right)
   \cdots R^{(i,i+1)}\!\left(\frac{z_i}{z_{i+1}};q\right).
 \label{eq:K}
\end{align}
Empty products are interpreted as the identity.
On the basis \eqref{eq:basis}, the diagonal twist acts by
\[
 \varkappa^{(\La_i-h^{(i)})/2}f^{\mathbf a}
 =\varkappa^{a_i}f^{\mathbf a},
\]
so only integer powers occur and no choice of branch is needed.
The parameter $p$ is the multiplicative step, and $\varkappa$ is
the twist; both are distinct from the KZ parameter $\ka$ below.
All the operators $K_i$ preserve $\mc M_r(q)$.

For an integer $s\ge0$, set
\[
 z^{(i,s)}=(z_1,\ldots,z_{i-1},p^s z_i,z_{i+1},\ldots,z_n).
\]
The trigonometric qKZ equations on $\mc M_r(q)$ are
\beq\label{eq:qkz}
 F(z^{(i,1)})=K_i(z;p,q,\varkappa)F(z),
 \qquad i=1,\ldots,n.
\eeq
These qKZ equations were introduced in \cite{FR92}; for the
present normalization see \cite[Section~3]{TV97}.
The trigonometric Yang--Baxter relations give the flatness
relations (see \cite[Theorem~3.13]{TV97})
\[
 K_i(z^{(j,1)})K_j(z)=K_j(z^{(i,1)})K_i(z),
 \qquad i,j=1,\ldots,n.
\]
For a positive integer $m$, the $m$-step qKZ transport in the
$i$th direction is the ordered product
\beq\label{eq:mtransport}
 \widetilde K_i^{(m)}(z)
 =K_i(z^{(i,m-1)})\cdots K_i(z^{(i,1)})K_i(z).
\eeq
Here the parameters of $K_i$ remain fixed, while only $z_i$ is
shifted.  In particular, a qKZ flat section satisfies
\beq\label{eq:m-qkz}
 F(z^{(i,m)})=\widetilde K_i^{(m)}(z)F(z).
\eeq

\subsection{The classical trigonometric KZ connection}\label{sec:kz}

For two $\slt$-Verma modules we use the normalized
classical trigonometric $r$-operator
\beq\label{eq:Omega}
 \Om_{\La_a,\La_b}(x)
 =\frac{1}{x-1}\left(
   x\,\mathbf e\ox \mathbf f+\mathbf f\ox \mathbf e
   +\frac{x+1}{4}(\mathbf h\ox \mathbf h-\La_a\La_b)\right).
\eeq
Our normalization is chosen so that
$\Om_{\La_a,\La_b}(x)$ annihilates
$\mathbf v_{\La_a}\ox\mathbf v_{\La_b}$.  For comparison with
\cite[Sections~2.1--2.2]{TV00}, the corresponding unnormalized
trigonometric $r$-operator, in the standard $\slt$ normalization, is
\[
 r^{\mathrm{TV}}_{\La_a,\La_b}(x)
 =\Om_{\La_a,\La_b}(x)
 +\frac{x+1}{4(x-1)}\La_a\La_b\,\id.
\]
Thus the difference is the explicit scalar term.  As above,
$\Om^{(i,j)}_{\La_i,\La_j}(x)$ acts on $\mathbf V$ in the
$i$th and $j$th factors and preserves every $\mathbf V_r$.

Let $Z=(Z_1,\ldots,Z_n)\in(\C^\times)^n$ with
$Z_i\ne Z_j$ for $i\ne j$.  Fix $\mu\in\C$.  The trigonometric
KZ Hamiltonians in the normalization used here are the
$\End(\mathbf V_r)$-valued functions
\beq\label{eq:H}
 H_i(Z;\mu)=\mu(\mathbf h^{(i)}-\La_i)
 +\sum_{j\ne i}\Om_{\La_i,\La_j}^{(i,j)}
       \left(\frac{Z_i}{Z_j}\right),\qquad i=1,\ldots,n.
\eeq
To explain the relation to \cite[Section~2.2]{TV00},
its Cartan parameter is specialized here to
$\lambda=\mu\mathbf h$.  With this choice, the unnormalized
Hamiltonian of \cite{TV00} is
\[
 H_i^{\mathrm{TV}}(Z;\mu)
 =\mu\mathbf h^{(i)}
 +\sum_{j\ne i}\bigl(r^{\mathrm{TV}}_{\La_i,\La_j}
 (Z_i/Z_j)\bigr)^{(i,j)}.
\]
The difference between this Hamiltonian and ours is scalar:
\[
 H_i(Z;\mu)=H_i^{\mathrm{TV}}(Z;\mu)
 -\left(\mu\La_i+\sum_{j\ne i}
 \frac{\La_i\La_j(Z_i+Z_j)}{4(Z_i-Z_j)}\right)\id.
\]
In particular, our normalization gives
$H_i(Z;\mu)(\mathbf v_{\La_1}\ox\cdots\ox\mathbf v_{\La_n})=0$.
For $\ka\in\C^\times$, the corresponding trigonometric KZ
equations for $\mathbf V_r$-valued functions $G(Z)$ are
\beq\label{eq:KZ}
 \ka Z_i\frac{\partial G}{\partial Z_i}
 =H_i(Z;\mu)G,\qquad i=1,\ldots,n.
\eeq
Equivalently, they define the flat connection
\beq\label{eq:nabla}
 \nabla_1(\ka,\mu)
 =\mathrm{d}-\frac1\ka\sum_{i=1}^n
 H_i(Z;\mu)\,\frac{\mathrm{d}Z_i}{Z_i}
\eeq
on the trivial bundle with fiber $\mathbf V_r$ over
$\{Z\in(\C^\times)^n:Z_i\ne Z_j\text{ for }i\ne j\}$.

\section{Root-of-unity limits of the qKZ connection}\label{sec:connection-limits}

\subsection{The root-of-unity limit}\label{sec:theorem}

Fix a positive integer $m$ and a primitive $m$th root of unity
$\zeta_m$.  Let $p$, $q$, and $\varkappa$ depend holomorphically
on $\delta$ near $0$ such that, as $\delta\to0$,
\beq\label{eq:scaling}
\begin{aligned}
 p&=\zeta_m\bigl(1+\delta+O(\delta^2)\bigr),\\
 q&=1-\frac{\delta}{2\ka}+O(\delta^2),\\
 \varkappa&=1-\frac{2\mu\delta}{\ka}+O(\delta^2).
\end{aligned}
\eeq
In particular, $p^m=1+m\delta+O(\delta^2)$.
Only the zeroth- and first-order terms in \eqref{eq:scaling}
enter the limit.  The twist tends to the identity.
Define
\[
 \pi_m(z)=(z_1^m,\ldots,z_n^m),\qquad
 \mc U_m=\{z\in(\C^\times)^n:z_i^m\ne z_j^m\ \text{for}\ i\ne j\}.
\]
\begin{thm}[Limit of the qKZ connection]\label{thm:limit}
Let $\La_1,\ldots,\La_n\in\C$, $\ka\in\C^\times$,
and $\mu\in\C$.  Under the identification
$\mc M_r(q)\simeq\mathbf V_r$, the $m$-step qKZ operators
\eqref{eq:mtransport} with parameters \eqref{eq:scaling}
have, for each fixed $z\in\mc U_m$, the expansion
\beq\label{eq:main-expand}
 \boxed{\quad
 \widetilde K_i^{(m)}(z)
 =\id+\frac{m\delta}{\ka}H_i(\pi_m(z);\mu)+O(\delta^2),
 \quad i=1,\ldots,n.\quad}
\eeq
The corresponding first-order differential connection on
$\mc U_m$ is
\beq\label{eq:nabla-m}
 \nabla_m(\ka,\mu)
 :=\mathrm{d}-\frac1\ka\sum_{i=1}^n
 H_i(\pi_m(z);\mu)\frac{\mathrm{d}z_i}{z_i}.
\eeq
It is the pullback of the trigonometric KZ connection:
\beq\label{eq:pullback}
 \boxed{\qquad
 \nabla_m(\ka,\mu)=\pi_m^*\nabla_1(m\ka,\mu).
 \qquad}
\eeq
\end{thm}

\begin{rem}
The first-order expansion of the $m$-step qKZ monodromy
for fixed $p=\zeta_m$ and $q\to1$ was studied in
\cite[Appendix~A]{TV26}. For the specialization
$\vartheta=1$ in that paper, the corresponding
cyclotomic Gaudin Hamiltonian is
\[
 \widetilde X_i(z)
 =\sum_{s=0}^{m-1}
 H_i(z_1,\ldots,\zeta_m^s z_i,\ldots,z_n;\mu)
 =mH_i(\pi_m(z);\mu),
\]
where the last equality follows from
Lemma~\ref{lem:average}.
Thus the first-order transport coefficient in
\eqref{eq:main-expand} agrees with that of \cite{TV26}.
In the present theorem the step $p$ also approaches
$\zeta_m$, so that $p^m=1+m\delta+O(\delta^2)$.
Consequently, this coefficient determines the
differential connection \eqref{eq:nabla-m} and
the pullback identity \eqref{eq:pullback}.
\end{rem}

Notice that the limiting differential connection
$\nabla_m(\ka,\mu)$ in \eqref{eq:nabla-m} depends only on
$m$, and not on the choice of the primitive $m$th root
$\zeta_m$.

A formal version of this theorem over fields of characteristic zero
is given in Subsection~\ref{app:formal-limit}.

\subsection{Proof of Theorem \ref{thm:limit}}\label{sec:proof}

\subsubsection{Semiclassical \texorpdfstring{$R$}{R}-matrix}

\begin{lem}[Semiclassical $R$-matrix]\label{lem:R-expand}
Fix $\La_a,\La_b\in\C$, a degree $r\ge0$, and
$x\in\C^\times\setminus\{1\}$. As $q\to1$,
\beq\label{eq:R-expand}
 R_{\La_a,\La_b}(x;q)
 =\id-2(q-1)\,\Om_{\La_a,\La_b}(x)+O((q-1)^2).
\eeq
This is an expansion on the degree-$r$ subspace of
$\mathbf M_{\La_a}\ox\mathbf M_{\La_b}$.
\end{lem}

\begin{proof}
On each fixed degree space, the entries of $R(x;q)$ are
rational in $x$, $q^2$, $q^{\La_a}$, and $q^{\La_b}$;
see \cite[Section~3]{TV97}.  They are holomorphic near
$(x_0,1)$ for every fixed $x_0\in\C^\times\setminus\{1\}$.
Consequently, their Taylor coefficients in $q-1$ are
holomorphic in $x$ near $x_0$.  Write
$R(x;q)=R_0+(q-1)B+O((q-1)^2)$.

Let $f_1=\mathbf f\ox1$, $f_2=1\ox\mathbf f$,
$h_1=\mathbf h\ox1$, and $h_2=1\ox\mathbf h$
act on $\mathbf M_{\La_a}\ox\mathbf M_{\La_b}$.
At $q=1$, \eqref{eq:R1}--\eqref{eq:R2} show that
$R_0$ commutes with $f_1+f_2$ and $f_1+xf_2$.
For $x\ne1$, it therefore commutes with $f_1$ and $f_2$
separately.  By \eqref{eq:R3} it is the identity on the
whole tensor product.

Comparing coefficients of $q-1$ in \eqref{eq:R1} and
\eqref{eq:R2} gives
\begin{align}
 [B,f_1+f_2]&=f_1h_2-h_1f_2,\label{eq:comm1}\\
 [B,f_1+xf_2]&=xh_1f_2-f_1h_2.\label{eq:comm2}
\end{align}
Using the $\slt$ relations, one checks that
$B=-2\Om_{\La_a,\La_b}(x)$ satisfies both identities; indeed,
\begin{align*}
 [-2\Om(x),f_1]
 &=\frac{x+1}{x-1}f_1h_2-\frac{2x}{x-1}h_1f_2,\\
 [-2\Om(x),f_2]
 &=\frac{x+1}{x-1}h_1f_2-\frac{2}{x-1}f_1h_2.
\end{align*}
The normalized operator $\Om(x)$ annihilates the tensor
product of highest-weight vectors.  By the normalization
\eqref{eq:R3}, $B$ also annihilates this vector.
The difference between $B$ and $-2\Om(x)$ commutes with
$f_1$ and $f_2$ and annihilates the highest-weight tensor.  It therefore vanishes
on every vector
$f_1^k f_2^l(\mathbf v_{\La_a}\ox\mathbf v_{\La_b})$.
This proves \eqref{eq:R-expand}.
\end{proof}

\subsubsection{Cyclotomic averaging}

\begin{lem}[Cyclotomic averaging]\label{lem:average}
For a primitive $m$th root of unity $\zeta_m$ and $x^m\ne1$,
\beq\label{eq:averaging}
 \boxed{\quad
 \sum_{s=0}^{m-1}\Om_{\La_a,\La_b}(\zeta_m^s x)
 =m\,\Om_{\La_a,\La_b}(x^m).
 \quad}
\eeq
This is an identity of operators on each degree subspace of
$\mathbf M_{\La_a}\ox\mathbf M_{\La_b}$.
\end{lem}

\begin{proof}
Applying $x\frac{\mathrm d}{\mathrm d x}$ to the logarithm of
$\prod_{s=0}^{m-1}(1-\zeta_m^s x)=1-x^m$
gives
\[
 \sum_{s=0}^{m-1}\frac{\zeta_m^s x}{\zeta_m^s x-1}
 =\frac{m x^m}{x^m-1}.
\]
Subtracting $m$ from both sides gives
\[
 \sum_{s=0}^{m-1}\frac1{\zeta_m^s x-1}
 =\frac{m}{x^m-1}.
\]
Adding these identities yields
\[
 \sum_{s=0}^{m-1}\frac{\zeta_m^s x+1}{\zeta_m^s x-1}
 =m\frac{x^m+1}{x^m-1}.
\]
Applying the three scalar identities to \eqref{eq:Omega}
proves \eqref{eq:averaging}.
\end{proof}

\subsubsection{Completion of the proof}

\begin{proof}[Proof of Theorem \ref{thm:limit}]
Since $q-1=-\delta/(2\ka)+O(\delta^2)$, Lemma
\ref{lem:R-expand} gives, for each fixed
$x\in\C^\times\setminus\{1\}$,
\beq\label{eq:local-R}
 R_{\La_a,\La_b}(x;q)
 =\id+\frac\delta\ka\Om_{\La_a,\La_b}(x)+O(\delta^2).
\eeq
By the integer-power action of the diagonal twist and the
last line of \eqref{eq:scaling},
\beq\label{eq:twist-expand}
 \varkappa^{(\La_i-h^{(i)})/2}
 =\id+\frac\delta\ka\,\mu(\mathbf h^{(i)}-\La_i)
 +O(\delta^2).
\eeq

Consider $K_i(z^{(i,s)})$ for $s=0,\ldots,m-1$.
For $j<i$, the corresponding $R$-matrix has argument
\[
 p^{s+1}\frac{z_i}{z_j}
 =\zeta_m^{s+1}\frac{z_i}{z_j}(1+O(\delta)),
\]
while for $j>i$ it has argument
\[
 p^s\frac{z_i}{z_j}
 =\zeta_m^s\frac{z_i}{z_j}(1+O(\delta)).
\]
Here ``argument'' means the spectral variable $x$ of the
$R$-matrix, namely $x=p^{s+1}z_i/z_j$ for $j<i$ and
$x=p^s z_i/z_j$ for $j>i$.  In either case, write
$x=x_0+O(\delta)$, where $x_0$ is the displayed
root-of-unity multiple of $z_i/z_j$.  Since
$z\in\mc U_m$, we have $x_0\ne1$, and hence
$\Om_{\La_i,\La_j}(x)=\Om_{\La_i,\La_j}(x_0)+O(\delta)$.
The first-order term of \eqref{eq:local-R} contains the
factor $\delta$; therefore this replacement changes the
$R$-matrix only by $O(\delta^2)$.  Consequently,
\begin{align}
 K_i(z^{(i,s)})
 &=\id+\frac\delta\ka\Biggl[
 \mu(\mathbf h^{(i)}-\La_i)
 +\sum_{j<i}\Om_{\La_i,\La_j}^{(i,j)}
       \left(\zeta_m^{s+1}\frac{z_i}{z_j}\right)\nonumber\\
 &\hspace{60pt}+
 \sum_{j>i}\Om_{\La_i,\La_j}^{(i,j)}
       \left(\zeta_m^s\frac{z_i}{z_j}\right)
 \Biggr]+O(\delta^2).
 \label{eq:one-step-expansion}
\end{align}

Multiply these expressions for $s=0,\ldots,m-1$ in the
order prescribed by \eqref{eq:mtransport}.  At first order,
the operator terms simply add; noncommutativity appears
only at order $\delta^2$.  For each $j\ne i$, the arguments
are precisely all the $m$ root-of-unity translates of $z_i/z_j$.
Lemma \ref{lem:average} therefore gives
\begin{align*}
 \widetilde K_i^{(m)}(z)
 &=\id+\frac{m\delta}\ka\Biggl[
 \mu(\mathbf h^{(i)}-\La_i)
 +\sum_{j\ne i}\Om_{\La_i,\La_j}^{(i,j)}
       \left(\left(\frac{z_i}{z_j}\right)^m\right)
 \Biggr]+O(\delta^2)\\
 &=\id+\frac{m\delta}\ka H_i(\pi_m(z);\mu)+O(\delta^2).
\end{align*}
This proves \eqref{eq:main-expand}.

Moreover, $p^m=1+m\delta+O(\delta^2)$ implies, on a
fixed holomorphic test function $F$,
\[
 F(z^{(i,m)})
 =F(z)+m\delta\,z_i\frac{\partial F}{\partial z_i}(z)
 +O(\delta^2).
\]
Comparison of the first-order shift and transport operators
identifies the differential operator
$z_i\partial_{z_i}-\ka^{-1}H_i(\pi_m(z);\mu)$,
and hence the connection \eqref{eq:nabla-m}.
Finally, using $Z_i=z_i^m$, we have
$\mathrm{d}Z_i/Z_i=m\,\mathrm{d}z_i/z_i$.  Thus
\[
 \pi_m^*\nabla_1(m\ka,\mu)
 =\mathrm{d}-\frac1\ka\sum_{i=1}^n
 H_i(\pi_m(z);\mu)\frac{\mathrm{d}z_i}{z_i}
 =\nabla_m(\ka,\mu),
\]
which is \eqref{eq:pullback}.  The theorem follows.
\end{proof}

\subsection{Formal root-of-unity limits over fields of characteristic zero}
\label{app:formal-limit}

\subsubsection{The formal limit theorem}

Let $K$ be a field of characteristic zero containing a primitive
$m$th root of unity $\zeta_m$.  We will repeat the calculation
of Subsection~\ref{sec:proof} over $K$, without taking an analytic
limit.  The symbol $\delta$ is now an indeterminate, not a complex
number tending to zero.

We use $K[[\delta]]$, the ring of formal power series in $\delta$.
In an expression
\[
 A(\delta)=A_0+\delta A_1+O(\delta^2),
\]
the symbol $O(\delta^2)$ denotes a multiple of $\delta^2$.
We will compare only the constant and linear coefficients.
All matrix calculations below are made on a fixed
finite-dimensional degree subspace.

For $c\in K$, put
\[
 (1+t)^c=\sum_{k=0}^{\infty}\binom{c}{k}t^k\in K[[t]].
\]
If $q\in1+\delta K[[\delta]]$, we define $q^c$ by substituting
$t=q-1$ in this binomial series.  In particular,
\[
 q^c=1+c(q-1)+O(\delta^2).
\]
The quantum integer $[c]_q$
has a removable singularity at $q=1$; after cancellation it is a
formal series with $[c]_q=c+O((q-1)^2)$.
Consequently the quantum Verma modules on each fixed degree can
be defined by the same formulas as in Subsection~\ref{sec:quantum},
with coefficients in $K[[\delta]]$.

Let $\mathbf V_r(K)$ denote the degree-$r$ subspace of the tensor
product of the $\slt$ Verma modules over $K$, with the basis
\eqref{eq:basis}.  We set
\[
 \mc U_m(K)=\{z\in(K^\times)^n:\ z_i^m\ne z_j^m
             \text{ for }i\ne j\},
 \qquad \pi_m(z)=(z_1^m,\ldots,z_n^m).
\]
The Hamiltonians $H_i$ and the connections $\nabla_1$ and
$\nabla_m$ are defined by the rational formulas
\eqref{eq:H}, \eqref{eq:nabla}, and \eqref{eq:nabla-m},
now with coefficients in $K$.

\begin{thm}[Formal limit of the qKZ connection]
\label{thm:formal-root-limit}
Let $\La_1,\ldots,\La_n,\mu\in K$ and $\ka\in K^\times$.
For a formal variable $\delta$, choose
$p,q,\varkappa\in K[[\delta]]$ such that
\beq\label{eq:formal-scaling}
 \begin{aligned}
 p&=\zeta_m\bigl(1+\delta+O(\delta^2)\bigr),\\
 q&=1-\frac{\delta}{2\ka}+O(\delta^2),\\
 \varkappa&=1-\frac{2\mu\delta}{\ka}+O(\delta^2).
 \end{aligned}
\eeq
Define the quantum Verma-module actions and normalized qKZ
transport as formal series, using the binomial-series definition above for
$q^c$ when $c\in K$.  Then, on each fixed-degree subspace,
\beq\label{eq:formal-transport}
 \boxed{\quad
 \widetilde K_i^{(m)}(z)
 =\id+\frac{m\delta}{\ka}H_i(\pi_m(z);\mu)
      +O(\delta^2),\qquad i=1,\ldots,n.
 \quad}
\eeq
This is an equality modulo $\delta^2$ of matrices in the
basis \eqref{eq:basis}.  The coefficients of the constant and
linear terms are rational functions of $z$ without poles on
$\mc U_m(K)$.
The resulting algebraic differential connection satisfies
\beq\label{eq:formal-pullback}
 \boxed{\quad
 \nabla_m(\ka,\mu)=\pi_m^*\nabla_1(m\ka,\mu).
 \quad}
\eeq
\end{thm}

\begin{proof}
First, the first-order binomial expansion and the second
line of \eqref{eq:formal-scaling} give
\[
 q^c=1-\frac{c\delta}{2\ka}+O(\delta^2),\qquad c\in K.
\]
Hence the quantum Verma-module matrices are regular at
$\delta=0$ and reduce to the classical $\slt$ matrices.

The normalized trigonometric $R$-matrix is also regular on each
fixed-degree subspace when its spectral parameter $x$ has
constant term $x_0\ne1$.  This can be seen directly from its
intertwining relations \eqref{eq:R1}--\eqref{eq:R3}.  At
$\delta=0$ the two lowering-operator combinations are
$f_1+f_2$ and $f_1+x_0f_2$.
Since $x_0-1\ne0$, these determine the actions of $f_1$
and $f_2$ separately, and the normalized matrix can be
constructed degree by degree.  The same induction determines
its coefficients over $K[[\delta]]$.

As in Lemma~\ref{lem:R-expand}, comparison of the
first-order coefficients in the intertwining relations gives
\[
\begin{aligned}
 R_{\La_a,\La_b}(x;q)
  &=\id-2(q-1)\Om_{\La_a,\La_b}(x_0)+O(\delta^2)\\
  &=\id+\frac{\delta}{\ka}\Om_{\La_a,\La_b}(x_0)
    +O(\delta^2),
\end{aligned}
\]
whenever $x=x_0+O(\delta)$ and $x_0\ne1$.
The twist factor similarly satisfies
\[
 \varkappa^{(\La_i-h^{(i)})/2}
 =\id+\frac{\mu\delta}{\ka}
       (\mathbf h^{(i)}-\La_i)+O(\delta^2).
\]
(The exponent of $\varkappa$ on each basis vector is an
integer, as in Subsection~\ref{sec:qkz}.)

In the $s$th one-step transport, $s=0,\ldots,m-1$,
the constant terms of the spectral parameters are
$\zeta_m^{s+1}z_i/z_j$ for $j<i$ and
$\zeta_m^sz_i/z_j$ for $j>i$.  They are all different
from $1$ because $z\in\mc U_m(K)$.
Multiplying the $m$ one-step transports and retaining only
the linear term in $\delta$, we add their first-order
contributions.  The cyclotomic identity
\[
 \prod_{s=0}^{m-1}(1-\zeta_m^s x)=1-x^m
\]
and its logarithmic derivative (an identity of rational
functions, requiring no formal logarithm) give the averaging
formula \eqref{eq:averaging} over $K$.  We obtain
\[
 \widetilde K_i^{(m)}(z)
 =\id+\frac{m\delta}{\ka}
 \left(\mu(\mathbf h^{(i)}-\La_i)
   +\sum_{j\ne i}\Om^{(i,j)}_{\La_i,\La_j}
       \left(\left(\frac{z_i}{z_j}\right)^m\right)\right)
       +O(\delta^2),
\]
which is \eqref{eq:formal-transport}.
Finally, $p^m=1+m\delta+O(\delta^2)$ gives the
first-order operator
$z_i\partial_{z_i}-\ka^{-1}H_i(\pi_m(z);\mu)$.
The pullback identity \eqref{eq:formal-pullback} follows
from $\mathrm d(z_i^m)/z_i^m=m\,\mathrm dz_i/z_i$.
\end{proof}

\subsubsection{Compatibility under composition of power maps}

The power maps satisfy $\pi_a\circ\pi_b=\pi_{ab}$.
Applying \eqref{eq:formal-pullback} twice therefore relates
the limiting connections for different positive integers
$a$ and $b$.  If $K$ contains a primitive $(ab)$th root of
unity, then on $\mc U_{ab}(K)$ we have
\beq\label{eq:formal-composition}
 \boxed{\quad
 \nabla_{ab}(\ka,\mu)
 =\pi_b^*\nabla_a(b\ka,\mu)
 =\pi_a^*\nabla_b(a\ka,\mu).
 \quad}
\eeq
For example,
\[
 \pi_b^*\nabla_a(b\ka,\mu)
 =\pi_b^*\pi_a^*\nabla_1(ab\ka,\mu)
 =\pi_{ab}^*\nabla_1(ab\ka,\mu)
 =\nabla_{ab}(\ka,\mu).
\]
The other equality follows by interchanging $a$ and $b$.
This identity is a consequence of pullback, rather than
another calculation of a root-of-unity limit.

\subsubsection{\texorpdfstring{$\ell$}{ell}-adic specialization}

The preceding theorem concerns formal first-order coefficients.
We now explain when it can also be regarded as a limit in an
$\ell$-adic field.

\begin{cor}[$\ell$-adic specialization]
\label{cor:ell-adic-limit}
Let $\ell$ be a prime and let $K$ be a finite extension
of $\mathbb Q_\ell$ containing $\zeta_m$ and the parameters
$\La_1,\ldots,\La_n,\ka,\mu$, where $\ka\ne0$.
Suppose that $p(\delta)$, $q(\delta)$, and
$\varkappa(\delta)$ are $K$-analytic functions for
$\delta$ in a sufficiently small neighborhood of $0$,
and their Taylor expansions have the constant and linear
terms specified in \eqref{eq:formal-scaling}.
Then, for every fixed $z\in\mc U_m(K)$,
\[
 \lim_{\delta\to0}
 \frac{\widetilde K_i^{(m)}(z;\delta)-\id}{\delta}
 =\frac{m}{\ka}H_i(\pi_m(z);\mu),
 \qquad i=1,\ldots,n,
\]
where the limit is taken in the $\ell$-adic topology.
The associated first-order differential connection is
$\nabla_m(\ka,\mu)$, and it satisfies
\eqref{eq:formal-pullback}.
\end{cor}

\begin{proof}
For any $c\in K$, the binomial series $(1+t)^c$ has a
positive $\ell$-adic radius of convergence.  Thus the powers
$q^c$ appearing in the quantum Verma-module matrices are
analytic for sufficiently small $\delta$.

For fixed $z\in\mc U_m(K)$, the constant terms of the
$R$-matrix spectral parameters are
$\zeta_m^s z_i/z_j$, none of which equals $1$.
The denominators occurring in the fixed-degree $R$-matrices
therefore have nonzero constant terms.  After shrinking
the $\ell$-adic neighborhood of $\delta=0$ if necessary,
the $R$-matrices and the finite $m$-step transport are analytic.
The first-order identity \eqref{eq:formal-transport} then
gives the asserted limit.

The argument also applies when $\ell$ divides $m$:
we are working in the characteristic-zero field $K$,
and nonzero denominators remain invertible on a sufficiently
small $\ell$-adic neighborhood.
\end{proof}

\section{Hypergeometric solutions of the \texorpdfstring{$\mathrm{\lowercase{q}}$KZ}{qKZ} equations}
\label{sec:hyper-qkz}

We recall the one-variable hypergeometric solutions of the
trigonometric qKZ equations from \cite{TV97}, using the quantum group
and the qKZ connection defined in Subsections~\ref{sec:quantum}--\ref{sec:qkz}.

\subsection{The phase and weight functions}

We specialize to $\La_1=\cdots=\La_n=1$.
Let $M(q)=M_1(q)$ be the Verma module with highest-weight vector $v$,
so that
\beq\label{eq:hyp-highest}
 ev=0,\qquad q^h v=qv.
\eeq
For $n\ge2$, put $\mc M(q)=M(q)^{\ox n}$.
The degree-one subspace $\mc M_1(q)$ has dimension $n$ and basis
\beq\label{eq:hyp-basis}
 v_j=v\ox\cdots\ox
 \underset{j\text{-th factor}}{fv}
 \ox\cdots\ox v,
 \qquad j=1,\ldots,n.
\eeq

Let $p\in\C$ with $0<|p|<1$.  Set
\beq\label{eq:hyp-theta-def}
 (u;p)_\infty=\prod_{s=0}^\infty(1-p^su),\qquad
 \tht_p(u)=(u;p)_\infty(p/u;p)_\infty(p;p)_\infty.
\eeq
The theta function satisfies
\beq\label{eq:hyp-theta}
 \tht_p(pu)=-u^{-1}\tht_p(u).
\eeq
In our conventions, the parameters $\eta$ and $\xi_a$ of
\cite{TV97} are $\eta=q^2$ and $\xi_a=q$, $a=1,\ldots,n$.
The phase function is
\beq\label{eq:hyp-phase}
 \Phi(t;z)=\prod_{a=1}^n
 \frac{(q^{-1}t/z_a;p)_\infty}
      {(qt/z_a;p)_\infty}.
\eeq
This is the specialization of \cite[(5.2)]{TV97} to one
integration variable.

The trigonometric weight functions are
\beq\label{eq:hyp-wj}
 w_j(t;z)=\frac{t}{t-qz_j}
 \prod_{a<j}\frac{qt-z_a}{t-qz_a},
 \qquad j=1,\ldots,n,
\eeq
and the corresponding $\mc M_1(q)$-valued weight function is
\beq\label{eq:hyp-w}
 w(t;z)=\sum_{j=1}^n w_j(t;z)v_j.
\eeq
These are the degree-one specializations of the trigonometric
weight functions in \cite[(2.20)]{TV97}.

Let $\varkappa\in\C^\times$ be the twist parameter of the qKZ
connection \eqref{eq:K}. For $i=1,\ldots,n$, put
\beq\label{eq:hyp-Wi}
 W_i(t;z)=
 \frac{\tht_p(\varkappa^{-1}q^{n+1-2i}t/z_i)}
      {\tht_p(q^{-1}t/z_i)}
 \prod_{a<i}
 \frac{\tht_p(qt/z_a)}{\tht_p(q^{-1}t/z_a)}.
\eeq
They are the degree-one elliptic weight functions in
\cite[(2.30)]{TV97}.

We use the following adjusting factors, corresponding to the choice
$c_a=1$ in the example following \cite[(2.32)]{TV97}:
\beq\label{eq:hyp-Yi}
 Y_i(z)=
 \left(\prod_{a<i}\frac{\tht_p(qz_a)}{\tht_p(z_a)}\right)
 \frac{\tht_p(\varkappa^{-1}q^{n+1-2i}z_i)}
      {\tht_p(z_i)}
 \left(\prod_{a>i}\frac{\tht_p(q^{-1}z_a)}{\tht_p(z_a)}\right).
\eeq
The quasiperiodicity relation \eqref{eq:hyp-theta} implies
\beq\label{eq:hyp-Ytransform}
 \frac{Y_i(z_1,\ldots,pz_a,\ldots,z_n)}{Y_i(z)}
 =\begin{cases}
 q^{-1},&a<i,\\
 \varkappa q^{2i-n-1},&a=i,\\
 q,&a>i.
 \end{cases}
\eeq
These are the multipliers required by \cite[(2.32)]{TV97}.
Set
\beq\label{eq:hyp-What}
 \widehat W_i(t;z)=Y_i(z)W_i(t;z).
\eeq

\subsection{Circle integrals and qKZ solutions}
\label{subsec:hyp-circle-integrals}

Assume
\beq\label{eq:hyp-circle-domain}
 0<|p|<|q|<1,\qquad |z_a|=1,\quad z_a\ne1,
 \qquad a=1,\ldots,n,
\eeq
with distinct $z_a$.  Under these assumptions the integrands below
have no poles on the unit circle.  The circle is oriented
counterclockwise.

For $i=1,\ldots,n$, define the $\mc M_1(q)$-valued function
\beq\label{eq:hyp-Psi}
 \boxed{\quad
 \Psi_i(z;p,q,\varkappa)=\frac{1}{2\pi\mathrm{i}}
 \int_{|t|=1}\Phi(t;z)\widehat W_i(t;z)
 w(t;z)\,\frac{dt}{t}.
 \quad}
\eeq
Its coordinates in the basis \eqref{eq:hyp-basis} form the matrix
\beq\label{eq:hyp-matrix}
 \Psi(z)=(\Psi_{ji}(z))_{j,i=1}^n,\qquad
 \Psi_{ji}(z)=\frac{1}{2\pi\mathrm{i}}\int_{|t|=1}
 \Phi(t;z)w_j(t;z)\widehat W_i(t;z)\frac{dt}{t}.
\eeq
The formulas \eqref{eq:hyp-Psi}--\eqref{eq:hyp-matrix} initially
define functions when \eqref{eq:hyp-circle-domain} holds.  As in
\cite[Section~5]{TV97}, they extend meromorphically in $z$ by
deforming the $t$-contour so that it continues to separate the
inward and outward poles.  Here and below, $\Psi_i$ denotes this
meromorphic continuation on $(\C^\times)^n$.
By \cite[Corollary~5.19]{TV97}, each $\Psi_i$ satisfies the
qKZ equations \eqref{eq:qkz}, including at the shifted arguments.
In \cite[(4.1)]{TV97}, the degree-one tensor-coordinate factor is
$q^{\La_{\mathrm{TV97}}}=q^{1/2}$, since
$\La_{\mathrm{here}}=1=2\La_{\mathrm{TV97}}$.
With our change of generators $f=Fq^H$, one has
$fv=q^{1/2}Fv$ on each highest-weight vector.
Thus this tensor-coordinate factor cancels exactly when the
solution is written in the basis \eqref{eq:hyp-basis}; no additional
common scalar is required in \eqref{eq:hyp-Psi}.

The integral formulas \eqref{eq:hyp-Psi} select the complex vector space
\beq\label{eq:hyp-S}
 \mc S_{\mathrm{hyp}}(p,q,\varkappa)
 =\operatorname{span}_{\C}\{\Psi_1,\ldots,\Psi_n\}.
\eeq

\subsection{Determinant formula}

The determinant formula of \cite[Theorem~5.9]{TV97}, with one
integration variable and $\xi_a=q$, gives the following result.

\begin{thm}\label{thm:hyp-det}
Assume \eqref{eq:hyp-circle-domain} and let $\varkappa\in\C^\times$.
Then
\beq\label{eq:hyp-det}
 \det\Psi(z)
 =\left(\prod_{i=1}^nY_i(z)\right)
   D_n(z;p,q,\varkappa),
\eeq
where
\begin{align}
 \label{eq:hyp-Dn}
 D_n(z;p,q,\varkappa)
 &\,=q^{n(n-1)/2}
 \prod_{a=1}^{n-1}\tht_p(\varkappa^{-1}q^{n-2a})
 \frac{(\varkappa^{-1}q^n;p)_\infty
       (p\varkappa q^n;p)_\infty}
      {(p;p)_\infty^{2n-1}(q^2;p)_\infty^n}
 \\[-2pt]
 &\qquad\times
 \prod_{1\le a<b\le n}
 \frac{(q^{-2}z_a/z_b;p)_\infty}
      {(q^{2}z_a/z_b;p)_\infty}.\notag
\end{align}
In particular, for generic $\varkappa\in\C^\times$ and generic
$z$ satisfying \eqref{eq:hyp-circle-domain}, the determinant is
nonzero.  Consequently, for these generic values of
$(\varkappa,z)$, the matrix $\Psi(z)$ is fundamental, and
$\dim_\C\mc S_{\mathrm{hyp}}(p,q,\varkappa)=n$.
\end{thm}

The factor $(2\pi\mathrm{i})^n$ in \cite[Theorem~5.9]{TV97} is absent from
\eqref{eq:hyp-det} because of the normalization
$(2\pi\mathrm{i})^{-1}$ in \eqref{eq:hyp-Psi}; the adjusting factors
$Y_i$ multiply the columns.  Formula \eqref{eq:hyp-det} follows
first under the genericity assumptions in \cite[Theorem~5.9]{TV97}.
Both sides extend holomorphically in $q$ within
$|p|<|q|<1$ and in $\varkappa\in\C^\times$; hence the identity
holds throughout the chamber \eqref{eq:hyp-circle-domain},
including values excluded by the genericity assumptions.
For fixed $p$ and $q$ with $0<|p|<|q|<1$, the displayed product is not identically zero as a function of
$(\varkappa,z)$.  Its denominator factors are nonzero under
\eqref{eq:hyp-circle-domain}.

\subsection{Fundamental matrices and quasiconstants}

Let $\mc P_p$ be the field of meromorphic functions
$c$ on $(\C^\times)^n$ satisfying
\beq\label{eq:hyp-quasiconstants}
 c(z_1,\ldots,pz_a,\ldots,z_n)=c(z),
 \qquad a=1,\ldots,n.
\eeq
The meromorphic solutions of the qKZ equations form a module over
$\mc P_p$.

\begin{prop}\label{prop:hyp-fundamental}
Let $F(z)$ and $G(z)$ be $n\times n$ matrices whose columns are
meromorphic solutions of the qKZ equations in $\mc M_1(q)$
on $(\C^\times)^n$.  A matrix of solutions is called
\emph{fundamental} if its determinant is not identically zero.
Assume that $F$ and $G$ are fundamental.
Then
\beq\label{eq:hyp-transition}
 G(z)=F(z)C(z),\qquad C(z)\in GL_n(\mc P_p).
\eeq
Conversely, for every $C(z)\in GL_n(\mc P_p)$, the matrix
$F(z)C(z)$ is another fundamental matrix of solutions.
Moreover, every meromorphic solution $I(z)$ has a unique expression
\beq\label{eq:hyp-solution-coordinates}
 I(z)=F(z)c(z),\qquad c(z)\in\mc P_p^n.
\eeq
\end{prop}

\begin{proof}
Use the shift notation $z^{(a,1)}$ introduced just before \eqref{eq:qkz}.
Since $F$ and $G$ satisfy the qKZ equations,
\[
 F(z^{(a,1)})=K_a(z)F(z),\qquad
 G(z^{(a,1)})=K_a(z)G(z).
\]
For $C=F^{-1}G$ we obtain
\[
 C(z^{(a,1)})=F(z^{(a,1)})^{-1}G(z^{(a,1)})
   =(K_a(z)F(z))^{-1}K_a(z)G(z)=C(z).
\]
Conversely, if $C(z^{(a,1)})=C(z)$, then $FC$ satisfies the qKZ
equations. Applying the same argument to the column $I$ in place
of $G$ proves \eqref{eq:hyp-solution-coordinates}.
\end{proof}

Thus, whenever a fundamental matrix exists, the module of
meromorphic solutions is free of rank $n$ over $\mc P_p$.
The space $\mc S_{\mathrm{hyp}}(p,q,\varkappa)$ in
\eqref{eq:hyp-S} is the particular complex vector space of qKZ
solutions considered in this paper, rather than the entire module
over $\mc P_p$.

\begin{example}[the case $n=2$]
We give the ingredients of the preceding construction explicitly.
In the basis $(v_1,v_2)$ the phase function and the two
trigonometric weights are
\begin{align*}
 \Phi(t;z)&=\frac{(q^{-1}t/z_1;p)_\infty(q^{-1}t/z_2;p)_\infty}
 {(qt/z_1;p)_\infty(qt/z_2;p)_\infty},\\
 w_1(t;z)&=\frac{t}{t-qz_1},\qquad
 w_2(t;z)=\frac{t(qt-z_1)}{(t-qz_1)(t-qz_2)}.
\end{align*}
The elliptic weight functions are
\begin{align*}
 W_1(t;z)&=\frac{\tht_p(\varkappa^{-1}qt/z_1)}
 {\tht_p(q^{-1}t/z_1)},\\
 W_2(t;z)&=\frac{\tht_p(\varkappa^{-1}q^{-1}t/z_2)}
 {\tht_p(q^{-1}t/z_2)}
 \frac{\tht_p(qt/z_1)}{\tht_p(q^{-1}t/z_1)},
\end{align*}
and the adjusting factors are
\begin{align}
 Y_1(z)&=\frac{\tht_p(\varkappa^{-1}qz_1)\tht_p(q^{-1}z_2)}
 {\tht_p(z_1)\tht_p(z_2)},\notag\\
 Y_2(z)&=\frac{\tht_p(qz_1)\tht_p(\varkappa^{-1}q^{-1}z_2)}
 {\tht_p(z_1)\tht_p(z_2)}.\label{eq:n2-qkz-adjust}
\end{align}
Thus the two columns $\Psi_1,\Psi_2$ are given simultaneously by
\beq\label{eq:n2-qkz-integrals}
 \Psi(z)=\frac{1}{2\pi\mathrm{i}}\int_{|t|=1}\Phi(t;z)
 \begin{pmatrix}
 \displaystyle\frac{Y_1W_1}{t-qz_1}&
 \displaystyle\frac{Y_2W_2}{t-qz_1}\\[9pt]
 \displaystyle\frac{Y_1W_1(qt-z_1)}{(t-qz_1)(t-qz_2)}&
 \displaystyle\frac{Y_2W_2(qt-z_1)}{(t-qz_1)(t-qz_2)}
 \end{pmatrix}\,dt,
\eeq
where $W_i=W_i(t;z)$ and $Y_i=Y_i(z)$. The factor $t$ in
$w_j$ has canceled the $t$ in $dt/t$.

The normalized trigonometric $R$-matrix on the degree-one subspace
of $M_1(q)\ox M_1(q)$ is
\beq\label{eq:n2-qkz-R}
 R(x;q)=\frac{1}{x-q^2}
 \begin{pmatrix}
 q(x-1)&1-q^2\\
 x(1-q^2)&q(x-1)
 \end{pmatrix}.
\eeq
It follows directly from the two intertwining relations
in Subsection~\ref{sec:qkz} and the highest-vector normalization.  If
\[
 a(x)=\frac{q(x-1)}{x-q^2},\qquad
 b(x)=\frac{1-q^2}{x-q^2},\qquad
 c(x)=\frac{x(1-q^2)}{x-q^2},
\]
then the qKZ transport matrices in the basis $(v_1,v_2)$ are
\beq\label{eq:n2-qkz-K}
 \begin{aligned}
 K_1(z)&=\begin{pmatrix}
 \varkappa a(z_1/z_2)&\varkappa b(z_1/z_2)\\
 c(z_1/z_2)&a(z_1/z_2)
 \end{pmatrix},\\[6pt]
 K_2(z)&=\begin{pmatrix}
 a(pz_2/z_1)&\varkappa c(pz_2/z_1)\\
 b(pz_2/z_1)&\varkappa a(pz_2/z_1)
 \end{pmatrix}.
 \end{aligned}
\eeq
In particular,
$\Psi(pz_1,z_2)=K_1(z)\Psi(z)$ and
$\Psi(z_1,pz_2)=K_2(z)\Psi(z)$ after meromorphic continuation.

Finally, Theorem~\ref{thm:hyp-det} becomes
\begin{align}
 \det\Psi(z)
 &=Y_1(z)Y_2(z)\,q\tht_p(\varkappa^{-1})
 \frac{(\varkappa^{-1}q^2;p)_\infty
 (p\varkappa q^2;p)_\infty}
 {(p;p)_\infty^3(q^2;p)_\infty^2}
 \notag\\[-2pt]
 &\quad\times
 \frac{(q^{-2}z_1/z_2;p)_\infty}
 {(q^2z_1/z_2;p)_\infty}.
 \label{eq:n2-qkz-det}
\end{align}
All these formulas use the same initial parameter domain and
contour prescription as in Subsection~\ref{subsec:hyp-circle-integrals}.
\end{example}

\section{Hypergeometric solutions of the trigonometric KZ equations}
\label{sec:hyper-kz}

We describe a family of hypergeometric integral solutions of the
trigonometric KZ equations \eqref{eq:KZ} in the degree-one subspace
$\mathbf V_1$ of $\mathbf M_1^{\ox n}$.  Under the identification
of Subsection~\ref{sec:quantum}, we use the basis $v_1,\ldots,v_n$
introduced in \eqref{eq:hyp-basis} also for $\mathbf V_1$.
The integral representation is the degree-one case of the construction
in \cite[Sections~2.2 and~4]{MaV02}, written in the normalization
\eqref{eq:Omega}--\eqref{eq:H}.

\subsection{Master function and interval integrals}
\label{subsec:kz-master-integrals}

For this section, we choose the real parameter domain
\beq\label{eq:kz-chamber}
 \ka<-n,\qquad 0<\mu<\frac18.
\eeq
Set
\beq\label{eq:kz-exponents}
 \alpha=\frac{n/2-2\mu}{\ka},\qquad b=-\frac1\ka.
\eeq
In particular, $\alpha>-1$ and $b>0$.

Choose a reference configuration $z^0=(z_1^0,\ldots,z_n^0)$
of distinct points on the unit circle whose arguments are ordered:
\beq\label{eq:kz-angle-order}
 z_a^0=e^{\mathrm{i}\vartheta_a},\qquad
 0<\vartheta_1<\vartheta_2<\cdots<\vartheta_n<2\pi.
\eeq
Let $\mc U\subset\mc U_1$ be a sufficiently small simply
connected neighborhood of $z^0$.
For $a=1,\ldots,n$ fix the holomorphic logarithm
$L_a(z)$ on $\mc U$ determined by
\[
 e^{L_a(z)}=z_a,\qquad L_a(z^0)=\mathrm{i}\vartheta_a.
\]
For $c\in\C$ we use the notation
$z_a^c=\exp(cL_a(z))$.

For $j=1,\ldots,n$ introduce the master function
\beq\label{eq:kz-master}
 \Phi_j(x;z)=x^\alpha(1-x)^b
 \prod_{\substack{a=1\\a\ne j}}^n
 \left(1-\frac{z_j}{z_a}x\right)^b,
 \qquad 0<x<1,
\eeq
and the $\mathbf V_1$-valued rational function
\beq\label{eq:kz-vector}
 g_j(x;z)=\sum_{k=1}^n\frac{z_j}{z_jx-z_k}\,v_k.
\eeq
Define
\beq\label{eq:kz-I}
 \boxed{\quad
 I_j(z;\ka,\mu)=
 z_j^\alpha\prod_{a=1}^n z_a^{-1/(2\ka)}
 \int_0^1\Phi_j(x;z)g_j(x;z)\,dx,
 \qquad j=1,\ldots,n.
 \quad}
\eeq

We now specify the branches in \eqref{eq:kz-master}--\eqref{eq:kz-I}.
For $0<x<1$, the factors $x^\alpha$ and $(1-x)^b$
are defined using the ordinary real logarithm of $x$ and $1-x$.
For $a\ne j$, the number $1-xz_j/z_a$ has positive real part
at $z=z^0$, for every $0\le x\le1$; this follows from the
fact that $z_j^0/z_a^0$ lies on the unit circle and is not $1$.
After making $\mc U$ smaller if necessary, the same is true
for all $z\in\mc U$.  Accordingly, we define
\[
 \left(1-\frac{z_j}{z_a}x\right)^b
 =\exp\left(b\,\operatorname{Log}
       \left(1-\frac{z_j}{z_a}x\right)\right),
 \qquad a\ne j,
\]
where $\operatorname{Log}$ is the principal logarithm.
It is the continuous logarithm whose value at $x=0$ is $0$.
The prefactor powers of $z_j$ and $z_a$ in \eqref{eq:kz-I}
are defined by the logarithms $L_a(z)$ fixed above.
Notice also that $(1-x)^b$ in \eqref{eq:kz-master}
is precisely the term with $a=j$ in
$\prod_{a=1}^n(1-z_jx/z_a)^b$.

The integrals \eqref{eq:kz-I} converge absolutely.
Indeed, their integrands are $O(x^\alpha)$ as $x\to0$,
while at $x\to1$ the $v_j$-component is
$O((1-x)^{b-1})$ and the other components are
$O((1-x)^b)$.

There is an equivalent contour-integral expression.
Let $\gamma_j$ denote the straight segment from $0$ to $z_j$,
oriented toward $z_j$.  Then
\beq\label{eq:kz-loaded}
 I_j(z;\ka,\mu)=
 \left(\prod_{a=1}^n z_a^{-1/(2\ka)}\right)
 \int_{\gamma_j}
 u^\alpha\prod_{a=1}^n
 \left(1-\frac{u}{z_a}\right)^b
 \left(\sum_{k=1}^n\frac{v_k}{u-z_k}\right)du.
\eeq
In this formula $\gamma_j$ is a \emph{loaded interval}:
in addition to its orientation, the multivalued factors
in the integrand have specified branches along the path.
Explicitly, parametrizing the interval by
\[
 u=z_jx,\qquad 0<x<1,
\]
we take
\[
 \begin{aligned}
 u^\alpha
 &=\exp\bigl(\alpha(L_j(z)+\log x)\bigr)
   =z_j^\alpha x^\alpha,\\
 \left(1-\frac{u}{z_j}\right)^b&=(1-x)^b,\\
 \left(1-\frac{u}{z_a}\right)^b
 &=\left(1-\frac{z_j}{z_a}x\right)^b,
 \qquad a\ne j,
 \end{aligned}
\]
with the branches on the right defined above.
Since $du=z_j\,dx$, this substitution transforms
\eqref{eq:kz-loaded} exactly into \eqref{eq:kz-I}.

\subsection{The KZ equations}

\begin{thm}\label{thm:kz-integrals}
For $\ka,\mu$ satisfying \eqref{eq:kz-chamber}, the functions
$I_1,\ldots,I_n$ defined by \eqref{eq:kz-I} are local holomorphic
solutions on $\mc U$ of the trigonometric KZ equations
\eqref{eq:KZ} in $\mathbf V_1$.
\end{thm}

\begin{proof}
We explain the scalar change of normalization in the degree-one
specialization of \cite[Corollary~4.2]{MaV02}.  Let
$H_i^{\mathrm{std}}$ denote its Hamiltonian with Cartan parameter
$\lambda=\mu\mathbf h$ and all highest weights equal to $1$.
By \eqref{eq:Omega}--\eqref{eq:H},
\[
 H_i(z;\mu)=H_i^{\mathrm{std}}(z;\mu)
 -\left(\mu+\frac14\sum_{a\ne i}
                  \frac{z_i+z_a}{z_i-z_a}\right)\id.
\]
On the simply connected neighborhood $\mc U$, choose compatible
branches and set
\[
 g(z)=\prod_{a=1}^n
 z_a^{\,(-\mu+(n-1)/4)/\ka}
 \prod_{1\le a<c\le n}(z_a-z_c)^{-1/(2\ka)}.
\]
Logarithmic differentiation gives
\[
 \ka z_i\frac{\partial\log g}{\partial z_i}
 =-\mu-\frac14\sum_{a\ne i}\frac{z_i+z_a}{z_i-z_a}.
\]
Thus multiplication by $g$ converts solutions for
$H_i^{\mathrm{std}}$ into solutions for $H_i$.

For completeness, the degree-one specialization of the master
function in \cite[(16)]{MaV02}, raised to the power $1/\ka$, is
\[
 \prod_{a<c}(z_a-z_c)^{1/(2\ka)}
 \prod_{a=1}^n z_a^{(\mu+(3-n)/4)/\ka}\,
 u^\alpha\prod_{a=1}^n(u-z_a)^b.
\]
The corresponding vector weight is
$\sum_k v_k/(u-z_k)$, up to an overall sign from the basis
convention. After multiplication by $g$, the factors
$(z_a-z_c)^{1/(2\ka)}$ cancel and the remaining scalar
prefactor is $\prod_a z_a^{1/(2\ka)}$.
Using $b=-1/\ka$ and
$(u-z_a)^b=(-z_a)^b(1-u/z_a)^b$ with compatible branches
therefore gives exactly \eqref{eq:kz-loaded}, up to a nonzero
constant independent of $z$. Such a constant does not change
the differential equations.  The asserted KZ equations follow
from \cite[Corollary~4.2]{MaV02}.
\end{proof}

\subsection{The space of KZ hypergeometric solutions}

Let $\mc I(z)=(I_{kj}(z))_{k,j=1}^n$ be the matrix of coordinates
of the solutions $I_j$ in the basis $v_1,\ldots,v_n$.
A matrix of local holomorphic solutions of the KZ equations is
called \emph{fundamental} if its determinant is not identically
zero on $\mc U$.
Set
\beq\label{eq:kz-solution-space}
 \mc S_{\mathrm{KZ}}(\ka,\mu)
 =\operatorname{span}_{\C}\{I_1,\ldots,I_n\}.
\eeq

\begin{prop}[Exact determinant formula]\label{prop:kz-basis}
For every $\ka,\mu$ satisfying \eqref{eq:kz-chamber}, the matrix
$\mc I(z)$ is fundamental. More precisely, on $\mc U$ one has
\beq\label{eq:kz-exact-det}
 \boxed{\quad
 \det\mc I(z)=(-1)^n
 \frac{\Gamma(\alpha+1)\Gamma(b)^n}
 {\Gamma(\alpha+nb+1)}
 \left(\prod_{a=1}^n z_a^{2\mu b}\right)
 \prod_{1\le a<c\le n}
 \left(1-\frac{z_a}{z_c}\right)^b
 \left(1-\frac{z_c}{z_a}\right)^b.
 \quad}
\eeq
Here $z_a^{2\mu b}=\exp(2\mu b L_a(z))$ and the factors
$(1-z_a/z_c)^b$, $(1-z_c/z_a)^b$ are defined using the
principal logarithms at the reference unit-circle configuration
and then continued holomorphically to $\mc U$.
In particular, at $z_a=e^{\mathrm{i}\vartheta_a}$,
\beq\label{eq:kz-det-circle}
 \det\mc I(z)=(-1)^n
 \frac{\Gamma(\alpha+1)\Gamma(b)^n}
 {\Gamma(\alpha+nb+1)}
 \left(\prod_{a=1}^n e^{2\mathrm{i}\mu b\vartheta_a}\right)
 \prod_{1\le a<c\le n}|z_a-z_c|^{2b}.
\eeq
Consequently,
$\dim_\C\mc S_{\mathrm{KZ}}(\ka,\mu)=n$
for every pair of parameters in \eqref{eq:kz-chamber}.
\end{prop}

\begin{proof}
We use the one-dimensional arrangement of points
\[
 0=t_0<t_1<\dots<t_n
\]
and the real intervals $\Delta_j=[t_{j-1},t_j]$.
Consider the matrix
\beq\label{eq:kz-V89-real-matrix}
 A_{kj}(t)=\int_{t_{j-1}}^{t_j}
 u^\alpha\prod_{a=1}^n|u-t_a|^b\,
 \frac{du}{u-t_k},\qquad k,j=1,\ldots,n.
\eeq
The specialization of \cite[Theorem~1.1]{V89} to this
arrangement, with exponents $(\alpha,b,\ldots,b)$, gives
\beq\label{eq:kz-V89-real-det}
 \det A(t)=(-1)^n
 \frac{\Gamma(\alpha+1)\Gamma(b)^n}
 {\Gamma(\alpha+nb+1)}
 \left(\prod_{a=1}^n t_a^{\alpha+b}\right)
 \prod_{1\le a<c\le n}(t_c-t_a)^{2b}.
\eeq
Let us explain the normalization. The hypergeometric
$1$-forms in \cite{V89} associated with the points $t_k$
contain the factor $b\,du/(u-t_k)$.
Dividing their determinant by $b^n$ changes
$\Gamma(b+1)^n$ to $\Gamma(b)^n$.
Our columns use the left-to-right orientations of the
intervals, whereas the internal orientations of \cite{V89}
are right-to-left; this gives the factor $(-1)^n$.
The branches on the real intervals contribute the same
phase on both sides of the determinant formula and can be
removed to give \eqref{eq:kz-V89-real-det}.
Although \cite[Theorem~1.1]{V89} is initially stated for
positive exponents, \eqref{eq:kz-V89-real-det} continues to
$\alpha>-1$, $b>0$ by analyticity of the integrals and of the
gamma factors. These inequalities hold in \eqref{eq:kz-chamber}.

We next compare \eqref{eq:kz-V89-real-matrix} with the loaded
radial intervals in \eqref{eq:kz-loaded}.
Choose the positive numbers $t_1<\dots<t_n$ above and deform
our reference configuration through
\[
 z_a(s)=t_a^s\exp\bigl(\mathrm{i}(1-s)\vartheta_a\bigr),
 \qquad 0\le s<1.
\]
All rays remain distinct before the limiting configuration
$s=1$, so the loaded integrals can be continued along this
deformation. As $s\to1^-$, the $j$th radial path passes
immediately above each earlier point $t_a$, $a<j$.
On the $h$th real subinterval the factors
$(1-u/t_a)^b$, $a<h$, therefore have argument $-\pi b$,
while the other factors have argument $0$.
Because $\alpha>-1$ and $b>0$, the local endpoint
singularities are integrable; splitting the intervals at the
points $t_a$ justifies taking these boundary values.
The limiting $j$th column is consequently
\beq\label{eq:kz-cycle-transition}
 I_j\longrightarrow
 \left(\prod_{a=1}^n t_a^{-b/2}\right)
 \sum_{h=1}^j e^{-\pi\mathrm{i}b(h-1)}A_{\bullet h}(t),
\eeq
where $A_{\bullet h}$ denotes the $h$th column of $A$.
The change-of-columns matrix is triangular; its diagonal
entries are $e^{-\pi\mathrm{i}b(j-1)}$. Hence
\beq\label{eq:kz-det-real-boundary}
 \det\mc I\longrightarrow
 \left(\prod_{a=1}^n t_a^{-nb/2}\right)
 e^{-\pi\mathrm{i}b n(n-1)/2}\det A(t).
\eeq

On the other hand, the KZ equations imply
\[
 \ka z_i\frac{\partial}{\partial z_i}\det\mc I
 =\tr H_i(z;\mu)\det\mc I,
 \qquad
 \tr H_i(z;\mu)=-2\mu-
 \sum_{j\ne i}\frac{z_i+z_j}{z_i-z_j}.
\]
The expression on the right side of \eqref{eq:kz-exact-det}
obeys the same first-order equations: this follows by
logarithmic differentiation and $b=-1/\ka$.
Thus the quotient of $\det\mc I$ by the nonzero product of
powers in \eqref{eq:kz-exact-det} is constant under analytic
continuation.

To determine the constant, note that along the preceding
deformation the product of powers tends to
\[
 e^{-\pi\mathrm{i}b n(n-1)/2}
 \left(\prod_{a=1}^n t_a^{2\mu b-(n-1)b}\right)
 \prod_{a<c}(t_c-t_a)^{2b}.
\]
Inserting \eqref{eq:kz-V89-real-det} into
\eqref{eq:kz-det-real-boundary}, and using
\[
 \alpha=\left(2\mu-\frac n2\right)b,
\]
identifies the constant as
$(-1)^n\Gamma(\alpha+1)\Gamma(b)^n/
\Gamma(\alpha+nb+1)$, proving
\eqref{eq:kz-exact-det}.
At the unit-circle configuration,
$(1-z_a/z_c)(1-z_c/z_a)=|z_a-z_c|^2$ and the two
chosen logarithms sum to the real logarithm of this
positive number, yielding \eqref{eq:kz-det-circle}.
All factors are nonzero in the stated domain.
\end{proof}

\begin{example}[the case $n=2$]
Put
\[
 \alpha=\frac{1-2\mu}{\ka},\qquad
 b=-\frac1\ka,\qquad
 r=\frac{z_1}{z_2},\quad s=\frac{z_2}{z_1}.
\]
The master functions and the vector-valued rational functions are
\begin{align*}
 \Phi_1(x;z)&=x^\alpha(1-x)^b(1-rx)^b,\\
 \Phi_2(x;z)&=x^\alpha(1-x)^b(1-sx)^b,\\
 g_1(x;z)&=\frac{v_1}{x-1}+\frac{z_1}{z_1x-z_2}v_2,\\
 g_2(x;z)&=\frac{z_2}{z_2x-z_1}v_1+\frac{v_2}{x-1}.
\end{align*}
Therefore the two loaded radial integral solutions are
\begin{align}
 I_1(z)&=z_1^\alpha(z_1z_2)^{b/2}
 \int_0^1x^\alpha(1-x)^b(1-rx)^b
 \left(\frac{v_1}{x-1}+\frac{z_1}{z_1x-z_2}v_2\right)dx,\notag\\
 I_2(z)&=z_2^\alpha(z_1z_2)^{b/2}
 \int_0^1x^\alpha(1-x)^b(1-sx)^b
 \left(\frac{z_2}{z_2x-z_1}v_1+\frac{v_2}{x-1}\right)dx.
 \label{eq:n2-kz-integrals}
\end{align}
Here $(z_1z_2)^{b/2}$ abbreviates
$\exp\bigl(\tfrac b2(L_1(z)+L_2(z))\bigr)$;
all other branches are those specified in Subsection~\ref{subsec:kz-master-integrals}.

The normalized classical trigonometric $r$-operator on
$\mathbf V_1$ and the two KZ Hamiltonians are
\beq\label{eq:n2-kz-H}
 \begin{gathered}
 \Om(r)=\frac1{r-1}
 \begin{pmatrix}-(r+1)/2&1\\r&-(r+1)/2\end{pmatrix},\\[4pt]
 H_1=\Om(r)+\begin{pmatrix}-2\mu&0\\0&0\end{pmatrix},\qquad
 H_2=-\Om(r)+\begin{pmatrix}0&0\\0&-2\mu\end{pmatrix}.
 \end{gathered}
\eeq
The two columns of $\mc I$ satisfy
$\ka z_i\partial_{z_i}\mc I=H_i\mc I$, $i=1,2$.

All four entries can also be written in terms of Gauss
hypergeometric functions. Define
\[
 \begin{aligned}
 F(u)&={}_2F_1(-b,\alpha+1;\alpha+b+1;u),\\
 G(u)&={}_2F_1(1-b,\alpha+1;\alpha+b+2;u),\\
 B_0&=\frac{\Gamma(\alpha+1)\Gamma(b)}
 {\Gamma(\alpha+b+1)},\qquad
 B_1=\frac{\Gamma(\alpha+1)\Gamma(b+1)}
 {\Gamma(\alpha+b+2)}.
 \end{aligned}
\]
Euler's beta-integral representation gives
\beq\label{eq:n2-kz-hypergeom}
 \boxed{\quad
 \mc I(z)=-(z_1z_2)^{b/2}
 \begin{pmatrix}
 z_1^\alpha B_0F(r)&z_2^\alpha sB_1G(s)\\[4pt]
 z_1^\alpha rB_1G(r)&z_2^\alpha B_0F(s)
 \end{pmatrix}.
 \quad}
\eeq
The functions ${}_2F_1$ in this formula are defined by their
Euler integrals along $0<x<1$, using the same branch of
$(1-ux)^b$ as in \eqref{eq:n2-kz-integrals}; no independent
analytic-continuation convention is needed.

In this case \eqref{eq:kz-det-circle} reads
\beq\label{eq:n2-kz-det}
 \boxed{\quad
 \det\mc I(z)=
 \frac{\Gamma(\alpha+1)\Gamma(b)^2}
 {\Gamma(\alpha+2b+1)}
 (z_1z_2)^{2\mu b}|z_1-z_2|^{2b}.
 \quad}
\eeq
The fractional power in \eqref{eq:n2-kz-det} is defined
using $L_1(z)+L_2(z)$, and the formula is stated for
$|z_1|=|z_2|=1$ with distinct points.
\end{example}

\section{Limits of hypergeometric solutions in arbitrary angular chambers}
\label{sec:hyper-limit}

\subsection{Angular chambers, loadings, and the main theorem}
\label{subsec:hyp-limit-all-chambers}

Throughout this section $n\ge2$.  We study the root-of-unity
limits of the $n$ qKZ solutions $\Psi_1,\ldots,\Psi_n$ defined by
the circle integrals \eqref{eq:hyp-Psi} in
Section~\ref{sec:hyper-qkz}.  Their coordinate matrix in the
basis $v_1,\ldots,v_n$ is denoted by $\Psi$.  We use throughout
the same phase function $\Phi$, trigonometric weight function $w$,
elliptic weight functions $W_i$, and adjusting factors $Y_i$.

Fix an integer $m\ge1$ and a primitive $m$th root of unity
$\zeta_m$.  Choose real parameters
\beq\label{eq:hyp-limit-real-parameters}
 \ka<-n,\qquad 0<\mu<\frac18.
\eeq
For $0<r<1$, set
\beq\label{eq:hyp-limit-scaling}
 p=\zeta_m r,\qquad
 q=r^{-1/(2\ka)},\qquad
 \varkappa=r^{-2\mu/\ka},
 \qquad r\longrightarrow1^-.
\eeq
Thus $p\to\zeta_m$ and $q,\varkappa\to1$.  Since
$0<-1/(2\ka)<1$, for every $0<r<1$ we have
\[
 0<|p|=r<q=r^{-1/(2\ka)}<1.
\]
In particular, the inequalities in \eqref{eq:hyp-circle-domain}
remain valid along the entire limiting path.

We now specify the configurations of the variables $z$, which
are held fixed as $r\to1^-$.  For each permutation $\si\in S_n$,
let
\beq\label{eq:hyp-limit-chamber-sigma}
 \mc D_{m,\si}=\left\{z_a=e^{\mathrm{i}\vartheta_a}:\
 0<\vartheta_{\si(1)}<\cdots<\vartheta_{\si(n)}
 <\frac{2\pi}{m}\right\}.
\eeq
For $z\in\mc D_{m,\si}$, the points $z_a$ are distinct, lie on
the unit circle, and are different from $1$.  Together with the
inequality above, this ensures that \eqref{eq:hyp-Psi} is given
by its original integral over the counterclockwise unit circle
$|t|=1$ for every $0<r<1$.

Put
\[
 Z_a=z_a^m,\qquad a=1,\ldots,n.
\]
Thus $Z=\pi_m(z)$, and $Z_1,\ldots,Z_n$ are distinct points on
the unit circle: their arguments $m\vartheta_a$ lie in $(0,2\pi)$.
For each configuration in $\mc D_{m,\si}$ we use the logarithms
\[
 L_a=\mathrm{i}m\vartheta_a,\qquad e^{L_a}=Z_a,
\]
continued holomorphically to a sufficiently small simply
connected neighborhood of that configuration.

Introduce the parameters
\beq\label{eq:hyp-limit-parameters}
 \lambda_m=\frac1{m\ka},\qquad
 \alpha_m=\frac{n/2-2\mu}{m\ka},\qquad
 b_m=-\lambda_m.
\eeq
The assumptions \eqref{eq:hyp-limit-real-parameters} imply
\[
 0<\frac n2-2\mu<\frac n2,\qquad |\ka|>n.
\]
Since $m\ge1$ and $\ka<0$, it follows that
\[
 b_m>0,\qquad
 -\frac12<m\alpha_m<0,\qquad
 -\frac12<\alpha_m<0.
\]

For each $j=1,\ldots,n$, define the $\mathbf V_1$-valued radial integral
\beq\label{eq:hyp-limit-radials-sigma}
 \begin{aligned}
 I_j^\si(Z;m\ka,\mu)
 ={}&\left(\prod_{a=1}^n Z_a^{-\lambda_m/2}\right)
 \int_{[0,Z_j]}u^{\alpha_m}
 \prod_{a=1}^n\left(1-\frac{u}{Z_a}\right)^{b_m}\\
 &\hspace{18mm}\times
 \left(\sum_{k=1}^n\frac{v_k}{u-Z_k}\right)du.
 \end{aligned}
\eeq
The interval $[0,Z_j]$ is oriented from $0$ to $Z_j$.
The branches of the integrand are fixed by the parametrization
$u=Z_jx$, $0<x<1$.  More precisely,
\beq\label{eq:hyp-limit-loadings-sigma}
 \begin{aligned}
 u^{\alpha_m}&=\exp\bigl(\alpha_m(L_j+\log x)\bigr),\\
 \left(1-\frac{u}{Z_j}\right)^{b_m}&=(1-x)^{b_m},\qquad
 Z_a^{-\lambda_m/2}=\exp(-\lambda_m L_a/2).
 \end{aligned}
\eeq
Here $\log x$ and $\log(1-x)$ are the real logarithms.  For
$a\ne j$, the logarithm of $1-xZ_j/Z_a$ is chosen continuously
from its value $0$ at $x=0$; at the specified unit-circle
configurations it is the principal logarithm.  These choices
also define holomorphic branches on sufficiently small
complex neighborhoods of the configurations.

The integrals converge absolutely.  As $x\to0$ their
integrands are $O(x^{\alpha_m})$.  As $x\to1$, the $v_j$-component
is $O((1-x)^{b_m-1})$, whereas the other components are
$O((1-x)^{b_m})$.  The inequalities above give
$\alpha_m>-1$ and $b_m>0$, so all these terms are integrable.
When $\si=\mathrm{id}$, the integrals agree with
\eqref{eq:kz-I}--\eqref{eq:kz-loaded}, with $\ka$ replaced by
$m\ka$.  Let
\[
 \mc I_\si(Z;m\ka,\mu)=(I_1^\si,\ldots,I_n^\si)
\]
be their coordinate matrix in the basis $v_1,\ldots,v_n$.

To state the limit theorem, for $i=1,\ldots,n$ put
\beq\label{eq:hyp-limit-di}
 d_i=\frac n2-i+1-2\mu.
\eeq
For distinct labels $j,a$, define
\beq\label{eq:hyp-limit-order-signs}
 \eps_{ja}^\si=
 \begin{cases}
  1,&\si^{-1}(j)<\si^{-1}(a),\\
 -1,&\si^{-1}(j)>\si^{-1}(a).
 \end{cases}
\eeq
Thus $\eps_{ja}^\si$ records whether the point $z_j$ precedes
or follows $z_a$ in the angular order of the chamber.

\begin{thm}[Limit in an arbitrary angular chamber]
\label{thm:hyper-limit-all-chambers}
\label{thm:st1-all-chambers}
Under \eqref{eq:hyp-limit-real-parameters}, for every $\si\in S_n$
and every $z\in\mc D_{m,\si}$, the qKZ matrix
$\Psi(z;p,q,\varkappa)$, defined by the circle integrals
\eqref{eq:hyp-Psi} and the coordinate formula
\eqref{eq:hyp-matrix}, satisfies
\beq\label{eq:hyp-limit-main-sigma}
 \boxed{\quad
 \lim_{r\to1^-}\Psi(z;p,q,\varkappa)
 =\mc I_\si(Z;m\ka,\mu)\,C_{m,\si}^{(n)}(\ka,\mu).
 \quad}
\eeq
The matrix $C_{m,\si}^{(n)}=(c_{ji}^{(m,\si)})$ is upper triangular.
Its entries are
\beq\label{eq:hyp-limit-C-sigma}
 \boxed{
 c_{ji}^{(m,\si)}=
 \begin{cases}
 \displaystyle
 \frac{\sin(\pi\lambda_m)}{\pi}
 \exp\!\left[\pi\mathrm{i}\lambda_m\left(
 2\mu+d_i\eps_{ji}^\si+
 \sum_{\substack{a<i\\a\ne j}}\eps_{ja}^\si\right)\right],&j<i,\\[12pt]
 \displaystyle
 \frac{\sin(\pi d_i\lambda_m)}{\pi}
 \exp\!\left[\pi\mathrm{i}\lambda_m\left(
 2\mu+\sum_{a<i}\eps_{ia}^\si\right)\right],&j=i,\\[10pt]
 0,&j>i.
 \end{cases}}
\eeq
In particular,
\beq\label{eq:hyp-limit-C-compare-sigma}
 \boxed{\quad C_{m,\si}^{(n)}(\ka,\mu)
       =C_{1,\si}^{(n)}(m\ka,\mu).\quad}
\eeq
The convergence is locally uniform in $\mc D_{m,\si}$. It also
holds uniformly on compact subsets of sufficiently small complex
neighborhoods, where $\Psi$ is understood by its meromorphic
continuation and the limiting integrals use the prescribed
holomorphic branches. The matrix $C_{m,\si}^{(n)}(\ka,\mu)$ is
invertible under \eqref{eq:hyp-limit-real-parameters}. Moreover,
Proposition~\ref{prop:kz-basis}, applied after relabelling the points,
shows that $\mc I_\si(Z;m\ka,\mu)$ is invertible for every
$z\in\mc D_{m,\si}$ and every pair $(\ka,\mu)$ satisfying
\eqref{eq:hyp-limit-real-parameters}. Therefore, for all these
parameters and configurations, the limiting columns form a
fundamental matrix of local flat sections of
$\nabla_m(\ka,\mu)$ in \eqref{eq:nabla-m}.
\end{thm}

The analytic preliminaries are collected in
Subsection~\ref{subsec:hyp-limit-analytic}.  The proof begins
in Subsection~\ref{subsec:hyp-limit-residues} and ends in
Subsection~\ref{subsec:hyp-limit-coefficients}.

\subsection{Canonical specialization and examples}
\label{subsec:hyp-limit-canonical}

For $\si=\mathrm{id}$, write $C_m^{(n)}=C_{m,\mathrm{id}}^{(n)}$
and $\mc I=\mc I_{\mathrm{id}}$. The chamber is
\beq\label{eq:hyp-limit-chamber}
 z_a=e^{\mathrm{i}\vartheta_a},\qquad
 0<\vartheta_1<\cdots<\vartheta_n<\frac{2\pi}{m}.
\eeq
\begin{cor}[Canonical-order specialization]
\label{cor:hyper-limit}
Under assumptions \eqref{eq:hyp-limit-real-parameters} and
\eqref{eq:hyp-limit-chamber}, the qKZ hypergeometric solutions
have limits as $r\to1^-$, and
\beq\label{eq:hyp-limit-main}
 \boxed{\quad
 \lim_{r\to1^-}\Psi(z;p,q,\varkappa)
 =\mc I(Z;m\ka,\mu)\,C_m^{(n)}(\ka,\mu).
 \quad}
\eeq
Here $p,q,\varkappa$ depend on $r$ as in
\eqref{eq:hyp-limit-scaling}, and $Z=(Z_1,\ldots,Z_n)$ is the
abbreviation
\[
 Z_a=z_a^m,\qquad a=1,\ldots,n.
\]
By \eqref{eq:hyp-limit-chamber}, the arguments of $Z_a$ are
$m\vartheta_a$, in increasing order between $0$ and $2\pi$.
The powers appearing in $\mc I(Z;m\ka,\mu)$ have the branches
specified in Section~\ref{sec:hyper-kz}, in particular
$\log Z_a=\mathrm{i}m\vartheta_a$.

Put $\lambda_m=1/(m\ka)$.  The matrix
$C_m^{(n)}=(c_{ji}^{(m)})_{j,i=1}^n$ is upper triangular,
with entries
\beq\label{eq:hyp-limit-C}
 \boxed{\quad
 c_{ji}^{(m)}=
 \begin{cases}
 \displaystyle
 \frac{e^{\pi\mathrm{i}(n/2+1-2j)\lambda_m}}{\pi}
       \sin(\pi\lambda_m),&j<i,\\[8pt]
 \displaystyle
 \frac{e^{\pi\mathrm{i}(2\mu-i+1)\lambda_m}}{\pi}
       \sin\!\bigl(\pi(n/2-i+1-2\mu)\lambda_m\bigr),&j=i,\\[8pt]
 0,&j>i.
 \end{cases}\quad}
\eeq
In particular,
\beq\label{eq:hyp-limit-C-compare}
 \boxed{\quad
 C_m^{(n)}(\ka,\mu)=C_1^{(n)}(m\ka,\mu).
 \quad}
\eeq
The convergence in \eqref{eq:hyp-limit-main} is locally uniform
in the chamber \eqref{eq:hyp-limit-chamber}, and the equality
extends to local holomorphic branches in a sufficiently small
simply connected neighborhood of each configuration.
The matrix $C_m^{(n)}$ is invertible under the stated real
parameter assumptions. Consequently, for every such pair
$(\ka,\mu)$ and every configuration in
\eqref{eq:hyp-limit-chamber}, the limits of
$\Psi_1,\ldots,\Psi_n$ form a fundamental matrix of local flat
sections of $\nabla_m(\ka,\mu)$ in \eqref{eq:nabla-m}.
\end{cor}

\begin{proof}
Specialize Theorem~\ref{thm:hyper-limit-all-chambers} to
$\si=\mathrm{id}$. For $j<i$, we have
$\eps_{ji}^{\mathrm{id}}=1$ and
$\sum_{a<i,\,a\ne j}\eps_{ja}^{\mathrm{id}}=i-2j$.
Since $2\mu+d_i+i-2j=n/2+1-2j$, the off-diagonal entries in
\eqref{eq:hyp-limit-C-sigma} reduce to \eqref{eq:hyp-limit-C}.
For $j=i$, the sum of signs is $-(i-1)$, giving its diagonal entries.
The loadings coincide with those in Section~\ref{sec:hyper-kz};
all other assertions are the corresponding specializations.
\end{proof}

\begin{example}
For $n=2$, the matrix in \eqref{eq:hyp-limit-C} is
\beq\label{eq:hyp-limit-C2}
 C_m^{(2)}(\ka,\mu)=\frac1\pi
 \begin{pmatrix}
 e^{2\pi\mathrm{i}\mu\lambda_m}
       \sin\bigl(\pi(1-2\mu)\lambda_m\bigr)
 &\sin(\pi\lambda_m)\\[4pt]
 0
 &-e^{\pi\mathrm{i}(2\mu-1)\lambda_m}
       \sin(2\pi\mu\lambda_m)
 \end{pmatrix},
 \qquad \lambda_m=\frac1{m\ka}.
\eeq
\end{example}

\begin{example}[The second column in an arbitrary ordering]
Put $\eps=\eps_{12}^\si$ and retain $d_2=n/2-1-2\mu$.
The coefficient formula gives
\beq\label{eq:hyp-limit-second-column-sigma}
 \begin{aligned}
 \lim_{r\to1^-}\Psi_2
 ={}&\frac{\sin(\pi\lambda_m)}{\pi}
 e^{\pi\mathrm{i}\lambda_m(2\mu+d_2\eps)} I_1^\si\\
 &+\frac{\sin(\pi d_2\lambda_m)}{\pi}
 e^{\pi\mathrm{i}\lambda_m(2\mu-\eps)} I_2^\si.
 \end{aligned}
\eeq
These coefficients depend only on whether $z_1$ precedes $z_2$.
More generally, formula~\eqref{eq:hyp-limit-C-sigma} shows that the
coefficients in column $i$ involve only comparisons between the
angular positions of $z_1,\ldots,z_i$. The radial integrals themselves
still depend on all $n$ points.
\end{example}

\subsection{Standard analytic limits and cyclotomic consequences}
\label{subsec:hyp-limit-analytic}

We collect here the analytic facts used in the proof of
Theorem~\ref{thm:hyper-limit-all-chambers}.  The first lemma is
an elementary limit from the theory of basic hypergeometric
functions. Its one-exponent form, with a real exponent, is
recorded in \cite[Section~1.3, formulas (1.3.18)--(1.3.19)]{GR04},
together with continuation in the variable to the slit plane.
We prove below the version with two complex exponents and
convergence uniform on compact sets that is needed here.
The following two lemmas are consequences
of this limit and the product identities.  The final lemma uses
the classical Jackson beta identity.  None of these statements
requires the residue representation of our qKZ solutions.

\subsubsection{The elementary product limit}

\begin{lem}[Classical limit of Pochhammer-product ratios]
\label{lem:hyp-elementary-products}
Let $A,B\in\C$ and $y\in\C\setminus[1,\infty)$.  Then
\beq\label{eq:hyp-elementary-P-limit}
 \lim_{h\to0^+}
 \frac{\displaystyle\prod_{k=0}^\infty
            (1-ye^{-h(k+A)})}
      {\displaystyle\prod_{k=0}^\infty
            (1-ye^{-h(k+B)})}
 =\exp\bigl((B-A)\operatorname{Log}(1-y)\bigr).
\eeq
Here $\operatorname{Log}$ is the principal logarithm.  The
convergence is uniform when $A,B$ range over compact subsets
of $\C$ and $y$ ranges over a compact subset of
$\C\setminus[1,\infty)$.
\end{lem}

\begin{proof}
We first calculate the limit, postponing the convergence
justifications until the end.
For sufficiently small $h>0$, define
\[
 L_h=\sum_{k=0}^\infty\bigl[
 \operatorname{Log}(1-ye^{-h(k+A)})
 -\operatorname{Log}(1-ye^{-h(k+B)})\bigr].
\]
The quotient under the limit sign in
\eqref{eq:hyp-elementary-P-limit} equals $\exp(L_h)$. For each
$k$ and each $s$ on the segment
joining $B$ to $A$,
\[
 \frac{d}{ds}\operatorname{Log}(1-ye^{-h(k+s)})
 =h\frac{ye^{-h(k+s)}}{1-ye^{-h(k+s)}}.
\]
Integrating from $B$ to $A$, and then summing over $k$, gives
\beq\label{eq:hyp-elementary-Lh}
 L_h=\int_B^A S_h(s)\,ds,
 \qquad
 S_h(s)=h\sum_{k=0}^\infty
 \frac{ye^{-h(k+s)}}{1-ye^{-h(k+s)}}.
\eeq
Put
\[
 g_y(u)=\frac{ye^{-u}}{1-ye^{-u}}.
\]
For fixed $s$, the sum
\[
 S_h(s)=h\sum_{k=0}^\infty g_y(hk+hs)
\]
is a Riemann sum for
\[
 I_h(s)=\int_0^\infty g_y(u+hs)\,du
       =-\operatorname{Log}(1-ye^{-hs}).
\]
The approximation by this Riemann sum becomes exact in the
limit $h\to0^+$.  Moreover, $hs\to0$ for every fixed $s$;
therefore
\beq\label{eq:hyp-elementary-Sh}
 \lim_{h\to0^+}S_h(s)
 =\lim_{h\to0^+}I_h(s)
 =-\operatorname{Log}(1-y).
\eeq
Thus $s$ does not occur in the limiting value of the sum.
Taking the limit in \eqref{eq:hyp-elementary-Lh}, we obtain
\[
 \lim_{h\to0^+}L_h
 =\int_B^A-\operatorname{Log}(1-y)\,ds
 =(B-A)\operatorname{Log}(1-y).
\]
Exponentiating gives \eqref{eq:hyp-elementary-P-limit}.

We now justify the operations above.  Let $A,B$ vary in fixed
compact subsets of $\C$ and $y$ in a compact subset of
$\C\setminus[1,\infty)$.  The points $s$ on the segments
joining $B$ and $A$ then also lie in a fixed compact subset
of $\C$. The set of all $ty$, with $0\le t\le1$ and $y$
in the chosen compact set, is compact and disjoint from
$[1,\infty)$. For sufficiently small $h$, its perturbations
$te^{-hs}y$ remain disjoint from this ray, uniformly for the
specified $s$. Consequently $1-te^{-hs}y$ avoids
$(-\infty,0]$ and is uniformly separated from zero. Taking
$t=e^{-hk}$ justifies the principal logarithms used above.
For every fixed sufficiently small $h>0$, each logarithm in
the summand defining $L_h$ is $O(e^{-hk})$ as $k\to\infty$,
since $\operatorname{Log}(1-w)=-w+O(w^2)$ at $w=0$.
The summands of $S_h(s)$ have the same geometric decay. Hence those series
converge absolutely; the latter series converges uniformly
along the integration segment.  This justifies integration
term by term.

Finally,
\[
 g_y'(u)=-\frac{ye^{-u}}{(1-ye^{-u})^2}.
\]
The same separation from zero shows that there is a constant
$C$, uniform over the compact sets under consideration, such
that
\[
 |g_y'(u+hs)|\le C e^{-u},\qquad u\ge0.
\]
By comparing the value at the left endpoint of each interval
$[hk,h(k+1)]$ with the integral over that interval, we get
\[
 \left|S_h(s)-I_h(s)\right|
 \le h\int_0^\infty|g_y'(u+hs)|\,du
 \le Ch.
\]
Also $-\operatorname{Log}(1-ye^{-hs})$ tends uniformly to
$-\operatorname{Log}(1-y)$ on the specified compact sets,
because $hs\to0$ uniformly there.  Thus the convergence in
\eqref{eq:hyp-elementary-Sh} is uniform along the segment,
which justifies taking the limit under the integral in
\eqref{eq:hyp-elementary-Lh} and proves the stated uniformity.
\end{proof}

\subsubsection{Cyclotomic products and theta functions}

We now return to the scaling in
\eqref{eq:hyp-limit-scaling} and introduce the positive real base
\[
 P=r^m=p^m.
\]
Then $P\to1^-$, $p=\zeta_m P^{1/m}$, and
\beq\label{eq:hyp-limit-scale-identity}
 q^2=P^{b_m},\qquad
 \varkappa q^{-n}=P^{\alpha_m},\qquad p^m=P.
\eeq
We write $\tht'_p(1)$ for the derivative of $\tht_p(u)$
with respect to $u$ at $u=1$.

\begin{lem}[Limits of Pochhammer products]
\label{lem:hyp-cyclotomic-products}
Let $p=\zeta_mP^{1/m}$ and $P\to1^-$.  Fix
$a\in\{0,\ldots,m-1\}$.  If $v^m\notin[1,\infty)$, then
\beq\label{eq:hyp-limit-products}
 \begin{aligned}
 \frac{(p^a v;p)_\infty}{(q^2p^a v;p)_\infty}
 &\longrightarrow (1-v^m)^{b_m},\\
 \frac{(p^{a+1}v;p)_\infty}
      {(pq^2p^a v;p)_\infty}
 &\longrightarrow (1-v^m)^{b_m},\\
 \frac{(p^{a+1}v;p)_\infty}
      {(q^2p^a v;p)_\infty}
 &\longrightarrow
 \frac{(1-v^m)^{b_m}}{1-\zeta_m^a v}.
 \end{aligned}
\eeq
Here $(1-v^m)^{b_m}=\exp(b_m\operatorname{Log}(1-v^m))$.
The limits are uniform on compact subsets of the stated
domain.  In particular, they are uniform for $v=\rho x$,
$0\le x\le1$, if $|\rho|=1$ and $\rho^m\ne1$.
\end{lem}

\begin{proof}
The exact factorization
\beq\label{eq:hyp-factorization}
 (u;p)_\infty=\prod_{j=0}^{m-1}(up^j;P)_\infty
\eeq
reduces the first quotient to
\[
 \prod_{j=0}^{m-1}
 \frac{(P^{(a+j)/m}\zeta_m^{a+j}v;P)_\infty}
      {(P^{(a+j)/m+b_m}\zeta_m^{a+j}v;P)_\infty}.
\]
Because $v^m\notin[1,\infty)$, each number
$\zeta_m^{a+j}v$ is in $\C\setminus[1,\infty)$.
Lemma~\ref{lem:hyp-elementary-products}, with $P=e^{-h}$,
therefore gives the limit
\[
 \exp\left(b_m\sum_{j=0}^{m-1}
    \operatorname{Log}(1-\zeta_m^{a+j}v)\right).
\]
On the indicated domain, the cyclotomic identity
$\prod_{j=0}^{m-1}(1-\zeta_m^{a+j}v)=1-v^m$
has the logarithmic form
\[
 \sum_{j=0}^{m-1}\operatorname{Log}(1-\zeta_m^{a+j}v)
 =\operatorname{Log}(1-v^m).
\]
Indeed, for $|v|<1$ both sides are given by the Taylor
series of the logarithms; the identity then extends
throughout the stated domain by analytic continuation.
This proves the first limit.  The second is the same
calculation with $a$ replaced by $a+1$.  Finally,
$(u;p)_\infty=(1-u)(pu;p)_\infty$ implies
\[
 \frac{(p^{a+1}v;p)_\infty}{(q^2p^av;p)_\infty}
 =\frac{1}{1-p^av}
   \frac{(p^av;p)_\infty}{(q^2p^av;p)_\infty},
\]
which proves the third limit.  Uniformity follows from
Lemma~\ref{lem:hyp-elementary-products} because the
factorization has only $m$ factors.
\end{proof}

For $0<P<1$, recall that the $P$-gamma function is defined by
\beq\label{eq:hyp-P-gamma-definition}
 \Gamma_P(s)=(1-P)^{1-s}
       \frac{(P;P)_\infty}{(P^s;P)_\infty}.
\eeq
Here $P^s=\exp(s\log P)$, with the real logarithm of $P$.
In this notation, $P=r^m$ is the positive real base; this is
not the $p$-adic gamma function.  The $P$-gamma function satisfies
\[
 \Gamma_P(1)=1,\qquad
 \Gamma_P(s+1)=\frac{1-P^s}{1-P}\Gamma_P(s),\qquad
 \lim_{P\to1^-}\Gamma_P(s)=\Gamma(s),
\]
where the limit is taken away from the poles of the classical
gamma function; see \cite[(5.18.4), (5.18.7), and
(5.18.10)]{DLMF}.

\medskip
\noindent\emph{Locally uniform convergence of the $P$-gamma function.}
We will use the more precise statement that
$\Gamma_P(s)\to\Gamma(s)$ uniformly on every compact subset of
$\C\setminus\{0,-1,-2,\ldots\}$. Here is a justification of
this uniformity from the classical limit above.
For $s=x+\mathrm{i}t$ with $x>0$, the product definition gives
\[
 |\Gamma_P(x+\mathrm{i}t)|\le\Gamma_P(x),
\]
because $|1-P^{x+k+\mathrm{i}t}|\ge1-P^{x+k}$ for each $k\ge0$.
Also, $\log\Gamma_P(x)$ is convex for $x>0$
\cite[Section~5.18(ii)]{DLMF}. Hence, for $0<a\le x\le b$,
\[
 \Gamma_P(x)\le\max\{\Gamma_P(a),\Gamma_P(b)\}.
\]
The right side is bounded as $P\to1^-$ by the classical
pointwise limit. Thus the holomorphic functions $\Gamma_P$
are locally bounded on $\operatorname{Re}s>0$. Montel's
theorem and the identity theorem, applied to their pointwise
limit on the positive real axis, give locally uniform
convergence to $\Gamma$ on that half-plane.

For a compact set $K\subset\C\setminus\{0,-1,-2,\ldots\}$,
choose an integer $N$ such that $\operatorname{Re}(s+N)>0$
for all $s\in K$. Iterating the recurrence gives
\[
 \Gamma_P(s)=\Gamma_P(s+N)
       \prod_{j=0}^{N-1}\frac{1-P}{1-P^{s+j}}.
\]
Each factor in the finite product tends uniformly on $K$
to $1/(s+j)$; its denominator is nonzero for $P$
sufficiently close to $1$. The asserted locally uniform
limit on $K$ follows. In particular, on each such compact set,
$\Gamma_P$ is holomorphic for all $P$ sufficiently close to $1$.

\begin{lem}[Limits of theta functions]
\label{lem:hyp-theta-limits}
Let $p=\zeta_mP^{1/m}$ and $P\to1^-$.  If
$u\in\C^\times$ and $u^m\notin\R_{\ge0}$, then, for
fixed $A,B\in\C$,
\beq\label{eq:hyp-limit-theta-ratio}
 \frac{\tht_p(P^A u)}{\tht_p(P^B u)}
 \longrightarrow
 \exp\bigl((B-A)\operatorname{Log}(-u^m)\bigr).
\eeq
The convergence is locally uniform in $u$ on this domain,
and uniform when $A,B$ also range over fixed
compact subsets of $\C$.  In addition, for $c\in\C\setminus\Z$,
\beq\label{eq:hyp-limit-theta-derivative}
 \frac{\tht_p(P^{-c})}{(1-P)\tht'_p(1)}
 \longrightarrow\frac{\sin(\pi c)}{\pi}.
\eeq
For fixed $m$ and $\zeta_m$, the latter convergence is
uniform for $c$ in compact subsets of $\C\setminus\Z$.
\end{lem}

\begin{proof}
Use the product definition
\[
 \tht_p(w)=(w;p)_\infty(p/w;p)_\infty(p;p)_\infty.
\]
By \eqref{eq:hyp-factorization} and
Lemma~\ref{lem:hyp-elementary-products}, the quotient
of the first products in \eqref{eq:hyp-limit-theta-ratio}
tends to
\[
 \exp\bigl((B-A)\operatorname{Log}(1-u^m)\bigr).
\]
The quotient of the products with arguments
$pP^{-A}/u$ and $pP^{-B}/u$ tends to
\[
 \exp\bigl((A-B)\operatorname{Log}(1-u^{-m})\bigr).
\]
Since $u^m\notin\R_{\ge0}$, the principal logarithms satisfy
\[
 \operatorname{Log}(1-u^m)-\operatorname{Log}(1-u^{-m})
 =\operatorname{Log}(-u^m).
\]
To check the branches, put $w=u^m$ and
$D=\C\setminus[0,\infty)$.  For $w\in D$, the three numbers
$1-w$, $1-w^{-1}$, and $-w$ avoid the negative real axis,
so their principal logarithms are holomorphic on $D$.
The identity
\[
 \frac{1-w}{1-w^{-1}}=-w
\]
shows that
\[
 \operatorname{Log}(1-w)-\operatorname{Log}(1-w^{-1})
                    -\operatorname{Log}(-w)
\]
takes values in $2\pi\mathrm{i}\Z$.  This expression is
continuous on the connected domain $D$, so it is constant.
At $w=-1$ it equals $\operatorname{Log}2-\operatorname{Log}2-
\operatorname{Log}1=0$.  Hence it is zero throughout $D$.
Multiplying the two limits now proves
\eqref{eq:hyp-limit-theta-ratio}.

For the second assertion, the exact identity
$\tht'_p(1)=-(p;p)_\infty^3$ gives
\[
 \frac{\tht_p(P^{-c})}{(1-P)\tht'_p(1)}
 =-\frac{(P^{-c};p)_\infty(pP^c;p)_\infty}
          {(1-P)(p;p)_\infty^2}.
\]
Factoring each product by \eqref{eq:hyp-factorization}
separates the factors whose arguments tend to $1$:
\[
 -\frac{(P^{-c};P)_\infty(P^{1+c};P)_\infty}
        {(1-P)(P;P)_\infty^2}.
\]
For each $j=1,\ldots,m-1$, the remaining factors form
\[
 \frac{(\zeta_m^jP^{j/m-c};P)_\infty}
      {(\zeta_m^jP^{j/m};P)_\infty}
 \frac{(\zeta_m^jP^{j/m+c};P)_\infty}
      {(\zeta_m^jP^{j/m};P)_\infty}.
\]
By Lemma~\ref{lem:hyp-elementary-products}, this
expression tends to $1$.  The first displayed quotient
can be rewritten, by the definition and recurrence of the
$P$-gamma function, as
\[
 -\frac{1-P^{-c}}{1-P}
  \frac{1}{\Gamma_P(1-c)\Gamma_P(1+c)}.
\]
Now use the locally uniform $P$-gamma limit established
above and the reflection identity
$\Gamma(1-c)\Gamma(1+c)=\pi c/\sin(\pi c)$.
Since $(1-P^{-c})/(1-P)\to-c$, the limit is
$\sin(\pi c)/\pi$.  The locally uniform versions of the
product and $P$-gamma limits give the claimed uniformity.
\end{proof}

\subsubsection{Jackson beta convergence}

The next statement is a uniform version of the classical
Jackson beta integral.  We recall the identity and the
limit of the $P$-gamma function from
\cite[(5.18.10)--(5.18.12)]{DLMF}.  The
additional continuous factor and the parameter-uniform
statement follow from a standard approximation argument.

\begin{lem}[Uniform Jackson--beta convergence]
\label{lem:hyp-Jackson-beta}
Let $a,b>0$ be real, and let $h_P:[0,1]\to\C$ be uniformly
bounded functions converging uniformly to a continuous
function $h$ as $P\to1^-$.  Then
\beq\label{eq:hyp-beta-lemma}
 \begin{aligned}
 &(1-P)\sum_{N=0}^\infty P^{Na}
 \frac{(P^{N+1};P)_\infty}{(P^{N+b};P)_\infty}
 h_P(P^N)\\
 &\hspace{15mm}\longrightarrow
 \int_0^1x^{a-1}(1-x)^{b-1}h(x)\,dx.
 \end{aligned}
\eeq
There is also a version with a vanishing first summand:
\beq\label{eq:hyp-beta-shifted}
 \begin{aligned}
 &(1-P)\sum_{N=0}^\infty P^{Na}
 \frac{(P^N;P)_\infty}{(P^{N+b};P)_\infty}
 h_P(P^N)\\
 &\hspace{15mm}\longrightarrow
 \int_0^1x^{a-1}(1-x)^b h(x)\,dx.
 \end{aligned}
\eeq
Both limits are uniform when $(a,b)$ ranges over a compact
subset of $(0,\infty)^2$ and when any auxiliary parameters
range over a compact set, provided the functions $h_P$
converge uniformly to a jointly continuous function of
$x$ and these parameters.
\end{lem}

\begin{proof}
Define the Jackson integral by
\beq\label{eq:hyp-limit-Jackson}
 \int_0^1 f(x)\,d_Px=(1-P)\sum_{N=0}^\infty P^Nf(P^N).
\eeq
The measures
\[
 d\nu_P(x)=x^{a-1}
 \frac{(Px;P)_\infty}{(P^bx;P)_\infty}\,d_Px
\]
are positive.  The Jackson beta identity
\cite[DLMF, (5.18.11)--(5.18.12)]{DLMF} gives, for
every nonnegative integer $\ell$,
\[
 \int_0^1 x^\ell\,d\nu_P(x)
 =\frac{\Gamma_P(a+\ell)\Gamma_P(b)}
        {\Gamma_P(a+b+\ell)}
 \longrightarrow B(a+\ell,b).
\]
Hence the integrals against polynomials converge to the
corresponding integrals against
$x^{a-1}(1-x)^{b-1}dx$.  Since the measures are positive
and their total masses remain bounded, approximation of
continuous functions by polynomials proves the first limit
with $h_P$ replaced by $h$.  The replacement of $h$ by
$h_P$ changes the integral by at most
$\|h_P-h\|_\infty\nu_P([0,1])$, which tends to zero.

For the second limit, the summand at $N=0$ is zero.
Set $N=M+1$ in the remaining sum.  The result is
\[
 P^a(1-P)\sum_{M=0}^\infty P^{Ma}
 \frac{(P^{M+1};P)_\infty}{(P^{M+b+1};P)_\infty}
 h_P(P^{M+1}).
\]
Apply the first limit with $b$ replaced by $b+1$
and $h_P(x)$ replaced by $h_P(Px)$.

Finally, the locally uniform $P$-gamma limit established
above implies uniform convergence of the moments and uniform
bounds on total mass for $(a,b)$ in compact subsets of
$(0,\infty)^2$.  Polynomial approximation in $x$ can
be chosen uniformly for jointly continuous functions of
$x$ and auxiliary parameters on compact sets.  The same
argument proves the stated uniformity for both limits.
\end{proof}

The analytic preliminaries are complete.  We now begin the
proof of Theorem~\ref{thm:hyper-limit-all-chambers} by
analyzing the poles of the qKZ circle integrands.

\subsection{Poles and exact residue strings}
\label{subsec:hyp-limit-residues}

We begin the proof of Theorem~\ref{thm:hyper-limit-all-chambers}
with the poles of the complete integrand, rather than the poles
of its individual factors. For $i=1,\ldots,n$ write
\[
 \om_i(t;z)=\Phi(t;z)Y_i(z)W_i(t;z)w(t;z)\,\frac{dt}{t},
\]
so that $\Psi_i(z)=(2\pi\mathrm{i})^{-1}\int_{|t|=1}\om_i(t;z)$.

\begin{lem}[Poles inside the unit circle]
\label{lem:hyp-inward-poles}
Assume \eqref{eq:hyp-limit-real-parameters} and
\eqref{eq:hyp-limit-scaling}, and let
$z\in\mc D_{m,\si}$. For each $i=1,\ldots,n$, the
complete vector-valued differential form $\om_i(t;z)$ has,
in the punctured disk $0<|t|<1$, exactly the simple poles
\beq\label{eq:hyp-limit-strings}
 t=qp^s z_j,\qquad s=0,1,\ldots,\qquad 1\le j\le i.
\eeq
The pole strings belonging to different labels $j$ are
disjoint. At $t=qz_j$, the $v_k$-component has zero residue
when $k<j$.

The circle integral equals the sum of the residues at
these poles:
\beq\label{eq:hyp-limit-residue-sum}
 \Psi_i(z)=\sum_{j=1}^i\sum_{s=0}^\infty
 \Res_{t=qp^sz_j}\om_i(t;z).
\eeq
The series in \eqref{eq:hyp-limit-residue-sum} converges
absolutely for each fixed $0<r<1$.
\end{lem}

\begin{proof}
First, the adjusting factor $Y_i(z)$ is finite and nonzero.
Indeed, every argument of a theta function in its numerator
or denominator is $c z_a$ for some positive real number $c$.
If $\tht_p(cz_a)=0$, then $cz_a=p^k$ for an integer $k$,
and therefore $c^mZ_a=P^k>0$. This contradicts
$Z_a\notin\R_{>0}$. Thus $Y_i$ neither cancels nor introduces
any pole in the variable $t$.

The denominator factors in the phase function
\eqref{eq:hyp-phase} have zeros only when
\[
 t=q^{-1}p^{-s}z_a,\qquad s=0,1,\ldots .
\]
All these points lie outside the unit circle. The numerator
of $\Phi$ has a simple zero at $t=qz_a$. Its other zeros
are $t=qp^{-s}z_a$, $s\ge1$, which are outside the unit
circle because $|p|<q$.

Each rational weight function $w_k$, defined by
\eqref{eq:hyp-wj}, can have poles inside the unit circle
only at $t=qz_a$, $a\le k$. The simple zero of $\Phi$
at each such point cancels the corresponding pole of $w_k$.
Thus $\Phi(t;z)w(t;z)$ is holomorphic in $0<|t|<1$.

Now consider the elliptic weight $W_i$ in
\eqref{eq:hyp-Wi}. Its denominator theta functions can
vanish only when
\[
 t=qp^s z_j,\qquad s\in\Z,\quad 1\le j\le i,
\]
since the zeros of $\tht_p(u)$ are $u=p^s$, $s\in\Z$.
The points with $s\ge0$ lie inside the unit circle.
Those with $s\le-1$ lie outside it, because
$q/|p|>1$.  If two of the inside points coincide, their
absolute values give $s=s'$, and then $z_j=z_{j'}$,
which implies $j=j'$. Thus the strings are disjoint, and
each zero of a denominator theta function is simple.

No numerator factor in $W_i$ cancels these zeros under our
parameter assumptions. Indeed, the relevant numerator
arguments at these points have moduli $|p|^{s-1/\ka}$
or $|p|^{s-d_i/\ka}$, where $d_i$ is defined in
\eqref{eq:hyp-limit-di}. We have
$0<-1/\ka<1$, $|d_i/\ka|<1$, and $d_i\ne0$.
Consequently, neither exponent can be an integer, as
would be required for a zero of $\tht_p$.

At an inside point with $s\ge1$, the vector
$\Phi(t;z)w(t;z)$ is regular and nonzero (its
$v_n$-component is nonzero). Therefore the pole of
$W_i$ is a simple pole of $\om_i$. At $s=0$, the zero
of $\Phi$ cancels the pole of $w_k$ when $k\ge j$,
leaving a nonzero regular value, and hence a simple pole
after multiplication by $W_i$. For $k<j$, the function
$w_k$ is regular at $t=qz_j$; its product with $\Phi$
vanishes there and cancels the pole of $W_i$.
This proves the assertions about poles and components.

It remains to justify the residue sum. Put $P=p^m=r^m$.
By theta quasiperiodicity \eqref{eq:hyp-theta} and
\eqref{eq:hyp-limit-scale-identity},
\[
 W_i(pt;z)=\varkappa q^{-n}W_i(t;z)
              =P^{\alpha_m}W_i(t;z).
\]
Iterating $mN$ times gives
\[
 W_i(P^Nt;z)=P^{Nm\alpha_m}W_i(t;z).
\]
Choose a radius $\rho$ such that
$q|p|^m<\rho<q|p|^{m-1}$. No pole lies on
$|t|=\rho P^N$, for any $N\ge0$. On the fixed
circle $|t|=\rho$, as $N\to\infty$ we have uniformly
\[
 \Phi(P^Nt;z)\longrightarrow1,
 \qquad w(P^Nt;z)=O(P^N).
\]
Since $dt/t$ is unchanged under this rescaling, the
integral on the small circle is
\[
 O\bigl(P^{N(1+m\alpha_m)}\bigr)\longrightarrow0,
\]
where $1+m\alpha_m>0$ by \eqref{eq:hyp-limit-parameters}
and the inequalities following it. Applying the residue
theorem to the annulus bounded by $|t|=1$ and
$|t|=\rho P^N$, then letting $N\to\infty$, proves
\eqref{eq:hyp-limit-residue-sum}.

Finally, as $s\to\infty$ the residue of $W_i(t;z)dt/t$
at $t=qp^sz_j$ is a constant times
$(\varkappa q^{-n})^s$, by the displayed quasiperiodicity.
At the same point $\Phi(t;z)\to1$ and
$w(t;z)=O(|p|^s)$. Hence the vector residue is
$O(|p|^{s(1+m\alpha_m)})$, a geometrically decreasing
bound. This proves absolute convergence.
\end{proof}

For $j\le i$, denote the contribution of the $j$th inward
string by
\[
 T_{ji}(P,z)=\sum_{s=0}^{\infty}
       \Res_{t=qp^sz_j}\om_i(t;z),
 \qquad P=p^m=r^m.
\]
Here $p,q,\varkappa$ depend on $P$ through the scaling
\eqref{eq:hyp-limit-scaling}. The preceding lemma gives
\[
 \Psi_i(z)=\sum_{j=1}^i T_{ji}(P,z).
\]
We next calculate the residues in each $T_{ji}$. Set
\[
 \rho_{ja}=\frac{z_j}{z_a}.
\]
For $j\le i$, define the constants $R_{ji}(p,z)$ by
\beq\label{eq:hyp-limit-Roff}
 \begin{aligned}
 R_{ji}(p,z)={}&\frac{\tht_p(q^2)}{\tht'_p(1)}
 \frac{\tht_p(\varkappa^{-1}q^{n-2i+2}\rho_{ji})}
      {\tht_p(\rho_{ji})}\\
 &\times\prod_{\substack{a<i\\a\ne j}}
 \frac{\tht_p(q^2\rho_{ja})}{\tht_p(\rho_{ja})},
 \qquad j<i;
 \end{aligned}
\eeq
and
\beq\label{eq:hyp-limit-Rdiag}
 R_{ii}(p,z)=
 \frac{\tht_p(\varkappa^{-1}q^{n-2i+2})}
      {\tht'_p(1)}
 \prod_{a<i}\frac{\tht_p(q^2\rho_{ia})}
                    {\tht_p(\rho_{ia})}.
\eeq

\begin{lem}[Exact residues along the inward strings]
\label{lem:hyp-exact-residues}
For $j\le i$ and $s\ge0$, the residue of the complete
integrand $\om_i(t;z)$ at $t=qp^sz_j$ is
\beq\label{eq:hyp-limit-full-residue}
 \Res_{t=qp^sz_j}\om_i(t;z)
 =Y_i(z)R_{ji}(p,z)(\varkappa q^{-n})^s
    \bigl[\Phi(qp^sz_j;z)w(qp^sz_j;z)\bigr],
\eeq
where the bracket denotes the value after cancellation
of removable factors.  Its $v_k$-component is given
explicitly by
\beq\label{eq:hyp-limit-phase-weight-string}
 \begin{aligned}
 &\Phi(qp^sz_j;z)w_k(qp^sz_j;z)
 =-q^{-(k-1)}p^s\rho_{jk}\\[-2pt]
 &\quad\times
 \prod_{a<k}
 \frac{(p^{s+1}\rho_{ja};p)_\infty}
      {(pq^2p^s\rho_{ja};p)_\infty}
 \frac{(p^{s+1}\rho_{jk};p)_\infty}
      {(q^2p^s\rho_{jk};p)_\infty}
 \prod_{a>k}
 \frac{(p^s\rho_{ja};p)_\infty}
      {(q^2p^s\rho_{ja};p)_\infty}.
 \end{aligned}
\eeq
Both formulas hold also for $s=0$, including the vanishing
of the $v_k$-component for $k<j$.
\end{lem}

\begin{proof}
First we compute the residues of $W_i(t;z)dt/t$.
The function $\tht_p(u)$ has a simple zero at $u=1$,
and
\[
 \Res_{u=1}\frac{du}{u\tht_p(u)}
 =\frac1{\tht'_p(1)}.
\]
If $j<i$, the pole of $W_i$ at $t=qz_j$ comes from the
factor $\tht_p(qt/z_j)/\tht_p(q^{-1}t/z_j)$.
Its residue, with respect to $dt/t$, is
$\tht_p(q^2)/\tht'_p(1)$. Evaluating the other factors
of $W_i$ at $t=qz_j$ gives precisely
$R_{ji}(p,z)$ in \eqref{eq:hyp-limit-Roff}.
If $j=i$, the pole comes from the denominator of the
leading theta ratio in $W_i$, and its residue is
$R_{ii}(p,z)$ in \eqref{eq:hyp-limit-Rdiag}.
For $s\ge0$, the quasiperiodicity already used in the
first lemma gives
\beq\label{eq:hyp-limit-residue-W}
 \Res_{t=qp^sz_j}W_i(t;z)\frac{dt}{t}
 =(\varkappa q^{-n})^s R_{ji}(p,z).
\eeq
Multiplying this identity by the regular value of
$Y_i\Phi w$ proves \eqref{eq:hyp-limit-full-residue}.

We now evaluate $\Phi w_k$ at a string point. For $s\ge1$,
substitution of $t=qp^sz_j$ into \eqref{eq:hyp-wj}
gives
\beq\label{eq:hyp-limit-w-on-string}
 \begin{aligned}
 w_k(qp^sz_j;z)
 ={}&-q^{-(k-1)}p^s\rho_{jk}
 \frac1{1-p^s\rho_{jk}}\\
 &\quad\times\prod_{a<k}
   \frac{1-q^2p^s\rho_{ja}}{1-p^s\rho_{ja}}.
 \end{aligned}
\eeq
The phase at the same point is
\[
 \Phi(qp^sz_j;z)=\prod_{a=1}^n
 \frac{(p^s\rho_{ja};p)_\infty}
      {(q^2p^s\rho_{ja};p)_\infty}.
\]
The elementary identity
\[
 (u;p)_\infty=(1-u)(pu;p)_\infty
\]
gives the two cancellation formulas
\[
 \begin{aligned}
 \frac{(u;p)_\infty}{(q^2u;p)_\infty}
 \frac{1-q^2u}{1-u}
 &=\frac{(pu;p)_\infty}{(pq^2u;p)_\infty},\\
 \frac{(u;p)_\infty}{(q^2u;p)_\infty}
 \frac1{1-u}
 &=\frac{(pu;p)_\infty}{(q^2u;p)_\infty}.
 \end{aligned}
\]
Apply the first formula with $u=p^s\rho_{ja}$ for
every $a<k$, and the second with $u=p^s\rho_{jk}$.
The factors with $a>k$ remain unchanged. Their product
is precisely \eqref{eq:hyp-limit-phase-weight-string}.

The expression on the right of
\eqref{eq:hyp-limit-phase-weight-string} is defined at
$s=0$. It is the value of the holomorphic product
$\Phi w_k$ after cancellation, so the equality extends
to $s=0$. When $k<j$, its last product contains
$(\rho_{jj};p)_\infty=(1;p)_\infty=0$, proving the
asserted vanishing. This completes the proof.
\end{proof}

\paragraph{Regrouping the residues as Jackson integrals.}
For use in the convergence argument, we now rewrite the
exact formulas of Lemma~\ref{lem:hyp-exact-residues}.
We reserve $r$ for the real radial parameter in
\eqref{eq:hyp-limit-scaling} and use $\nu$ for residue classes
modulo $m$. Every $s\ge0$ has a unique representation
\beq\label{eq:hyp-limit-index-decomp}
 s=mN+\nu,\qquad 0\le \nu<m,\quad N\ge0,
 \qquad x=P^N.
\eeq
Then $p^s=p^{\nu}x$ and
$(\varkappa q^{-n})^s=P^{\nu\alpha_m}x^{m\alpha_m}$.
Define the $v_k$-component of the kernel by
\beq\label{eq:hyp-limit-kernel}
 \begin{aligned}
 B_{jk}^{(\nu,P)}(x)
 ={}&q^{-(k-1)}\rho_{jk}
 \prod_{a<k}
 \frac{(p^{\nu+1}x\rho_{ja};p)_\infty}
      {(pq^2p^{\nu}x\rho_{ja};p)_\infty}\\[-1pt]
 &\times
 \frac{(p^{\nu+1}x\rho_{jk};p)_\infty}
      {(q^2p^{\nu}x\rho_{jk};p)_\infty}
 \prod_{a>k}
 \frac{(p^{\nu}x\rho_{ja};p)_\infty}
      {(q^2p^{\nu}x\rho_{ja};p)_\infty},
 \end{aligned}
\eeq
and put $B_j^{(\nu,P)}(x)=\sum_{k=1}^n
B_{jk}^{(\nu,P)}(x)v_k$. Formula
\eqref{eq:hyp-limit-phase-weight-string} becomes
\beq\label{eq:hyp-limit-value}
 \Phi(qp^{mN+\nu}z_j;z)w(qp^{mN+\nu}z_j;z)
 =-p^{\nu}x\,B_j^{(\nu,P)}(x).
\eeq
Consequently, \eqref{eq:hyp-limit-full-residue} is
\beq\label{eq:hyp-limit-exact-residue}
 \Res_{t=qp^{mN+\nu}z_j}\om_i(t;z)
 =-Y_i(z)R_{ji}(p,z)P^{\nu\alpha_m}p^\nu
 x^{m\alpha_m+1}B_j^{(\nu,P)}(x).
\eeq
Summing \eqref{eq:hyp-limit-exact-residue} over $N\ge0$
and $\nu=0,\ldots,m-1$, and using the Jackson measure
\eqref{eq:hyp-limit-Jackson}, we obtain, for each $j\le i$,
the exact equality
\beq\label{eq:hyp-limit-exact-Jackson}
 \begin{aligned}
 T_{ji}(P,z)
 &=\sum_{\nu=0}^{m-1}\sum_{N=0}^{\infty}
   \Res_{t=qp^{mN+\nu}z_j}\om_i(t;z)\\
 &=-\frac{Y_i(z)R_{ji}(p,z)}{1-P}
 \sum_{\nu=0}^{m-1}P^{\nu\alpha_m}p^\nu
 \int_0^1x^{m\alpha_m}B_j^{(\nu,P)}(x)\,d_Px.
 \end{aligned}
\eeq
Thus the contribution of the $j$th inward string to the
circle integral is expressed exactly as a finite sum of
Jackson integrals. Formula \eqref{eq:hyp-limit-exact-Jackson}
holds for every $0<P<1$ under our assumptions; it is not
an asymptotic formula.

\subsection{Convergence of the residue strings}
\label{subsec:hyp-limit-convergence}

We now prove that each Jackson integral in
\eqref{eq:hyp-limit-exact-Jackson} converges to the
corresponding ordinary integral. The statement is as follows.

\begin{prop}[Limit of each residue string]
\label{prop:hyp-string-limit}
For $z\in\mc D_{m,\si}$, $j=1,\ldots,n$, and
$\nu=0,\ldots,m-1$,
\beq\label{eq:hyp-string-integral-limit}
 \begin{aligned}
 &\int_0^1 x^{m\alpha_m}
       B_j^{(\nu,P)}(x)\,d_Px\\
 &\quad\longrightarrow
 \int_0^1 x^{m\alpha_m}
 \prod_{a=1}^n(1-\rho_{ja}^m x^m)^{b_m}
 \left(\sum_{k=1}^n
   \frac{\rho_{jk}}{1-\zeta_m^\nu\rho_{jk}x}v_k\right)dx.
 \end{aligned}
\eeq
For $\nu=0$, the $k=j$ component on the right is
interpreted as the integrable function with endpoint
behavior $(1-x)^{b_m-1}$. Convergence is locally
uniform in the stated parameter and configuration chamber.
\end{prop}

\begin{proof}
We fix $j,\nu$ and prove convergence of each $v_k$-component.
The product limits in Lemma~\ref{lem:hyp-cyclotomic-products}
give, for $0\le x<1$,
\beq\label{eq:hyp-limit-kernel-limit}
 B_{jk}^{(\nu,P)}(x)\longrightarrow
 \prod_{a=1}^n(1-\rho_{ja}^m x^m)^{b_m}
 \frac{\rho_{jk}}{1-\zeta_m^\nu\rho_{jk}x}.
\eeq
We cannot yet pass to the limit in the Jackson integrals:
when $\nu=0$ and $k=j$, the expression on the right is
singular at $x=1$. We therefore identify this singularity.

Since $\rho_{jj}=1$, the factor with $a=j$ in the
product in \eqref{eq:hyp-limit-kernel-limit} is
$(1-x^m)^{b_m}$. Factor
\[
 1-x^m=(1-x)(1+x+\cdots+x^{m-1})
\]
and define
\beq\label{eq:hyp-limit-Gj}
 G_j(x)=
 (1+x+\cdots+x^{m-1})^{b_m}
 \prod_{a\ne j}(1-\rho_{ja}^m x^m)^{b_m}.
\eeq
The branches of all powers are determined by continuous
continuation from their value $1$ at $x=0$. Because
$\rho_{ja}^m\ne1$ for $a\ne j$, the function $G_j$ is
continuous and nonzero on the whole interval $[0,1]$.
Thus the limiting expression in
\eqref{eq:hyp-limit-kernel-limit} becomes
\[
 \begin{cases}
 (1-x)^{b_m-1}G_j(x),&\nu=0,\ k=j,\\[3pt]
 (1-x)^{b_m}G_j(x)
       \dfrac{\rho_{jk}}{1-\zeta_m^\nu\rho_{jk}x},
       &(\nu,k)\ne(0,j).
 \end{cases}
\]
The first function is integrable at $x=1$, because
$b_m>0$. In the other cases the displayed denominator
does not vanish on $[0,1]$, so the limit extends
continuously to that interval.

To justify convergence of the Jackson integrals, it
remains to identify the corresponding singular factors
in the \emph{exact} kernels for $P<1$. We need the
following auxiliary lemma.

\begin{lem}[Endpoint factors of the residue kernels]
\label{lem:hyp-endpoint}
For fixed $j,k,\nu$, the kernel $B_{jk}^{(\nu,P)}(x)$ from
\eqref{eq:hyp-limit-kernel} admits the following
factorizations on $0\le x\le1$.

\emph{(a)} If $\nu=0$ and $k=j$, then
\beq\label{eq:hyp-endpoint-diagonal}
 B_{jj}^{(0,P)}(x)=
 \frac{(Px;P)_\infty}{(P^{b_m}x;P)_\infty}
 h_{jj}^{(0,P)}(x),
 \qquad h_{jj}^{(0,P)}(x)\longrightarrow G_j(x).
\eeq

\emph{(b)} If $\nu\ge1$, or $\nu=0$ and $k>j$, then
\beq\label{eq:hyp-endpoint-regular}
 B_{jk}^{(\nu,P)}(x)=
 \frac{(Px;P)_\infty}{(P^{b_m+1}x;P)_\infty}
 h_{jk}^{(\nu,P)}(x),
\eeq
with
\beq\label{eq:hyp-endpoint-h-limit}
 h_{jk}^{(\nu,P)}(x)\longrightarrow
 G_j(x)\frac{\rho_{jk}}{1-\zeta_m^\nu\rho_{jk}x}.
\eeq

\emph{(c)} If $\nu=0$ and $k<j$, then
\beq\label{eq:hyp-endpoint-shifted}
 B_{jk}^{(0,P)}(x)=
 \frac{(x;P)_\infty}{(P^{b_m}x;P)_\infty}
 \widehat h_{jk}^{(P)}(x),
\eeq
where
\[
 \widehat h_{jk}^{(P)}(x)\longrightarrow
 G_j(x)\frac{\rho_{jk}}{1-\rho_{jk}x}.
\]
Here $B_{jk}^{(0,P)}(1)=0$.

All the functions $h_{jk}^{(\nu,P)}$ and
$\widehat h_{jk}^{(P)}$ are uniformly bounded and
converge uniformly on $[0,1]$, locally uniformly in the
configuration in $\mc D_{m,\si}$ and the stated real parameters.
\end{lem}

\begin{proof}[Proof of the auxiliary lemma]
We factor the Pochhammer products in
\eqref{eq:hyp-limit-kernel} using
\eqref{eq:hyp-factorization}. The resulting ratios have
arguments of the form $P^A\zeta_m^h\rho_{ja}x$,
where $0\le h<m$ and $0\le x\le1$.
For $a\ne j$, the equality $\zeta_m^h\rho_{ja}=1$
would imply $z_j^m=z_a^m$, which is excluded in the chamber.
For $a=j$, it holds only when $h=0$.

Write the exact factorization in the form
\[
 B_{jk}^{(\nu,P)}(x)
 =q^{-(k-1)}\rho_{jk}\,
 E_{jk}^{(\nu,P)}(x)H_{jk}^{(\nu,P)}(x).
\]
Here $E_{jk}^{(\nu,P)}$ is the ratio for which $a=j$ and
$h=0$, and $H_{jk}^{(\nu,P)}$ is the product of all the
remaining ratios. Separating the cases $\nu=0$ and $\nu\ge1$,
and then comparing $k$ with $j$ in
\eqref{eq:hyp-limit-kernel}, gives
\[
 \begin{array}{c|c}
 \text{case}&E_{jk}^{(\nu,P)}(x)\\ \hline
 \nu=0,\ k=j
    &\displaystyle\frac{(Px;P)_\infty}
                            {(P^{b_m}x;P)_\infty}\\[9pt]
 \nu\ge1\text{ or }(\nu=0,k>j)
    &\displaystyle\frac{(Px;P)_\infty}
                            {(P^{b_m+1}x;P)_\infty}\\[9pt]
 \nu=0,\ k<j
    &\displaystyle\frac{(x;P)_\infty}
                            {(P^{b_m}x;P)_\infty}
 \end{array}
\]
In cases (a) and (b), set
$h_{jk}^{(\nu,P)}=q^{-(k-1)}\rho_{jk}H_{jk}^{(\nu,P)}$;
in case (c), set
$\widehat h_{jk}^{(P)}=q^{-(k-1)}\rho_{jk}H_{jk}^{(0,P)}$.
These definitions prove the three exact factorizations,
including their prefactors.

It remains to prove uniform convergence of the remaining
factors. Fix a compact subset $\mc K\subset\mc D_{m,\si}$.
Each ratio in $H_{jk}^{(\nu,P)}$ has
$\zeta_m^h\rho_{ja}\ne1$. Since $|\rho_{ja}|=1$,
all points $\zeta_m^h\rho_{ja}x$, with $z\in\mc K$
and $0\le x\le1$, form a compact subset of
$\C\setminus[1,\infty)$. Lemma~\ref{lem:hyp-elementary-products}
therefore applies uniformly to every such ratio.
There are only finitely many ratios, and the prefactor
$q^{-(k-1)}\rho_{jk}$ tends uniformly to $\rho_{jk}$.
Consequently the functions $h_{jk}^{(\nu,P)}$ and
$\widehat h_{jk}^{(P)}$ converge uniformly on $[0,1]$,
locally uniformly in the configuration and the real parameters.

For $0\le x<1$, the elementary product limit
\eqref{eq:hyp-elementary-P-limit} gives the limits
$(1-x)^{b_m-1}$ in case (a) and
$(1-x)^{b_m}$ in cases (b) and (c).
Dividing the full pointwise limit
\eqref{eq:hyp-limit-kernel-limit} by these factors
gives exactly the limits stated for the remaining
functions in (a)--(c). Those limiting functions are
continuous on $[0,1]$, so the uniform convergence
of the remaining products also identifies their values
at $x=1$. This proves the assertions. In case (c),
$(x;P)_\infty=0$ at $x=1$, giving
$B_{jk}^{(0,P)}(1)=0$.
\end{proof}

We now complete the proof of the proposition. Set
$a=m\alpha_m+1>0$. Apply the uniform Jackson--beta
convergence lemma, Lemma~\ref{lem:hyp-Jackson-beta},
to the factorizations above. In case (a) take the beta
parameter $b=b_m$; in case (b) take $b=b_m+1$.
For case (c), use the shifted beta formula
\eqref{eq:hyp-beta-shifted} with $b=b_m$. It treats the
vanishing first Jackson summand at $x=1$ correctly;
equivalently, we may use the identity
\beq\label{eq:hyp-endpoint-zero-identity}
 \frac{(x;P)_\infty}{(P^{b_m}x;P)_\infty}
 =(1-x)\frac{(Px;P)_\infty}{(P^{b_m}x;P)_\infty}.
\eeq
Multiplying the resulting beta weights by the uniform
limits of the remaining factors gives
\eqref{eq:hyp-string-integral-limit} in all cases.
The uniform statements in the two auxiliary results
imply local uniformity with respect to the parameters
and the configuration. The factor $x^{m\alpha_m}$,
including its dependence on $\mu$, is retained throughout.
\end{proof}

It remains to combine the $m$ residue classes.  As
$P\to1^-$, we have $P^{\nu\alpha_m}p^\nu\to\zeta_m^\nu$.
The cyclotomic partial-fraction identity is
\beq\label{eq:hyp-limit-cyclotomic-sum}
 \sum_{\nu=0}^{m-1}
 \frac{\zeta_m^\nu\rho}{1-\zeta_m^\nu\rho x}
 =\frac{m\rho^m x^{m-1}}{1-\rho^m x^m}.
\eeq
Thus the minus sign in the exact residue formula
\eqref{eq:hyp-limit-exact-Jackson}, followed by the
substitution $y=x^m$, gives the integral
\beq\label{eq:hyp-limit-Jj}
 \begin{aligned}
 \mc J_j(Z;m\ka,\mu)
 ={}&\int_0^1y^{\alpha_m}
 \prod_{a=1}^n\left(1-\frac{Z_j}{Z_a}y\right)^{b_m}\\[-2pt]
 &\quad\times
 \left(\sum_{k=1}^n\frac{Z_j}{Z_jy-Z_k}v_k\right)dy,
 \qquad Z_a=z_a^m.
 \end{aligned}
\eeq
Each factor $(1-Z_jy/Z_a)^{b_m}$ is defined by continuous
continuation from its value $1$ at $y=0$; the factor
$y^{\alpha_m}$ uses the real logarithm. These are precisely
the branches prescribed in \eqref{eq:hyp-limit-loadings-sigma};
for the canonical chamber they agree with those of
Section~\ref{sec:hyper-kz}.

By Lemma~\ref{lem:hyp-theta-limits}, the coefficient limit
\[
 a_{ji}(z)=\lim_{P\to1^-}
             \frac{Y_i(z)R_{ji}(p,z)}{1-P},\qquad j\le i,
\]
exists locally uniformly in the chamber. It will be calculated
in the next subsection. The exact identity
\eqref{eq:hyp-limit-exact-Jackson}, the proposition, and the
finite sum over $\nu$ now give the equality
\beq\label{eq:hyp-limit-jstring}
 \lim_{P\to1^-}T_{ji}(P,z)
 =a_{ji}(z)\mc J_j(Z;m\ka,\mu),\qquad j\le i.
\eeq
By \eqref{eq:hyp-limit-radials-sigma}, the normalization relating
$\mc J_j$ and $I_j^\si$ is
\beq\label{eq:hyp-limit-KZ-factor}
 I_j^\si(Z;m\ka,\mu)=S_j(Z)\mc J_j(Z;m\ka,\mu),
 \qquad
 S_j(Z)=Z_j^{\alpha_m}
 \prod_{a=1}^nZ_a^{-\lambda_m/2}.
\eeq
Consequently the coefficient of $I_j^\si$ in the limit of
$\Psi_i$ must be
\beq\label{eq:hyp-limit-coefficient-ratio}
 c_{ji}^{(m,\si)}=\frac{a_{ji}(z)}{S_j(Z)}
 =\lim_{P\to1^-}
 \frac{Y_i(z)R_{ji}(p,z)}{(1-P)S_j(Z)},
 \qquad j\le i.
\eeq

\subsection{Transition coefficients and proof of the main theorem}
\label{subsec:hyp-limit-coefficients}

\begin{proof}[Proof of Theorem~\ref{thm:hyper-limit-all-chambers}]
\noindent\emph{The limit and its coefficients.}
By \eqref{eq:hyp-limit-residue-sum} and
\eqref{eq:hyp-limit-jstring}, the limit of the $i$th column is
$\sum_{j\le i}a_{ji}(z)\mc J_j$. It remains to calculate
$a_{ji}(z)/S_j(Z)$, which is the coefficient of $I_j^\si$.
Recall $d_i$ from \eqref{eq:hyp-limit-di}.
The theta limits in Lemma~\ref{lem:hyp-theta-limits}
give
\beq\label{eq:hyp-limit-Y-limit}
 Y_i(z)\longrightarrow
 \prod_{a<i}(-Z_a)^{\lambda_m/2}
 (-Z_i)^{(d_i-1/2)\lambda_m}
 \prod_{a>i}(-Z_a)^{-\lambda_m/2}.
\eeq
The powers of negative arguments here and below mean
$\exp(c\operatorname{Log}(-u))$ with the principal logarithm,
as specified in Lemma~\ref{lem:hyp-theta-limits}.
For $j<i$, that lemma and the exact residue
\eqref{eq:hyp-limit-Roff} give
\beq\label{eq:hyp-limit-Roff-limit}
 \frac{R_{ji}(p,z)}{1-P}\longrightarrow
 \frac{\sin(\pi\lambda_m)}{\pi}
 \left(-\frac{Z_j}{Z_i}\right)^{d_i\lambda_m}
 \prod_{\substack{a<i\\a\ne j}}
 \left(-\frac{Z_j}{Z_a}\right)^{\lambda_m};
\eeq
for $j=i$,
\beq\label{eq:hyp-limit-Rdiag-limit}
 \frac{R_{ii}(p,z)}{1-P}\longrightarrow
 \frac{\sin(\pi d_i\lambda_m)}{\pi}
 \prod_{a<i}\left(-\frac{Z_i}{Z_a}\right)^{\lambda_m}.
\eeq

We now compare these limits with $S_j(Z)$ in
\eqref{eq:hyp-limit-KZ-factor}.  The powers of the
$Z_a$ in the product
$Y_iR_{ji}/(1-P)$ are exactly the powers in $S_j(Z)$.
This is checked directly: for $j<i$, the factors from
$R_{ji}$ contain $Z_j^{(d_i+i-2)\lambda_m}$ and
the remaining factors from $Y_i$ reduce the other
$Z_a$ to exponent $-\lambda_m/2$; since the factor from $Y_i$ also contributes
$Z_j^{\lambda_m/2}$, the total exponent of $Z_j$ is
$(d_i+i-3/2)\lambda_m
 =\alpha_m-\lambda_m/2$, as required.  The diagonal
case follows by the same calculation, with
$d_i+i-1=n/2-2\mu$.
Thus \eqref{eq:hyp-limit-coefficient-ratio} is independent
of the points $Z_a$, apart from a constant phase.

In $\mc D_{m,\si}$, the relevant logarithms satisfy
\beq\label{eq:hyp-limit-branch-logs-sigma}
 \begin{aligned}
 \operatorname{Log}(-Z_a)&=L_a-\pi\mathrm{i},\\
 \operatorname{Log}\left(-\frac{Z_j}{Z_a}\right)
 &=L_j-L_a+\pi\mathrm{i}\eps_{ja}^\si,\qquad j\ne a.
 \end{aligned}
\eeq
For the second equality, if $j$ precedes $a$, the imaginary part
of $L_j-L_a$ lies in $(-2\pi,0)$; adding $\pi\mathrm{i}$ gives
the principal logarithm. If $j$ follows $a$, subtracting
$\pi\mathrm{i}$ does so. These identities hold on the small
complex neighborhoods as well, with the logarithms fixed above.

The sum of the exponents of the negative arguments in
\eqref{eq:hyp-limit-Y-limit} is
\beq\label{eq:hyp-limit-Y-phase-sum}
 \lambda_m\left(\frac{i-1}{2}+d_i-\frac12-\frac{n-i}{2}\right)
 =-2\mu\lambda_m.
\eeq
Thus $Y_i$ contributes the constant phase
$e^{2\pi\mathrm{i}\mu\lambda_m}$. For $j<i$, the residue factors
in \eqref{eq:hyp-limit-Roff-limit} contribute
\beq\label{eq:hyp-limit-residue-phase-off-sigma}
 \exp\!\left[\pi\mathrm{i}\lambda_m\left(
 d_i\eps_{ji}^\si+
 \sum_{\substack{a<i\\a\ne j}}\eps_{ja}^\si\right)\right],
\eeq
whereas for $j=i$, \eqref{eq:hyp-limit-Rdiag-limit} contributes
\beq\label{eq:hyp-limit-residue-phase-diag-sigma}
 \exp\!\left[\pi\mathrm{i}\lambda_m
                 \sum_{a<i}\eps_{ia}^\si\right].
\eeq
After cancellation of the powers of $Z_a$ against $S_j$, these
phases give exactly \eqref{eq:hyp-limit-C-sigma}. No string with
$j>i$ occurs, so those entries are zero.

Together with \eqref{eq:hyp-limit-jstring} and
$\Psi_i=\sum_{j\le i}T_{ji}$, this coefficient calculation
proves \eqref{eq:hyp-limit-main-sigma}, locally uniformly on
each unit-circle chamber $\mc D_{m,\si}$.
The identity \eqref{eq:hyp-limit-C-compare-sigma} follows
because the coefficient formula depends on $m$ and $\ka$
only through $\lambda_m=1/(m\ka)$.

\medskip
\noindent\emph{Extension to complex neighborhoods.}
For the local holomorphic assertion, fix $z^0\in\mc D_{m,\si}$
and choose a sufficiently small simply connected complex
neighborhood $\mc U$ such that $z_a^m$ and $(z_j/z_a)^m$,
$j\ne a$, avoid $\R_{\ge0}$ throughout $\mc U$.
Fix a compact subset $\mc K\subset\mc U$.
For $a\ne j$ and $0\le h<m$, the arguments
$\zeta_m^h(z_j/z_a)x$, $z\in\mc K$, $0\le x\le1$,
stay in a compact subset of $\C\setminus[1,\infty)$.
The same holds for $a=j$ and $h=1,\ldots,m-1$.
Hence Lemma~\ref{lem:hyp-elementary-products} applies
uniformly to the corresponding Pochhammer ratios.
The remaining ratios, with $a=j$ and $h=0$, are
handled by Lemmas~\ref{lem:hyp-Jackson-beta}
and~\ref{lem:hyp-endpoint}.

For each fixed $0<P<1$ sufficiently close to $1$,
the series of residues converges absolutely and uniformly
for $z\in\mc K$. Indeed, write $s=mN+\nu$, with
$0\le \nu<m$. In \eqref{eq:hyp-limit-exact-residue},
$x=P^N\to0$ as $N\to\infty$. The remaining product
factors are bounded for $z\in\mc K$ and all sufficiently
large $N$. Thus the residues along each inward string
are bounded by
$C(P,\mc K)P^{N(1+m\alpha_m)}$.
Since $1+m\alpha_m>0$ and $P<1$, these bounds form
a convergent geometric series. This establishes
uniform convergence in $z$ for each fixed $P$;
it does not by itself establish uniform convergence as
$P\to1^-$. That stronger conclusion follows from
the preceding Jackson-integral estimates.
The restrictions on $\mc U$ also prevent collisions of
distinct pole strings and zeros of the coefficient
denominators.
The residue sum agrees with the original circle integral on a
possibly $P$-dependent neighborhood of $z^0$ where the circle
still separates its poles. It consequently defines its
holomorphic continuation to $\mc U$, equivalently obtained by
deforming the contour to separate the same pole strings.
A single undeformed circle for all $P$ and all $z\in\mc U$
is not required. The uniform estimates above then prove
convergence on compact subsets of this fixed $\mc U$.
The limit is the right side of \eqref{eq:hyp-limit-main-sigma}
with the loadings \eqref{eq:hyp-limit-loadings-sigma}.

\medskip
\noindent\emph{Nonvanishing and the KZ equations.}
The number $n/2-i+1$ is an integer or a half-integer, whereas
$0<2\mu<1/4$. Thus $d_i\ne0$. Moreover,
$|d_i|<n/2$ and $|\ka|>n$, so
\[
 0<|d_i\lambda_m|<\frac1{2m}.
\]
All diagonal sine factors in \eqref{eq:hyp-limit-C-sigma}
are therefore nonzero, and $C_{m,\si}^{(n)}$ is invertible.

The local differential identities underlying
Theorem~\ref{thm:kz-integrals} apply to the loaded integrals
\eqref{eq:hyp-limit-radials-sigma} in any angular order, so
these are KZ solutions with parameter $m\ka$.
Apply Proposition~\ref{prop:kz-basis} to the ordered points
$Z_{\si(1)},\ldots,Z_{\si(n)}$ and the correspondingly
permuted basis $v_{\si(1)},\ldots,v_{\si(n)}$.
This merely permutes simultaneously the rows and columns of
the radial solution matrix, and its determinant is nonzero
for every parameter pair in
\eqref{eq:hyp-limit-real-parameters}, since $m\ka<-n$.
Thus $\mc I_\si(Z;m\ka,\mu)$ is invertible, and so is
$\mc I_\si C_{m,\si}^{(n)}$. Finally,
\eqref{eq:pullback} identifies the pullback of this KZ
matrix as a matrix of local flat sections of
$\nabla_m(\ka,\mu)$. This completes the proof.
\end{proof}

\paragraph{Canonical phase calculation.}
For comparison with the original coefficient formula, the logarithm
identities \eqref{eq:hyp-limit-branch-logs-sigma} in the canonical
chamber \eqref{eq:hyp-limit-chamber} read
\beq\label{eq:hyp-limit-branch-logs}
 \begin{aligned}
 \operatorname{Log}(-Z_a)
  &=\log Z_a-\pi\mathrm{i},\\
 \operatorname{Log}\left(-\frac{Z_j}{Z_a}\right)
  &=\log Z_j-\log Z_a+\pi\mathrm{i},&&j<a,\\
 \operatorname{Log}\left(-\frac{Z_j}{Z_a}\right)
  &=\log Z_j-\log Z_a-\pi\mathrm{i},&&j>a.
 \end{aligned}
\eeq
The powers of $-Z_a$ in \eqref{eq:hyp-limit-Y-limit}
contribute the factor $e^{2\pi\mathrm{i}\mu\lambda_m}$.
For $j<i$, the factors in
\eqref{eq:hyp-limit-Roff-limit} contribute
$e^{\pi\mathrm{i}(d_i+i-2j)\lambda_m}$.
The combined phase is
\[
 e^{\pi\mathrm{i}(n/2+1-2j)\lambda_m},
\]
which gives the first line of \eqref{eq:hyp-limit-C}.
For $j=i$, the factors in
\eqref{eq:hyp-limit-Rdiag-limit} contribute
$e^{-\pi\mathrm{i}(i-1)\lambda_m}$; together with the
phase from $Y_i$ this gives
$e^{\pi\mathrm{i}(2\mu-i+1)\lambda_m}$,
which is the second line of \eqref{eq:hyp-limit-C}.

\subsection{Branch normalization and the power map}
\label{subsec:hyp-limit-normalizations}

\begin{rem}
The phases in \eqref{eq:hyp-limit-C-sigma} result from the
specified loadings and the angular order, through
\eqref{eq:hyp-limit-branch-logs-sigma}. For the canonical chamber,
these are precisely the logarithms of Section~\ref{sec:hyper-kz}
and the identities \eqref{eq:hyp-limit-branch-logs}, producing
\eqref{eq:hyp-limit-C}. The local identity between the limiting
solution matrix and the radial KZ matrix has identical analytic
continuations on its two sides along any path on which those
continuations are defined. This assertion concerns the limiting
functions; it does not assert an interchange of analytic
continuation with the root-of-unity limit, nor does it compare
independently chosen canonical bases in different chambers.
If the branches or normalizations of the loaded intervals are
changed independently, the coefficient matrix must be changed
by the corresponding constant change of basis.
The formulas here are for the prescribed loadings, not arbitrary
unrelated branch choices. No continuation between different
angular chambers was used in the direct-limit proof.
\end{rem}

\begin{rem}
The $m$ classes $s=mN+\nu$ correspond to the $m$ rays of
the residue strings ending at $\zeta_m^{\nu}z_j$.  Their sum
is converted by \eqref{eq:hyp-limit-cyclotomic-sum} and
$y=x^m$ into the KZ interval ending at $Z_j=z_j^m$.
This is the solution-level counterpart of the power map
in the connection identity \eqref{eq:pullback}.
\end{rem}

\clearpage
\section{Inward analytic continuation of the distinguished bases}
\label{sec:inward-formal}

Throughout this section, $m,\ka,\mu$ are fixed, and we write
$B_\si(z)=B_{m,\si}(z;\ka,\mu)$.
For every angular chamber $\mc D_{m,\si}$, Theorem~\ref{thm:hyper-limit-all-chambers}
provides a distinguished fundamental matrix $B_\si$ of limiting
hypergeometric solutions. We show that these matrices are identified
by inward analytic continuation along the right weak Bruhat order on
$S_n$. Equivalently, they determine a distinguished global flat frame
on the space obtained by joining the chambers along the prescribed
inward paths. The radial integral matrices $\mc I_\si$ enter the
proof as auxiliary bases in which the continuation can be calculated.

\subsection{Distinguished bases and the right weak Bruhat order}
\label{subsec:if-distinguished-order}

Fix integers $n\ge2$, $m\ge1$ and real parameters
$\ka<-n$, $0<\mu<1/8$, as in
\eqref{eq:hyp-limit-real-parameters}. We use the angular chambers
$\mc D_{m,\si}$ of \eqref{eq:hyp-limit-chamber-sigma}. In one-line
notation $\si=(\si(1),\ldots,\si(n))$, the corresponding angular
order is
\[
 0<\vartheta_{\si(1)}<\cdots<\vartheta_{\si(n)}<2\pi/m.
\]
The distinguished matrix in this chamber is
\beq\label{eq:if-distinguished-limit}
 B_\si(z)=\lim_{r\to1^-}\Psi(z;p,q,\varkappa),
 \qquad z\in\mc D_{m,\si},
\eeq
with the scaling \eqref{eq:hyp-limit-scaling} and the local
holomorphic continuation supplied by
Theorem~\ref{thm:hyper-limit-all-chambers}. Its columns are a basis
of local flat sections of $\nabla_m(\ka,\mu)$ for every parameter
pair in the stated range. We keep their numerical labels fixed.

Two chambers are neighboring if their angular orders differ by
interchanging two consecutive entries. Let $a,b\in\{1,\ldots,n\}$
be the numerical labels of these two points, named so that $a<b$.
We orient the edge from the chamber in which $z_a$ precedes $z_b$
to the chamber in which $z_b$ precedes $z_a$. Thus, for a position
$k\in\{1,\ldots,n-1\}$,
\beq\label{eq:if-neighbors}
 \si(k)=a,\qquad \si(k+1)=b,\qquad a<b,
 \qquad \si'=\si s_k,
\eeq
where $s_k=(k,k+1)$ acts on positions by right multiplication.
The directed exchange is
\[
 (\ldots,a,b,\ldots)\longrightarrow(\ldots,b,a,\ldots),\qquad a<b.
\]

Let $G_n$ be the graph with vertices $S_n$ and these neighboring
pairs as edges. Define the inversion number by
\beq\label{eq:if-inversion-length}
 \ell(\si)=\#\{(i,j):1\le i<j\le n,\ \si(i)>\si(j)\}.
\eeq
We use two relations on permutations. We write
$\si\lessdot_R\si'$ if there is \emph{one} arrow
$\si\longrightarrow\si'$ in the graph $G_n$.
We write $\si\le_R\tau$ if $\tau$ can be reached from $\si$
by following \emph{zero or more} arrows. In particular,
$\si\le_R\si$. The first relation is called a \emph{covering
relation}, and the second is a partial order: the
\emph{right weak Bruhat order} on $S_n$; see
\cite[Section~3.1]{BB05}. Explicitly, for a single arrow,
\beq\label{eq:if-weak-cover}
 \si\lessdot_R\si s_k
 \quad\Longleftrightarrow\quad
 \si(k)<\si(k+1)
 \quad\Longleftrightarrow\quad
 \ell(\si s_k)=\ell(\si)+1.
\eeq
Every arrow increases the inversion number, so a directed
cycle is impossible. The definition gives
\beq\label{eq:if-weak-reachability}
 \si\le_R\tau\quad\Longleftrightarrow\quad
 \text{there is a directed path in $G_n$ from $\si$ to $\tau$},
\eeq
where a path of length zero is allowed. The graph with these
arrows is the \emph{Hasse diagram} of the order: it displays
precisely the one-step relations, while longer comparisons are
obtained by following successive arrows.
For example, when $n=3$,
\[
 123\lessdot_R213\lessdot_R231\lessdot_R321,
 \qquad 123\le_R321,
\]
but $213$ and $132$ are incomparable.

\subsection{Inward paths, the space \texorpdfstring{$X$}{X}, and the main theorem}
\label{subsec:if-space-X}

For the cover \eqref{eq:if-neighbors}, the \emph{inward path}
$\gm_{\si,\si'}$ moves $z_a$ slightly inside the unit circle,
past the fixed point $z_b$, and back to the unit circle. All other
points remain fixed during this local exchange.

Here is a representative of its path class. Put $Z_j=z_j^m$.
Near the common wall choose $\eta>0$ so small that the interval
$(m\vartheta_b-\eta,m\vartheta_b+\eta)$ lies in $(0,2\pi)$
and contains no other marked angle. Keep $z_j$ fixed for $j\ne a$
and set
\beq\label{eq:if-path}
 \begin{aligned}
 \xi(\rho)&=\eta\exp\bigl(-\mathrm{i}(\pi/2+\pi\rho)\bigr),
                         &&0\le\rho\le1,\\
 z_a(\rho)&=z_b\exp\bigl(\xi(\rho)/m\bigr),
 &Z_a(\rho)&=Z_b\exp\bigl(\xi(\rho)\bigr).
 \end{aligned}
\eeq
Initially $Z_a/Z_b=e^{-\mathrm{i}\eta}$ and finally
$Z_a/Z_b=e^{\mathrm{i}\eta}$. Between these endpoints $|z_a|<1$.
Choosing $\eta$ sufficiently small keeps the moving point away
from all other marked points. Paths inside the initial and final
chambers can be attached to this local exchange. Since each chamber
is contractible, this specifies a path class between any chosen
endpoint configurations. We also write $\gm_{\mathrm{in}}$ for
the local path.

\begin{example}[$n=2$]
The graph has one arrow $12\lessdot_R21$. Its inward path
moves $z_1$ inside the fixed point $z_2$. Reversing this path
moves $z_1$ back along the same arc.
\end{example}

For every pair of neighboring chambers, choose one inward
exchange path of the form \eqref{eq:if-path}, oriented as above.
We choose these paths near different configurations at the
respective chamber boundaries, with no intersections between
connecting arcs outside the chambers.
Let $X$ be the union of all angular chambers and these connecting
arcs, regarded as a subspace of
\[
 \mc U_m=\{z\in(\C^\times)^n:z_i^m\ne z_j^m\text{ for }i\ne j\}.
\]
Thus each edge of $G_n$ is represented by one arc joining its two
chambers. The arc may be followed in either direction; following
it backward is simply the reverse of the chosen path. We denote
the inclusion by $\iota:X\longrightarrow\mc U_m$.

\begin{thm}[Distinguished bases and inward continuation along the weak order]
\label{thm:if-main}
\label{prop:if-weak-coherence}
Suppose $\si\le_R\tau$. Let $\gm$ be a path obtained by following
the prescribed inward exchanges along any directed chain
\[
 \si=\si_0\lessdot_R\si_1\lessdot_R\cdots
       \lessdot_R\si_N=\tau,
\]
where, after each exchange, we may move within the resulting
chamber to the starting configuration of the next exchange.
Such a move stays entirely in that one chamber; it does not cross
from one chamber to another. Then
\beq\label{eq:if-main}
 \boxed{\operatorname{Cont}_{\gm}B_\si=B_\tau.}
\eeq
This is an equality of fundamental matrices of holomorphic solution
germs. In particular, the identification
is independent of the chosen directed chain. The assertion holds
for every $n\ge2$, $m\ge1$ and every parameter pair in
\eqref{eq:hyp-limit-real-parameters}.

Moreover, the matrices $B_\si$, continued along the chosen connecting
paths, determine a global flat frame of the local system of horizontal
sections of $\nabla_m(\ka,\mu)$ restricted to $X$.
\end{thm}

For example, in the chain $123\lessdot_R213\lessdot_R231$
when $n=3$, the first exchange moves $z_1$ past $z_2$.
We can then move the configuration inside $\mc D_{m,213}$,
without altering its angular order, before making the exchange
of $z_1$ and $z_3$.

The last assertion concerns a local system on $X$, which is not an
open complex domain. The path-independence statement concerns the
specified exchanges and paths within the chambers, not arbitrary
paths in $\mc U_m$. The reverse of an edge is always understood
as the reverse of that same chosen path.

\subsection{Proof of the continuation theorem}
\label{subsec:if-proof}

We first prove the theorem for a cover
$\si\lessdot_R\si'=\si s_k$, with $a=\si(k)<b=\si(k+1)$.
To calculate this continuation, we use the auxiliary radial
hypergeometric bases from Section~\ref{sec:hyper-limit}. Abbreviate
\beq\label{eq:if-exponents}
 \la=\lambda_m=\frac1{m\ka},\qquad
 \al=\alpha_m=\frac{n/2-2\mu}{m\ka},\qquad
 b_m=-\la,\qquad d_i=\frac n2-i+1-2\mu.
\eeq
In particular, $b_m>0$ and $\al>-1$.

For each chamber let $L_j=\mathrm{i}m\vartheta_j$, with the
local holomorphic extensions fixed in Section~\ref{sec:hyper-limit}. Write
\beq\label{eq:if-form}
 \begin{aligned}
 K(Z)&=\exp\left(-\frac{\la}{2}\sum_{j=1}^n L_j\right),\\
 \om(u;Z)&=u^\al\prod_{j=1}^n
 \left(1-\frac{u}{Z_j}\right)^{b_m}
 \left(\sum_{j=1}^n\frac{v_j}{u-Z_j}\right)du.
 \end{aligned}
\eeq
Thus $b_m$ is the common exponent of the factors $(1-u/Z_j)$.
The auxiliary matrices are
\beq\label{eq:if-radials}
 I_j^\si(Z)=K(Z)\int_{[0,Z_j]}\om(u;Z),\qquad
 \mc I_\si=(I_1^\si,\ldots,I_n^\si).
\eeq
We suppress the fixed parameters $m\ka,\mu$. Each interval is
oriented toward $Z_j$ and has the loading
\eqref{eq:hyp-limit-loadings-sigma}: with $u=xZ_j$, $0<x<1$,
we take $\log u=L_j+\log x$, the real power $(1-x)^{b_m}$,
and the logarithm of $1-xZ_j/Z_k$, $k\ne j$, continued from
$0$ at $x=0$.

Theorem~\ref{thm:hyper-limit-all-chambers} gives
\beq\label{eq:if-input}
 B_\si(z)=\mc I_\si(Z)C_\si,\qquad
 C_\si=C_{m,\si}^{(n)}(\ka,\mu).
\eeq
Here $C_\si$ is constant in $z$ and invertible. We write
$c_{ji}^\si$ for its entries and use the signs $\eps_{jk}^\si$
from \eqref{eq:hyp-limit-order-signs}.

The two lemmas below compare the factors in \eqref{eq:if-input}.
Put
\beq\label{eq:if-t}
 t=\exp(2\pi\mathrm{i}\la)=\exp(-2\pi\mathrm{i}b_m).
\eeq
Let $T_{ab}$ be the identity matrix except in rows and columns
$a,b$, where, in the order $(a,b)$, its block is
\beq\label{eq:if-block}
 \boxed{T_{ab}\big|_{\{a,b\}}=
 \begin{pmatrix}1&1-t\\0&t\end{pmatrix}.}
\eeq

\subsubsection{The coefficient matrices}

\begin{lem}\label{lem:if-algebra}
The coefficient matrices satisfy
\beq\label{eq:if-algebra}
 C_{\si'}=T_{ab}C_\si.
\eeq
\end{lem}

\begin{proof}
Only $\eps_{ab}$ and $\eps_{ba}$ change under the interchange.
They change from $1,-1$ to $-1,1$, respectively. All other signs
are unchanged because $a,b$ are adjacent in angular order.
Moreover, for every $k\ne a,b$,
\beq\label{eq:if-equal-signs}
 \eps_{ak}^\si=\eps_{bk}^\si.
\eeq
Thus all rows except $a,b$ are unchanged. In columns $i<b$,
row $a$ is unchanged and row $b$ is zero.

For column $b$, put
\beq
 E=\exp\!\left[\pi\mathrm{i}\la\left(
 2\mu+\sum_{\substack{k<b\\k\ne a}}\eps_{ak}^\si\right)\right].
\eeq
Formula \eqref{eq:hyp-limit-C-sigma} gives
\beq\label{eq:if-column-b}
 \begin{aligned}
 c_{ab}^\si&=\frac{E}{\pi}\sin(\pi\la)e^{\pi\mathrm{i}d_b\la},
 &c_{ab}^{\si'}&=\frac{E}{\pi}\sin(\pi\la)e^{-\pi\mathrm{i}d_b\la},\\
 c_{bb}^\si&=\frac{E}{\pi}\sin(\pi d_b\la)e^{-\pi\mathrm{i}\la},
 &c_{bb}^{\si'}&=\frac{E}{\pi}\sin(\pi d_b\la)e^{\pi\mathrm{i}\la}.
 \end{aligned}
\eeq
Consequently,
\beq
 \begin{aligned}
 c_{bb}^{\si'}&=t c_{bb}^\si,\\
 c_{ab}^{\si'}-c_{ab}^\si
 &=-\frac{2\mathrm{i}E}{\pi}\sin(\pi\la)\sin(\pi d_b\la)
   =(1-t)c_{bb}^\si.
 \end{aligned}
\eeq

For every column $i>b$, \eqref{eq:if-equal-signs} and
\eqref{eq:hyp-limit-C-sigma} imply
\beq
 c_{ai}^\si=t c_{bi}^\si,\qquad
 c_{ai}^{\si'}=t^{-1}c_{ai}^\si,\qquad
 c_{bi}^{\si'}=t c_{bi}^\si.
\eeq
Hence
\beq
 c_{ai}^{\si'}=c_{ai}^\si+(1-t)c_{bi}^\si.
\eeq
\nopagebreak[4]
These are precisely the row operations in \eqref{eq:if-algebra}.
\end{proof}

\subsubsection{Transport of the loaded radial intervals}

\begin{lem}\label{lem:if-topology}
For the inward path \eqref{eq:if-path},
\beq\label{eq:if-topology}
 \operatorname{Cont}_{\gm_{\mathrm{in}}}\mc I_\si
 =\mc I_{\si'}T_{ab}.
\eeq
Equivalently, writing primes for the canonically loaded final integrals,
\beq\label{eq:if-path-components}
 \begin{aligned}
 \operatorname{Cont}_{\gm_{\mathrm{in}}}I_a^\si&=I_a',\\
 \operatorname{Cont}_{\gm_{\mathrm{in}}}I_b^\si&=(1-t)I_a'+tI_b',\\
 \operatorname{Cont}_{\gm_{\mathrm{in}}}I_j^\si&=I_j',\qquad j\ne a,b.
 \end{aligned}
\eeq
\end{lem}

\begin{proof}
We calculate the continuation of the radial integrals in the
$Z$-plane. The explicit path
\eqref{eq:if-path} shows that the same argument applies to every $m$.

\medskip\noindent
\textbf{1. The short loaded interval.}
Choose a small disk $D$ centered at $Z_b$ which excludes $0$, all
$Z_j$ with $j\ne a,b$, and their radial intervals. Choose $\eta$
so small that the path of $Z_a$ and all chords $[Z_a,Z_b]$ lie in $D$.
At the initial configuration define
\beq\label{eq:if-chord}
 V_{ab}=K(Z)\int_{[Z_a,Z_b]}\om(u;Z).
\eeq
The chord is oriented from $Z_a$ to $Z_b$ and carries the branch
induced from the triangle with vertices $0,Z_a,Z_b$.

There is a single branch of the integrand in the interior of this
triangle agreeing with both radial loadings. In particular, a branch
of $\log u$ exists in the angular sector between the two nearby rays.
Cauchy's theorem, after deleting small neighborhoods of the vertices,
gives
\beq\label{eq:if-chord-difference}
 V_{ab}=I_b^\si-I_a^\si.
\eeq
The omitted arc at $0$ contributes $O(\epsilon^{\al+1})$,
and each arc at $Z_a,Z_b$ contributes $O(\epsilon^{b_m})$.
They tend to zero because $\al>-1$ and $b_m>0$.
The same argument in the final configuration gives
\beq\label{eq:if-final-chord}
 V_{ab}'=I_b'-I_a'.
\eeq

\medskip\noindent
\textbf{2. The branch multiplier, with its sign fixed.}
Put
\beq
 w=Z_a/Z_b,\qquad \dl=Z_a-Z_b,
 \qquad u=Z_b+(1-s)\dl,\quad 0<s<1.
\eeq
On the moving chord,
\beq\label{eq:if-local-factors}
 1-\frac{u}{Z_a}=s(1-w^{-1}),\qquad
 1-\frac{u}{Z_b}=(1-s)(1-w).
\eeq
The positive factors $s,1-s$ acquire no phases. Thus we need only
track the logarithms of
\beq
 f_1(w)=1-w^{-1},\qquad f_2(w)=1-w.
\eeq
Initially $w=e^{-\mathrm{i}\eta}$; finally
$w=e^{\mathrm{i}\eta}$.
For $0<\rho<1$, $|w(\rho)|<1$, so
$\operatorname{Re}f_2(w(\rho))>0$.
Its continued logarithm is the canonical final logarithm.

The function $f_1(w(\rho))$ crosses the negative real axis exactly
once, from below to above: at $\rho=1/2$,
$w=e^{-\eta}$ and $f_1=1-e^\eta<0$.
For $\eta$ sufficiently small, the imaginary part of $w$ changes
sign only at this point, and
$\operatorname{Im}(1-w^{-1})=\operatorname{Im}(w)/|w|^2$.
The continuously chosen argument therefore decreases through $-\pi$.
Relative to the canonical final logarithm, we obtain exactly
\beq\label{eq:if-log-comparison}
 \begin{aligned}
 \bigl[\log f_1\bigr]_{\mathrm{cont}}
   &=\operatorname{Log}f_1(e^{\mathrm{i}\eta})-2\pi\mathrm{i},\\
 \bigl[\log f_2\bigr]_{\mathrm{cont}}
   &=\operatorname{Log}f_2(e^{\mathrm{i}\eta}).
 \end{aligned}
\eeq
Hence the continued product of the two factors in
\eqref{eq:if-local-factors}, each raised to exponent $b_m$,
is $e^{-2\pi\mathrm{i}b_m}=t$ times its canonical final value.

No other factor contributes an additional multiplier.
The moving chord lies in $D$, where $u^\al$ has a single-valued
branch agreeing with its prescribed endpoint branches.
For $j\ne a,b$, the factors $(1-u/Z_j)^{b_m}$ retain their canonical
branches: the interior of every moving chord lies in $|u|<1$,
where $\operatorname{Re}(1-u/Z_j)>0$.
The prefactor $K(Z)$ also retains its prescribed logarithms.
The rational vector factor and $du$ are single-valued.

After the fixed parametrization by $s$, the endpoint behavior is
bounded by a constant times $s^{b_m-1}(1-s)^{b_m-1}$ along the
compact continuation path. Thus the chord integral admits the
claimed analytic continuation, and the branch comparison gives
\beq\label{eq:if-chord-transport}
 \operatorname{Cont}_{\gm_{\mathrm{in}}}V_{ab}=tV_{ab}'.
\eeq
This also proves directly that the multiplier is independent of
$\al$, and hence of $\mu$.

\medskip\noindent
\textbf{3. The radial intervals.}
The moving segment $[0,Z_a(\rho)]$ meets no fixed marked point.
For $k\ne a$ and $0<x<1$,
\beq
 \operatorname{Re}\left(1-\frac{xZ_a(\rho)}{Z_k}\right)>0.
\eeq
Its loading therefore arrives at the canonical final loading.
The factors $u^\al$, $(1-x)^{b_m}$, and $K(Z)$ agree with their
prescribed final branches as well. Hence
\beq
 \operatorname{Cont}_{\gm_{\mathrm{in}}}I_a^\si=I_a'.
\eeq
For $j\ne a,b$, keep the radial segment $[0,Z_j]$ fixed.
The disk $D$ misses this segment, so the changing factor
$1-xZ_j/Z_a(\rho)$ is nonzero on the entire parameter rectangle
$0\le x,\rho\le1$. Its continuous logarithm normalized by value
$0$ at $x=0$ ends at the canonical final logarithm.
All other factors keep their branches. Thus these integrals
continue to $I_j'$.

Finally, use \eqref{eq:if-chord-difference},
\eqref{eq:if-final-chord}, and \eqref{eq:if-chord-transport}:
\beq
 \begin{aligned}
 \operatorname{Cont}_{\gm_{\mathrm{in}}}I_b^\si
 &=I_a'+tV_{ab}'\\
 &=I_a'+t(I_b'-I_a')\\
 &=(1-t)I_a'+tI_b'.
 \end{aligned}
\eeq
The three component formulas prove \eqref{eq:if-topology}.
\end{proof}

\subsubsection{Completion of the proof}

\begin{proof}[Proof of Theorem~\ref{thm:if-main}]
For a cover $\si\lessdot_R\si'$, the matrix $C_\si$ is independent
of $z$. Lemmas~\ref{lem:if-algebra} and~\ref{lem:if-topology} give
\beq
 \begin{aligned}
 \operatorname{Cont}_{\gm_{\si,\si'}}B_\si
 &=\left(\operatorname{Cont}_{\gm_{\si,\si'}}\mc I_\si\right)C_\si\\
 &=\mc I_{\si'}T_{ab}C_\si
  =\mc I_{\si'}C_{\si'}=B_{\si'}.
 \end{aligned}
\eeq
These are identities of solution germs. The loaded integrals vary
holomorphically near the endpoint configurations, and their prescribed
branches continue throughout each angular chamber. After an exchange, points may be moved to another configuration
of the same chamber before the next exchange: its angles remain
in the same order throughout. Analytic continuation within a chamber
has no monodromy. Replacing an exchange by a homotopic path in the
nonsingular configuration space also preserves the equality.

For a directed chain from $\si$ to $\tau$, apply this cover identity
successively. At each step the continued matrix is precisely the
distinguished matrix of the next chamber. The final matrix is
therefore $B_\tau$, regardless of the chosen chain. This proves
\eqref{eq:if-main} for all comparable permutations.

Reversing one of the chosen paths gives the inverse identification
and carries $B_{\si'}$ back to $B_\si$. Extend each column of
$B_\si$ along every incident connecting path. The cover identities
ensure agreement with the prescribed columns in the neighboring
chambers. The interiors of the connecting paths are disjoint, and
the matrices are already single-valued along each contractible
chamber. These sections consequently give a global flat frame on
$X$. Equivalently, every edge identification has the identity matrix
in the distinguished bases, so the product around every closed edge
path is the identity.
\end{proof}

\subsection{Topology of \texorpdfstring{$X$}{X} and the permutohedron}
\label{subsec:if-topology}

The angular chambers are contractible. In $X$ they are joined by
one prescribed path for each pair of neighboring chambers. Thus
$X$ has the homotopy type of the graph $G_n$. We shall prove that
the inclusion of $X$ into the larger configuration space $\mc U_m$
can be continuously deformed to a constant map.

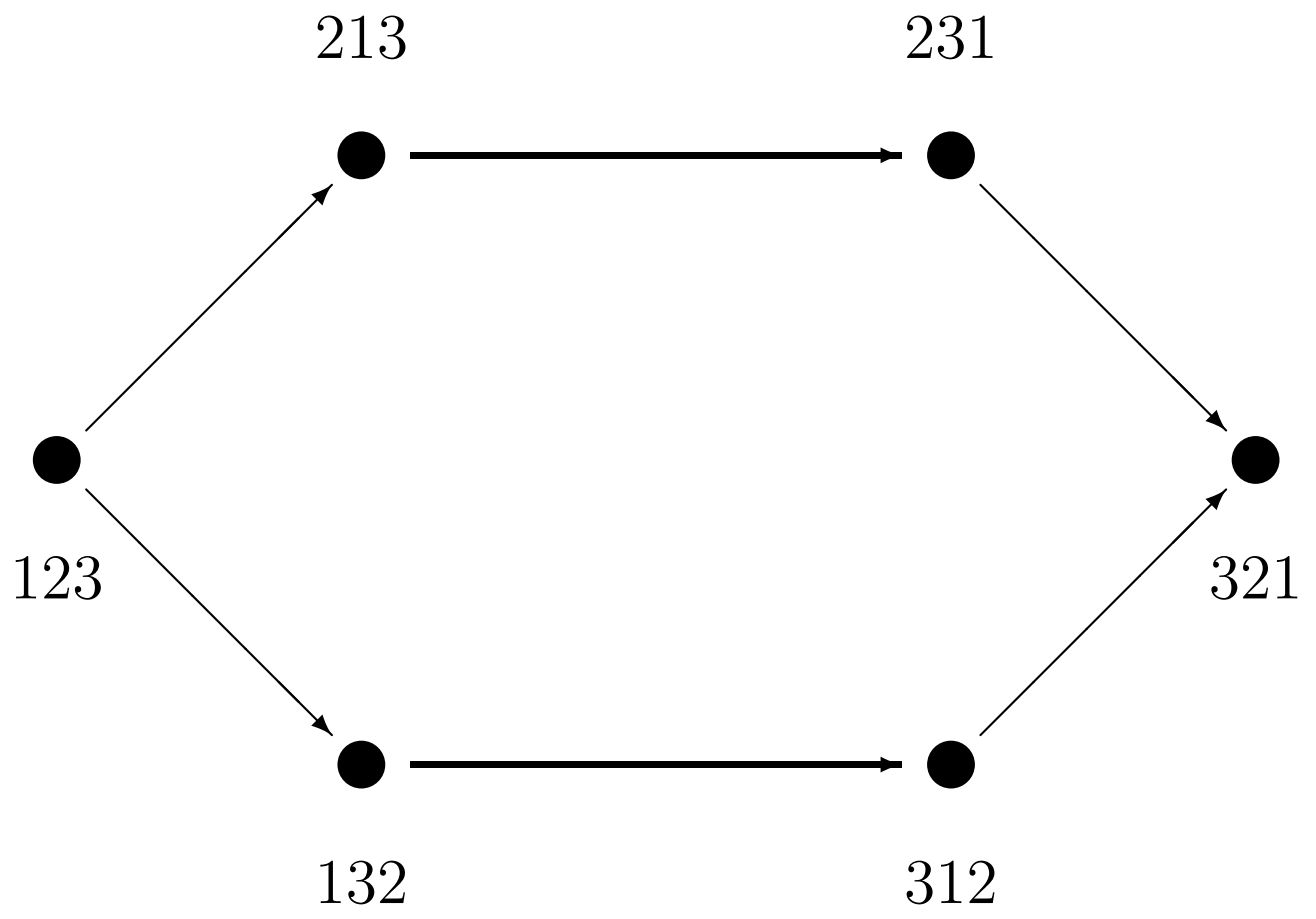
\begin{figure}[htbp]
\centering
\begingroup
\setlength{\unitlength}{0.65mm}
\begin{picture}(158,93)
 \linethickness{1.3pt}
 \put(20,48){\vector(1,1){25}}
 \put(53,76){\vector(1,0){50}}
 \put(111,73){\vector(1,-1){25}}
 \put(20,42){\vector(1,-1){25}}
 \put(53,14){\vector(1,0){50}}
 \put(111,17){\vector(1,1){25}}
 \linethickness{0.9pt}
 \put(17,45){\circle*{5.0}}
 \put(48,76){\circle*{5.0}}
 \put(108,76){\circle*{5.0}}
 \put(139,45){\circle*{5.0}}
 \put(48,14){\circle*{5.0}}
 \put(108,14){\circle*{5.0}}
 \put(17,33){\makebox(0,0){$123$}}
 \put(48,88){\makebox(0,0){$213$}}
 \put(108,88){\makebox(0,0){$231$}}
 \put(139,33){\makebox(0,0){$321$}}
 \put(48,2){\makebox(0,0){$132$}}
 \put(108,2){\makebox(0,0){$312$}}
\end{picture}
\endgroup
\caption{The oriented graph $G_3$. Each vertex represents an
angular chamber, and each arrow is a prescribed inward exchange.}
\label{fig:if-hexagon}
\end{figure}

\begin{example}[the graph for $n=3$]
The six angular chambers correspond to the six vertices of
Figure~\ref{fig:if-hexagon}. Each arrow indicates a prescribed
inward exchange: two consecutive labels $a<b$ exchange places
by moving $z_a$ inside $z_b$. The two directed paths from $123$
to $321$ form the two sides of the hexagon.
\end{example}

This loop can be contracted in the larger configuration space.

For general $n$, the graph $G_n$ can be identified with the
one-skeleton of the permutohedron $\Pi_n$. One realization of
$\Pi_n$ is the convex hull of the points
\beq\label{eq:if-permutohedron}
 \mathbf p_\si=(\si^{-1}(1),\ldots,\si^{-1}(n)),
 \qquad \si\in S_n.
\eeq
See \cite[Section~1.1]{KZ16}.

\begin{prop}[Contraction in the configuration space]
\label{prop:if-topology-X}
The inclusion $\iota:X\longrightarrow\mc U_m$ is homotopic to
a constant map. In particular,
\beq\label{eq:if-pi1-image}
 \iota_*\bigl(\pi_1(X)\bigr)=\{1\}.
\eeq
\end{prop}

\begin{proof}
We describe one continuous deformation of the \emph{whole space}
$X$. At each stage the deformation acts on every configuration in
$X$, including all points along every connecting path. The
deformation takes place in $\mc U_m$; it is not required to stay
inside $X$.

The observation behind the proof is simple. In an angular
chamber no two points have the same argument. On a connecting
inward path, if two points have the same argument, then they are
the two points being exchanged. The point with the smaller
numerical label lies closer to the origin, because it is the one
moved inward. Thus, in every configuration belonging to $X$,
\emph{two points with the same argument are ordered by their
numerical labels, with the smaller label closer to the origin}.

First, keep the arguments of all $z_j$ fixed and change their
distances from the origin continuously, until the distance of
$z_j$ is $j/(n+1)$. Change each distance \emph{linearly} from its
initial value to this final value. No collision can occur:
points having different arguments remain separated; points
having equal arguments begin and end with their distances in the
same strict order, which a linear interpolation preserves.
In particular, for $n=3$ we move $z_1,z_2,z_3$ to the three
different radii $1/4,1/2,3/4$, respectively. The labels
$1,2,3$ here refer to the original marked points
$z_1,z_2,z_3$, not to vertices of the graph $G_3$.

Second, keep these different radii fixed and move the arguments
of all points continuously to the common value $\pi/m$.
The points cannot meet during this step, because their radii
are different. At the end we reach the same configuration
$z_j=\frac{j}{n+1}e^{\mathrm{i}\pi/m}$, independently of the
initial configuration in $X$.

Throughout the deformation the arguments remain in the sector
$0<\arg z_j<2\pi/m$. In this sector, if $z_a^m=z_b^m$, then
the two points have the same argument and the same radius.
The preceding argument excludes this possibility, so the
entire deformation stays in $\mc U_m$. It ends at a fixed
configuration, proving that $\iota$ is homotopic to a constant
map. The contraction is performed in $\mc U_m$, not in $X$.
\end{proof}

The proposition explains why a flat connection on $\mc U_m$
has trivial monodromy when restricted to $X$. The additional
content of Theorem~\ref{thm:if-main} is that the particular
matrices $B_\si$ constructed as root-of-unity limits agree under
the prescribed continuations.

\clearpage
\section{Functorial compatibility of root-of-unity limits}
\label{sec:admissible-composition}
\label{sec:functorial-compatibility}

The preceding sections describe the limiting connections and their
distinguished hypergeometric bases at roots of unity. We now collect
their compatibility properties. Recall that $\Psi(z;p,q,\varkappa)$
is the $n\times n$ coordinate matrix of the qKZ circle-integral
solutions $\Psi_1,\ldots,\Psi_n$ in the basis
$v_1,\ldots,v_n$; see \eqref{eq:hyp-Psi}--\eqref{eq:hyp-matrix}
in Section~\ref{sec:hyper-qkz}. Multiplication of root orders is
reflected in coordinatewise power maps and rescaling of $\ka$;
the right weak Bruhat order describes the inward continuations
identifying the distinguished bases in different angular chambers.

\subsection{Admissible chambers and power maps}
\label{subsec:functorial-chambers}

For the hypergeometric solutions, fix $n\ge2$ and write
\beq\label{eq:common-real-domain}
 \mc A_n=\left\{(\ka,\mu)\in\R^2:\
              \ka<-n,\quad 0<\mu<\frac18\right\}.
\eeq
This is the parameter domain of Theorem~\ref{thm:hyper-limit-all-chambers}
for every positive integer $m$. We use the angular chambers
$\mc D_{m,\si}$ from \eqref{eq:hyp-limit-chamber-sigma},
\[
 \mc D_{m,\si}=\left\{z_a=e^{\mathrm{i}\vartheta_a}:\
 0<\vartheta_{\si(1)}<\cdots<\vartheta_{\si(n)}
 <\frac{2\pi}{m}\right\},
\]
and put
\beq\label{eq:admissible-domain}
 \mc D_m=\mc D_{m,\mathrm{id}}.
\eeq
Recall that $\pi_b(z)=(z_1^b,\ldots,z_n^b)$.

\begin{prop}[Chambers under power maps]
\label{prop:admissible-domains}
For every $\si\in S_n$, the chambers satisfy
\beq\label{eq:nested-domains-sigma}
 \mc D_{1,\si}\supset\mc D_{2,\si}\supset\cdots.
\eeq
For positive integers $a,b$, the power map restricts to a
real-analytic bijection
\beq\label{eq:domain-power-map-sigma}
 \pi_b:\mc D_{ab,\si}\longrightarrow\mc D_{a,\si}.
\eeq
For positive integers $m_1,\ldots,m_d$, with
$M=\max(m_1,\ldots,m_d)$, one has
\beq\label{eq:domain-finite-intersection-sigma}
 \bigcap_{h=1}^d\mc D_{m_h,\si}=\mc D_{M,\si}\ne\varnothing.
\eeq
\end{prop}

\begin{proof}
The chamber inequalities have upper bound $2\pi/m$, which gives
both the inclusions and the intersection formula. Multiplying all
arguments by $b$ carries the inequalities for $\mc D_{ab,\si}$
to those for $\mc D_{a,\si}$. Conversely, dividing those arguments
by $b$ selects a unique coordinatewise $b$th root in
$\mc D_{ab,\si}$. Both maps preserve the angular order $\si$.
\end{proof}

In particular, $\pi_m$ identifies $\mc D_{m,\si}$ with
$\mc D_{1,\si}$. The proposition also specifies a common chamber
on which to compare any finite collection of root-of-unity limits.

\subsection{Functoriality of the limiting connections}
\label{subsec:functorial-connections}

For connections, we use the full complex parameter range:
\[
 \La_1,\ldots,\La_n,\mu\in\C,\qquad \ka\in\C^\times.
\]
The following identity holds on every fixed-degree space of
Sections~\ref{sec:connections}--\ref{sec:connection-limits}.

\begin{prop}[Composition of limiting connections]
\label{prop:connection-composition}
For all positive integers $a,b$, on $\mc U_{ab}$ one has
\beq\label{eq:connection-composition}
 \boxed{\quad
 \nabla_{ab}(\ka,\mu)
 =\pi_b^*\nabla_a(b\ka,\mu)
 =\pi_a^*\nabla_b(a\ka,\mu).
 \quad}
\eeq
\end{prop}

\begin{proof}
Theorem~\ref{thm:limit} and $\pi_a\circ\pi_b=\pi_{ab}$ give
\[
 \pi_b^*\nabla_a(b\ka,\mu)
 =\pi_b^*\pi_a^*\nabla_1(ab\ka,\mu)
 =\pi_{ab}^*\nabla_1(ab\ka,\mu)
 =\nabla_{ab}(\ka,\mu).
\]
Interchanging $a$ and $b$ gives the second equality. The power
maps here are restricted to the corresponding nonsingular
configuration spaces.
\end{proof}

The same composition identity holds for the algebraic limiting
connections over fields of characteristic zero, as established in
\eqref{eq:formal-composition}.

\subsection{Distinguished bases at roots of unity}
\label{subsec:functorial-solutions}

We return to the degree-one solutions with
$\La_1=\cdots=\La_n=1$ and $(\ka,\mu)\in\mc A_n$.
To compare different root orders, we display the parameters in the
notation of Section~\ref{sec:inward-formal}: write
$B_{m,\si}(z;\ka,\mu)$ for the distinguished matrix denoted there
by $B_\si$ when $m,\ka,\mu$ are fixed. Also write
$B_m=B_{m,\mathrm{id}}$ for the canonical chamber.

For $0<r<1$, set
\beq\label{eq:common-radial-parameters}
 q_\ka(r)=r^{-1/(2\ka)},\qquad
 \varkappa_{\ka,\mu}(r)=r^{-2\mu/\ka}.
\eeq
For each primitive $m$th root $\zeta_m$ and each
$z\in\mc D_{m,\si}$, Theorem~\ref{thm:hyper-limit-all-chambers}
gives
\beq\label{eq:Lm-formula-sigma}
 \begin{aligned}
 B_{m,\si}(z;\ka,\mu)
 &=\lim_{r\to1^-}
   \Psi\bigl(z;\zeta_m r,q_\ka(r),\varkappa_{\ka,\mu}(r)\bigr)\\
 &=\mc I_\si(z^m;m\ka,\mu)\,
                  C_{1,\si}^{(n)}(m\ka,\mu).
 \end{aligned}
\eeq
The convergence is locally uniform in the chamber and on the
small complex neighborhoods supplied by that theorem. For every
parameter pair in $\mc A_n$, the matrix $B_{m,\si}$ is a
fundamental matrix of flat sections of $\nabla_m(\ka,\mu)$.
The powers in the radial integrals use the logarithms
$\log(z_j^m)=\mathrm{i}m\vartheta_j$ prescribed in
\eqref{eq:hyp-limit-loadings-sigma}.

\begin{lem}[Independence of the primitive root]
\label{lem:primitive-independence}
For fixed $m$, $(\ka,\mu)\in\mc A_n$, and
$z\in\mc D_{m,\si}$, the limit in
\eqref{eq:Lm-formula-sigma} is the same for every primitive
$m$th root of unity $\zeta_m$.
\end{lem}

\begin{proof}
The expression
$\mc I_\si(z^m;m\ka,\mu)C_{1,\si}^{(n)}(m\ka,\mu)$
depends on $m,\ka,\mu,z$ and the prescribed branches. It is the
limit for every choice of $\zeta_m$.
\end{proof}

\begin{cor}[Composition of distinguished bases]
\label{cor:solution-composition}
Let $a,b$ be positive integers and $(\ka,\mu)\in\mc A_n$.
For $\si\in S_n$ and $z\in\mc D_{ab,\si}$,
\beq\label{eq:solution-composition-sigma}
 \boxed{\quad
 B_{ab,\si}(z;\ka,\mu)
 =B_{a,\si}(z^b;b\ka,\mu)
 =B_{b,\si}(z^a;a\ka,\mu).
 \quad}
\eeq
In particular,
\beq\label{eq:Lm-to-L1-sigma}
 B_{m,\si}(z;\ka,\mu)=B_{1,\si}(z^m;m\ka,\mu),
 \qquad z\in\mc D_{m,\si}.
\eeq
\end{cor}

\begin{proof}
By Proposition~\ref{prop:admissible-domains},
$z^b\in\mc D_{a,\si}$ and $z^a\in\mc D_{b,\si}$.
The parameter pairs $(b\ka,\mu)$ and $(a\ka,\mu)$ belong to
$\mc A_n$. Formula \eqref{eq:Lm-formula-sigma} expresses all
three matrices as
\[
 \mc I_\si(z^{ab};ab\ka,\mu)\,
                    C_{1,\si}^{(n)}(ab\ka,\mu).
\]
The prescribed logarithm of the $j$th coordinate of $z^{ab}$
is $\mathrm{i}ab\vartheta_j$ in all three expressions.
The other powers in each radial integral are continued from
their value $1$ at its initial endpoint. Thus both the integrals
and their normalizations agree. Taking $a=1$, $b=m$ gives
\eqref{eq:Lm-to-L1-sigma}.
\end{proof}

\begin{cor}[Limits at finitely many roots of unity]
\label{cor:finite-roots}
Fix $(\ka,\mu)\in\mc A_n$, $\si\in S_n$, and positive integers
$m_1,\ldots,m_d$. Put $M=\max(m_1,\ldots,m_d)$ and choose a
primitive $m_h$th root $\zeta_{m_h}$ for each $h$.
For $z\in\mc D_{M,\si}$ define
\beq\label{eq:finite-roots-matrices}
 \Psi_h(r;z)=
 \Psi\bigl(z;\zeta_{m_h}r,q_\ka(r),\varkappa_{\ka,\mu}(r)\bigr),
 \qquad 1\le h\le d.
\eeq
Then all these matrices have limits, and
\beq\label{eq:finite-roots-sigma}
 \lim_{r\to1^-}\Psi_h(r;z)
 =B_{m_h,\si}(z;\ka,\mu)
 =B_{1,\si}(z^{m_h};m_h\ka,\mu),
 \qquad 1\le h\le d.
\eeq
Each limiting matrix is fundamental throughout the stated
parameter domain. The convergence is locally uniform on the
common chamber.
\end{cor}

\begin{proof}
The inclusion
$\mc D_{M,\si}\subset\mc D_{m_h,\si}$ follows from
Proposition~\ref{prop:admissible-domains}. Apply
\eqref{eq:Lm-formula-sigma} and \eqref{eq:Lm-to-L1-sigma}
for each of the finitely many indices $h$.
\end{proof}

Thus a finite collection of roots can be approached with the same
configuration $z$ and the same functions $q_\ka(r)$ and
$\varkappa_{\ka,\mu}(r)$. The resulting bases are expressed by
the single family $B_{1,\si}$ evaluated at the corresponding
power-map variables and rescaled parameters. All formulas above
include the canonical case $\si=\mathrm{id}$.

\subsection{Compatibility with inward continuation}
\label{subsec:functorial-continuation}

By Theorem~\ref{thm:if-main}, the distinguished matrices are
identified by inward continuation along the right weak Bruhat
order. Explicitly, for $\si\le_R\tau$ and a directed chain of
prescribed exchanges represented by a path $\gm$,
\beq\label{eq:functorial-weak-continuation}
 \operatorname{Cont}_{\gm}B_{m,\si}(\,\cdot\,;\ka,\mu)
 =B_{m,\tau}(\,\cdot\,;\ka,\mu).
\eeq
The power maps preserve both this angular order and the
prescribed inward path classes.

\begin{cor}[Power maps and inward continuation]
\label{cor:continuation-power-compatibility}
Let $d,e\ge1$, $(\ka,\mu)\in\mc A_n$, and $\si\le_R\tau$.
Let $\gm$ be a path from $\mc D_{de,\si}$ to
$\mc D_{de,\tau}$ obtained by following a directed chain of
prescribed inward exchanges, with connecting paths in its
successive chambers. Then $\pi_e\circ\gm$ follows the same
chain of angular orders at level $d$, and
\beq\label{eq:continuation-power-compatibility}
 \begin{aligned}
 \operatorname{Cont}_{\gm}B_{de,\si}(\,\cdot\,;\ka,\mu)
 &=\left(
   \operatorname{Cont}_{\pi_e\circ\gm}
      B_{d,\si}(\,\cdot\,;e\ka,\mu)
   \right)\circ\pi_e\\
 &=B_{de,\tau}(\,\cdot\,;\ka,\mu).
 \end{aligned}
\eeq
These are equalities of holomorphic solution germs in the final
chamber. The same assertion holds with $d,e$ interchanged.
\end{cor}

\begin{proof}
First consider one exchange of neighboring labels $a<b$.
In the local path \eqref{eq:if-path} at level $de$,
\[
 z_a=z_b\exp\bigl(\xi/(de)\bigr).
\]
Raising the coordinates to the $e$th power gives
\[
 z_a^e=z_b^e\exp\bigl(\xi/d\bigr),
\]
which is the prescribed inward path at level $d$, with the
same orientation. Proposition~\ref{prop:admissible-domains}
identifies the initial and final chambers. It also sends each
connecting path inside a chamber to a path inside the
corresponding chamber. This proves the path assertion for
every directed chain.

By \eqref{eq:solution-composition-sigma}, the initial solution
germs satisfy
\[
 B_{de,\si}(\,\cdot\,;\ka,\mu)
 =B_{d,\si}(\,\cdot\,;e\ka,\mu)\circ\pi_e.
\]
Continue this identity along the two paths and apply
Theorem~\ref{thm:if-main}. The composition identity in the
final chamber then gives
\eqref{eq:continuation-power-compatibility}.
\end{proof}

For each fixed $m$, Section~\ref{sec:inward-formal} constructs
a space $X$ from the angular chambers and the chosen connecting
arcs. Theorem~\ref{thm:if-main} identifies the matrices
$B_{m,\si}$ as a global flat frame over this space.
Corollary~\ref{cor:continuation-power-compatibility} shows that
the continuation identifications of these distinguished frames
are compatible with the power-map identities.

\subsection{Power-map compatibility of the limiting connections and solutions}
\label{subsec:habiro-type}

The results can be summarized by two identities:
\beq\label{eq:functorial-summary}
 \boxed{\quad
 \begin{aligned}
 \nabla_m(\ka,\mu)&=\pi_m^*\nabla_1(m\ka,\mu),\\
 B_{m,\si}(z;\ka,\mu)&=B_{1,\si}(z^m;m\ka,\mu).
 \end{aligned}
 \quad}
\eeq
The first holds for arbitrary complex weights and
$\ka\ne0$, $\mu\in\C$, on every fixed-degree space over
$\mc U_m$. The second holds for the degree-one solutions with
$\La_1=\cdots=\La_n=1$, $(\ka,\mu)\in\mc A_n$, and
$z\in\mc D_{m,\si}$, with the branches fixed in Section~\ref{sec:hyper-limit}.
It identifies the complete normalized fundamental matrices
throughout this parameter domain.

The transformations of variables and parameters
\beq\label{eq:power-parameter-functoriality}
 F_m:(z,\ka,\mu)\longmapsto(z^m,m\ka,\mu)
 \qquad\text{satisfy}\qquad
 F_a\circ F_b=F_{ab}=F_b\circ F_a.
\eeq
Proposition~\ref{prop:connection-composition} and
Corollary~\ref{cor:solution-composition} express this composition
law for the limiting connections and distinguished bases.
Lemma~\ref{lem:primitive-independence} identifies the limits
at all primitive roots of a fixed order.
Corollary~\ref{cor:finite-roots} places any finite collection
of root orders on a common angular chamber. Finally,
Theorem~\ref{thm:if-main} and
Corollary~\ref{cor:continuation-power-compatibility} identify
the bases along the right weak Bruhat order in a manner
compatible with these power maps.

\begin{rem}[A comparison of qKZ fundamental matrices]
Fix $m\ge1$, $(\ka,\mu)\in\mc A_n$, $\si\in S_n$,
$z\in\mc D_{m,\si}$, and a primitive $m$th root $\zeta_m$.
For $0<r<1$, put $p_r=\zeta_m r$, $q_r=q_\ka(r)$, and
$\varkappa_r=\varkappa_{\ka,\mu}(r)$. For $r$ sufficiently
close to $1$, define
\[
 \mc R_m(r;z)=
 \Psi(z;p_r,q_r,\varkappa_r)\,
 \Psi(z^m;p_r^{m^2},q_r^m,\varkappa_r^m)^{-1}.
\]
Then, locally uniformly for $z\in\mc D_{m,\si}$,
\[
 \lim_{r\to1^-}\mc R_m(r;z)=\id.
\]
Indeed, with $s=r^{m^2}$ one has
$q_r^m=s^{-1/(2m\ka)}$ and
$\varkappa_r^m=s^{-2\mu/(m\ka)}$; hence both
matrices in the quotient tend to the same invertible matrix
$B_{1,\si}(z^m;m\ka,\mu)$, by
\eqref{eq:Lm-formula-sigma} and \eqref{eq:Lm-to-L1-sigma}.
The power $p_r^{m^2}$ arises because $m$ shifts
$z_i\mapsto p_rz_i$, followed by $Z_i=z_i^m$,
give $Z_i\mapsto p_r^{m^2}Z_i$.
\end{rem}

Similar power-change phenomena in root-of-unity expansions
of $q$-hypergeometric functions were observed in \cite{GW25}.
In the present setting, the identities
\eqref{eq:functorial-summary} describe the compatibility of
limiting trigonometric KZ connections and their distinguished
normalized bases of flat sections. The coefficient matrices
$C_{1,\si}^{(n)}(m\ka,\mu)$ in
\eqref{eq:Lm-formula-sigma} give their exact normalization in
the radial KZ bases. Together with
\eqref{eq:continuation-power-compatibility}, the power-map
identities and inward continuations organize these limits into
one compatible family of differential connections with
distinguished hypergeometric bases.

\section*{Acknowledgments}
This paper was developed through a series of conversations with GPT-5.6 Sol. 
The author formulated the main questions, proposed examples leading to the discovery of general results, and originated the statements of Theorems \ref{thm:limit}, \ref{thm:hyper-limit-all-chambers}, and \ref{prop:if-weak-coherence} together with the key ideas of their proofs. Detailed proofs were developed with the assistance of the AI and subsequently verified and edited by the author. The interaction with the AI was similar in several respects to collaboration with a coauthor. Nevertheless, the author bears sole responsibility for the content of the paper, including all statements, proofs, and conclusions.
\smallskip

This work was supported in part by Simons Foundation grant TSM-00012774.
The author thanks Vitaly Tarasov for useful discussions.


\begin{thebibliography}{TVInv97}
\bibitem[BB05]{BB05}
A. Bj\"orner and F. Brenti,
\textit{Combinatorics of Coxeter Groups},
Graduate Texts in Mathematics, vol.~231,
Springer, New York, 2005, Section~3.1.

\bibitem[FR92]{FR92}
I. B. Frenkel and N. Yu. Reshetikhin,
\textit{Quantum affine algebras and holonomic difference equations},
Comm. Math. Phys. \textbf{146} (1992), 1--60.

\bibitem[GW25]{GW25}
S. Garoufalidis and C. Wheeler,
\textit{Explicit classes in Habiro cohomology},
2025, \texttt{arXiv:2505.19885v1}.

\bibitem[GR04]{GR04}
G. Gasper and M. Rahman,
\textit{Basic Hypergeometric Series}, second edition,
Encyclopedia of Mathematics and its Applications, vol.~96,
Cambridge University Press, Cambridge, 2004.

\bibitem[KZ16]{KZ16}
R. M. Kaufmann and Y. Zhang,
\textit{Permutohedral structures on $E_2$-operads},
Forum Math. \textbf{29} (2017), no.~6, 1371--1411,
\url{https://doi.org/10.1515/forum-2016-0052}.

\bibitem[MaV02]{MaV02}
Y. Markov and A. Varchenko,
\textit{Hypergeometric solutions of trigonometric KZ equations satisfy
dynamical difference equations}, Adv. Math. \textbf{166} (2002), no.~1,
100--147.

\bibitem[MV24]{MV24}
E. Mukhin and A. Varchenko,
\textit{Solutions of the $\mathfrak{sl}_2$ qKZ
equations modulo an integer},
J. London Math. Soc. \textbf{109} (2024),
no.~4, e12884.

\bibitem[DLMF]{DLMF}
NIST Digital Library of Mathematical Functions,
\textit{Section 5.18: $q$-Gamma and $q$-Beta Functions},
\url{https://dlmf.nist.gov/5.18}.

\bibitem[TVInv97]{TVInv97}
V. Tarasov and A. Varchenko,
\textit{Geometry of $q$-hypergeometric functions
as a bridge between Yangians and quantum
affine algebras},
Invent. Math. \textbf{128} (1997), 501--588.

\bibitem[TV97]{TV97}
V. Tarasov and A. Varchenko,
\textit{Geometry of q-hypergeometric functions, quantum affine algebras and
elliptic quantum groups}, Ast\'erisque \textbf{246} (1997), 1--135.

\bibitem[TV00]{TV00}
V. Tarasov and A. Varchenko,
\textit{Difference equations compatible with trigonometric KZ differential
equations}, Internat. Math. Res. Notices (2000), no.~15,
801--829; \texttt{arXiv:math/0002132}.

\bibitem[TV26]{TV26}
V. Tarasov and A. Varchenko,
\textit{Monodromy eigenvectors for difference equations with
root-of-unity step},
2026, \texttt{arXiv:2609.17742v2}.

\bibitem[V89]{V89}
A. N. Varchenko,
\textit{The Euler beta-function, the Vandermonde determinant,
Legendre's equation, and critical values of linear functions on a
configuration of hyperplanes. I},
Izv. Akad. Nauk SSSR Ser. Mat. \textbf{53} (1989), no.~6,
1206--1235 (in Russian); English translation:
Math. USSR-Izv. \textbf{35} (1990), no.~3, 543--571.
\end{thebibliography}
\end{document}